\documentclass[runningheads]{llncs}

\usepackage[utf8]{inputenc}
\usepackage[style=authoryear-icomp,maxbibnames=4,style=numeric,backend=bibtex,giveninits=true]{biblatex}
\usepackage{lmodern}
\usepackage[final]{microtype}
\usepackage{ellipsis, mparhack, ragged2e} \usepackage[l2tabu, orthodox]{nag} \usepackage{amsmath,amsfonts,amssymb}
\usepackage{xcolor}

\usepackage{hyperref}

\usepackage{caption}
\usepackage{subcaption}

\usepackage{algorithm}
\usepackage[noend]{algpseudocode}

\usepackage{tikz}
\usetikzlibrary{positioning,backgrounds}

\usepackage{pgfplots}
\usepackage{booktabs}

\usepackage{cancel}

\usepackage{todonotes}

\usepackage{graphicx}
\usepackage{mathtools}
\usepackage{hyperref}
\usepackage[capitalise]{cleveref}
\usepackage[thinc]{esdiff}
\usepackage[section]{placeins}
\usepackage{enumitem}

\usepackage[misc,geometry]{ifsym}
\newcommand{\id}{\mathbf{I}}
\newcommand{\vecbeta}{\boldsymbol{\beta}}
\newcommand{\vecalpha}{\boldsymbol{\alpha}}

\newcommand{\dbtilde}[1]{\tilde{\raisebox{0pt}[0.85\height]{$\tilde{#1}$}}}

\newcommand{\R}{\mathbb{R}}

\newcommand{\inputspace}{\mathcal{X}}

\newcommand{\NN}{N}
\newcommand{\matr}[2]{#1^{(#2)}}
\newcommand{\weight}[3]{w^{(#1)}(#2, #3)}
\newcommand{\weighta}[3]{{\tilde w}^{(#1)}(#2, #3)}

\newcommand{\vecx}{\mathbf{x}}
\newcommand{\vecy}{\mathbf{y}}
\newcommand{\vecz}{\mathbf{z}}

\newcommand{\z}{z}
\newcommand{\vech}{\mathbf{h}}
\newcommand{\h}{h}
\newcommand{\vecb}{\mathbf{b}}

\newcommand{\bias}[2]{b^{(#1)}_{#2}}

\newcommand{\vecone}{\mathbf{1}}

\newcommand{\basis}[1]{B^{(#1)}}
\newcommand{\lcoeff}[3]{\alpha_{#1,#2}^{(#3)}}
\newcommand{\repl}[1]{I^{(#1)}}

\DeclareMathOperator{\Var}{Var}
\DeclareMathOperator{\Mean}{Mean}

\crefname{algocf}{alg.}{algs.}
\Crefname{algocf}{Algorithm}{Algorithms}

\def\linna{LiNNA}
\makeatletter
\renewcommand\section{\@startsection{section}{1}{\z@}	{-8\p@ \@plus -4\p@ \@minus -4\p@}	{6\p@ \@plus 4\p@ \@minus 4\p@}	{\normalfont\large\bfseries\boldmath
		\rightskip=\z@ \@plus 8em\pretolerance=10000 }}
\renewcommand\subsection{\@startsection{subsection}{2}{\z@}	{-8\p@ \@plus -4\p@ \@minus -4\p@}	{6\p@ \@plus 4\p@ \@minus 4\p@}	{\normalfont\normalsize\bfseries\boldmath
		\rightskip=\z@ \@plus 8em\pretolerance=10000 }}
\renewcommand\subsubsection{\@startsection{subsubsection}{3}{\z@}	{-4\p@ \@plus -4\p@ \@minus -4\p@}	{-1.5em \@plus -0.22em \@minus -0.1em}	{\normalfont\normalsize\bfseries\boldmath}}
\renewcommand{\paragraph}{	\@startsection{paragraph}{4}	{\z@}{0ex \@plus 1ex \@minus .2ex}{-1em}	{\normalfont\normalsize\bfseries}}
\makeatother

\begin{document}
		
\title{Syntactic vs Semantic Linear Abstraction and Refinement of Neural Networks
\thanks{This research was funded in part by the German Research Foundation (DFG) project 427755713 \emph{GoPro}, the German Federal Ministry of Education and Research (BMBF) within the project \emph{SEMECO Q1} (03ZU1210AG), and the DFG research training group \emph{ConVeY} (GRK 2428).}}

\titlerunning{Syntactic vs Semantic Abstraction and Refinement of Neural Networks}

\author{Calvin Chau\orcidID{0000-0002-3437-0240}\inst{1} \and
Jan K\v{r}et\'{i}nsk\'{y}\orcidID{0000-0002-8122-2881}\inst{2,3} \and
Stefanie Mohr\orcidID{0000-0002-8630-3218}\inst{2}}

\authorrunning{Chau et al.}

\institute{Technische Universität Dresden, Dresden, Germany \and
	Technical University of Munich, Munich, Germany \and
	Masaryk University, Brno, Czech Republic\\
	\email{\{mohr, kretinsky\}@in.tum.de}\\
	\email{calvin.chau@tu-dresden.de}}
	
	\maketitle              

\begin{abstract}
Abstraction is a key verification technique to improve scalability.
However, its use for neural networks is so far extremely limited. 
Previous approaches for abstracting classification networks replace several neurons with one of them that is similar enough. 
We can classify the similarity as defined either syntactically (using quantities on the connections between neurons) or semantically (on the activation values of neurons for various inputs).
Unfortunately, the previous approaches only achieve moderate reductions, when implemented at all. 
In this work, we provide a more flexible framework, where a neuron can be replaced with a \emph{linear combination} of other neurons, improving the reduction. 
We apply this approach both on syntactic and semantic abstractions, and implement and evaluate them experimentally. 
Further, we introduce a refinement method for our abstractions, allowing for finding a better balance between reduction and precision.
\end{abstract}

\section{Introduction}\label{sec:intro}
\subsubsection{Neural Network Abstractions}
Abstraction is a key instrument for understanding complex systems and analyzing complex problems across all disciplines, including computer science.
Abstraction of complex systems, such as neural networks (NN), results in smaller systems, which are not only producing equivalent outputs (such as in distillation~\cite{hinton2015distilling}), but additionally can be mapped to the original system, providing a strong link between the individual parts of the two systems.
Consequently, abstraction find various applications.
For instance, the smaller (abstract) networks are more understandable and the strong link between the behaviours of the abstract and the original network allows for better explainability of the original behaviour, too;
smaller networks are more efficient in resource usage during runtime;
smaller networks are easier to verify.
Again, with no formal link between the original network and, say, a distilled or pruned one, verifying the smaller one is of no use to verifying the original one. 
In contrast, for abstractions, the verification guarantee can be in principle transfered to the original network, be it via lifting a counterexample or a proof of correctness.

Altogether, abstractions of neural networks are a key concept worth investigating \emph{eo ipso}, subsequently offering various applications.
However, currently it is still very under-developed.
For defining an abstraction, we need a transformation linking the original neurons to those in the abstraction.
Equivalently, we need a notion of the \emph{similarity of neurons}, to identify a good representative of a group of neurons. 
The difficulty in contrast to, e.g., predicate abstraction of programs is that neurons have no inner structure such as values of variables stored in a state.
On the one hand, approaches based on bisimilarity \cite{Prabhakar22} offer a solution focusing on the \emph{``syntax''} of neurons: the weights of the incoming connections.
The quantities give rise to an equivalence akin to probabilistic bisimulation.
On the other hand, in search of a stronger tool, approaches such as \cite{deepabstract} try to identify \emph{``semantics''} of the neurons. 
For instance, given a vector of inputs to the network, the \emph{I/O semantics} of a neuron \cite{deepabstract} is the vector of activation values of this neuron obtained on these inputs.
This represents a finite-dimensional approximation of the actual semantics of a neuron as a computational device.
Either way, replacing several neurons with one that is very similar yields only moderate savings on size if the abstract network is supposed to be similar, i.e., yield mostly the same predictions and ensure a tight connection between the similar neurons.

\subsubsection{Our Contribution}
We focus on studying abstraction irrespective of the use case (verification, smaller networks, explainability), to establish a better principal understanding of this crucial, yet in this context underdeveloped technique.
First, we explore a richer abstraction scheme, where a group of neurons can be represented not only by a chosen neuron but also by a \emph{linear combination of neurons}.
Thus instead of keeping exactly one representative per group, we can ``reuse'' the chosen representatives in many linear combinations; in other words, the representatives can attain many roles, partially representing many groups, which reduces their required count.
We provide several algorithms to do so, ranging from resource-intensive algorithms aiming to show the limits of the approach to efficient heuristics approximating the former ones quite closely.
We apply these algorithms to the semantic approach of \cite{deepabstract} as well as to the syntactic, bisimulation-like approach similar to \cite{Prabhakar22} not implemented previously.
Experimental results confirm the \emph{greater power of this linear-combination} approach; further, they provide insight into the \emph{advantages of semantic similarity over the syntactic} one, pointing out the more advantageous future research directions.

Further, we provide a formal link between the concrete and abstract neurons by proving an error bound induced by the abstraction, showing the abstraction is valid and (approximately) simulates the original network.
We show the bound is better than the one based on bisimulation.
While still not very practical, the experiments show that even on unseen data, the error is always closely bounded by the error on the data used for generating the abstraction, and mostly even a lot smaller.
This empirical version of the concept of error could thus enable the transfer of reasoning about the abstraction to the original network in a yet much tighter way.

In addition, we suggest \emph{abstraction-refinement} procedures to better fine-tune the trade-off between the precision and the size of the abstraction.
The experiments reveal that a more aggressive abstraction followed by a refinement provides better results than a direct, moderate abstraction.
Hence involving our refinement in the abstraction process improves the resulting quality, opening new lines of attack on efficient neural network abstractions.

\subsubsection{Summary}  Our contribution can be summarized as follows:
\begin{itemize}
	\item We define abstractions of neural networks with (approximate) equivalences being linear equations over semantics of neurons. 
	We provide a theoretical bound on the induced error, see Thm.~\ref{th:bisim-error}.
	We reflect this idea also on the syntactic, bisimulation-based abstraction.
	\item We implement both approaches and compare them mutually as well as to their previous, special cases with equivalences being (approximate) identities. 
	We perform the experiments on a number of standard benchmarks, such as MNIST, CIFAR, or FashionMNIST, concluding advantages of semantic over syntactic approaches and of linear over identity-based ones.
	\item We introduce an abstraction-refinement procedure and also evaluate its benefits experimentally.
\end{itemize}

\subsubsection{Related Work}
There are various approaches for verification of NN, however, we are \textbf{not} presenting another verifier. Instead, we introduce an approach that is \textbf{orthogonal to verification} and could be integrated with an existing verifier. 
Therefore, we do not compare our approach to any verification tool and refer the interested reader to the Verification of Neural Networks Competition \cite{DBLP:journals/corr/abs-2301-05815} for an overview of existing approaches \cite{marabou, deeppoly, xu2021fast, wang2021beta}.

Network compression techniques share many similarities with abstraction \cite{cheng2017survey} and either focus on reducing the memory footprint \cite{xue2013restructuring, huang2017densely} or computation time of the model \cite{gong2014compressing}, but in contrast, do not provide any formal relation to the original network, rendering them inappropriate for understanding redundancies or verification.
Knowledge distillation is a prominent technique, which can reduce networks by a significant amount, but completely loses any connection to the original network \cite{hinton2015distilling}, and can thus not be used in verification.
There is some progress in using abstract domains for scalable verification, like \cite{deeppoly, DBLP:conf/iclr/SinghGPV19, tran2021robustness}, but they do not produce an abstracted NN for verification. Instead, they apply abstraction only tightly entangled together with the verification algorithm. These approaches also try to generate a more scalable verification, however, the key difference is that they do not return an actual abstracted network that could be reused or manually inspected.
Katz et al.~\cite{katz_abstraction} introduce an abstraction scheme for NN, in which they decompose neurons into several parts, before merging them again to obtain an over-approximation of the original network. 
However, their approach is limited to networks with one output neuron. For networks with more output neurons, the property to be verified needs to be baked into the network, making the approach significantly less flexible. Additionally, this tight entanglement of specification and neural network does not allow for retrieving the abstraction later and reusing it for anything else than to verify that specific property. This strongly contrasts our generic and usage-agnostic abstraction and their property-restricted abstractions. 

Some other works use abstraction after representing a neural network as an interval neural network \cite{DBLP:conf/nips/PrabhakarA19}, or more generally, by using more complex abstract domains \cite{sotoudeh2020abstract}. 
While theoretically interesting, the practicality of these works has not been investigated.
There are two approaches that we consider to be the closest to our work: a bisimulation-based approach \cite{Prabhakar22}, and \emph{DeepAbstract} \cite{deepabstract}, which we will more closely introduce in the preliminaries, and compare to in the experiments.

\section{Preliminaries}\label{sec:prelims}
In this work, we focus on classification feedforward neural networks. Such a neural network $\NN$ consists of several layers $1,2,\dots,L$, with $1$ being the \emph{input layer}, $L$ being the \emph{output layer} and $2, \dots, L - 1$ being the \emph{hidden layers}.
Each layer $\ell$ contains $n_\ell$ \emph{neurons}.
Neurons of one layer are connected to neurons of the previous and next layers by means of weighted connections.
Associated with every layer $\ell$ that is not an output layer is a \emph{weight matrix} $\matr{W}{\ell} = (\weight{\ell}{i}{j}) \in \R^{n_{\ell+1} \times n_\ell}$
where $\weight{\ell}{i}{j}$ gives the weights of the connections to the $i^{th}$ neuron in layer $\ell+1$ from the $j^{th}$ neuron in layer $\ell$. 
We use the notation $\smash{\matr{W}{\ell}_{i,*} = [\weight{\ell}{i}{1}, \dots, \weight{\ell}{i}{n_\ell}]}$ to denote the incoming weights of neuron $i$ in layer $\ell+1$ and 

$\smash{\matr{W}{\ell}_{*,j} = [\weight{\ell}{1}{j}, \dots, \weight{\ell}{n_{\ell+1}}{j}]^\intercal}$ to denote the outgoing weights of neuron $j$ in layer $\ell$. 
Note that $\matr{W}{\ell}_{i,*}$ and $\matr{W}{\ell}_{*,j}$ correspond to the $i^{th}$ row and $j^{th}$ column of $\smash{\matr{W}{\ell}}$ respectively.
A vector $\vecb^{(\ell)}=[\bias{\ell}{1},\dots,\bias{\ell}{n_\ell}] \in \R^{n_\ell}$ called \emph{bias} is also associated with each hidden layer $\ell$.
The input and output of a neuron $i$ in layer $\ell$ is denoted by $\h^{(\ell)}_i$ and $\z^{(\ell)}_i$ respectively. 
We call $\vech^{\ell} = [\h^{(\ell)}_1, \dots, \h^{(\ell)}_{n_\ell}]^\intercal$ the vector of \emph{pre-activations} and $\vecz^{\ell} = [\z^{(\ell)}_1, \dots, \z^{(\ell)}_{n_\ell}]^\intercal$ the vector of \emph{activations} of layer $\ell$.
The neuron takes the input $\vech^{\ell}$, and applies an \emph{activation function} $\phi: \R \to \R$ element-wise on it. The output is then calculated as $\vecz^{\ell}=\phi(\vech^{\ell})$, where standard activation functions include tanh, sigmoid, or ReLU \cite{maas2013rectifier}. We assume that the activation function is Lipschitz continuous, which in particular holds for the aforementioned functions \cite{virmaux2018lipschitz}.
In a feedforward neural network, information flows strictly in one direction: from layer $\ell_m$ to layer $\ell_n$ where $\ell_m < \ell_n$. 
For an $n_1$-dimensional input $\vecx \in \inputspace$ from some input space $\inputspace \subseteq \R^{n_1}$, the output $\vecy \in \R^{n_L}$ of the neural network $\NN$, also written as $\vecy = \NN(\vecx)$ is iteratively computed as:
\begin{align}
	\vech^{(0)} = \vecz^{(0)} &= \vecx\notag\\
	\vech^{(\ell+1)} &= \matr{W}{\ell} \vecz^{(\ell)} + \vecb^{(\ell+1)} \label{eq:forward-h}\\
	\vecz^{(\ell+1)} &= \phi (\vech^{(\ell+1)}) \label{eq:forward-z}\\
	\vecy &= \vecz^{(L)}\notag
\end{align} 
where $\phi(\vecx)$ is the column vector obtained by applying $\phi$ component-wise to $\vecx$.
We abuse the notation and write $\vecz^{(\ell)}(\vecx)$, when we want to specify that the output of layer $\ell$ is computed by starting with $\vecx$ as input to the network.

\subsection{Syntactic and Semantic Abstractions}
We are interested in a general abstraction scheme that is not only useful for verification, but also for revealing redundancies, while keeping a formal link to the original network.
We distinguish between two types of abstraction: semantic and syntactic.
Syntactic abstraction makes use of the weights of the network, the syntactic information, and allows for overapproximation guarantees that are not restricted to specific inputs.
However, as we shall see in the experiments, the semantic abstraction can capture the behavior of the original network on typical input data much more accurately than its syntactic counterpart. This comes at the cost of a more challenging error analysis. 

\paragraph{Semantic Information}\label{sec:semantic-information} In line with \emph{DeepAbstract} \cite{deepabstract}, we will create the semantic information based on a set of inputs, the \emph{I/O set}, $X = \{\vecx_1,\dots,\vecx_n \} \subseteq \inputspace$, which is typically a subset of the training dataset.
We use the inputs $\vecx_j\in X$, feed them to the network and store the output values $\{\vecz^{(\ell)}(\vecx_j)\}_{\vecx_j\in X}$ of a layer $\ell$ in a matrix $\mathbf{Z}^{(\ell)}=(z_i^{(\ell)}(\vecx_j))_{i,j}$. 
Note that the columns are the $\vecz^{(\ell)}(\vecx_j)$ and the rows, denoted as $\mathbf{Z}^{(\ell)}_{i,*}$, correspond to the values one neuron $i$ produces for all inputs $\vecx_j$. 
We refer to the vector $\mathbf{Z}_{j,*}^{(\ell)}$ as the \emph{semantics} of neuron $i$.
This collection of matrices $\mathbf{Z}^{(\ell)}$ for all layers contains the semantic information of the network.

\subsubsection{DeepAbstract} 
Since we will compare our approach to \emph{DeepAbstract} \cite{deepabstract}, we will give a concise description of the idea of their work. 
First, it generates the semantic information $\mathbf{Z}$. 
For one layer $\ell$, it clusters the rows of the matrix 
by using standard clustering techniques, e.g. k-means clustering \cite{DBLP:books/lib/Bishop07}.
Each cluster is considered to be a group of neurons that have similar semantics and similar behavior. 
Thus, only one group representative is chosen to remain and the rest is replaced by the representatives. 
\subsubsection{Bisimulation} 
The idea of \cite{Prabhakar22} is to apply the notion of bisimulation to NN.
A bisimulation declares two neurons as equivalent if they agree on their incoming weights, biases, and activation functions.
Additionally, the paper introduces a $\delta$-bisimulation that allows neurons to be equivalent only up to $\delta$, i.e. two neurons $i,j$ of layer $\ell$ with the same activation function are considered to be $\delta$-bisimilar, if for all $k$ :  $|\weight{\ell-1}{i}{k}-\weight{\ell-1}{j}{k}|\leq\delta$ and $|\bias{\ell}{i}-\bias{\ell}{j}|\leq\delta$.

\section{Linear Abstraction}\label{sec:abstraction}
Our abstraction of a NN is based on the idea that huge NN in their practical application are usually trained with more neurons than necessary. Since there are techniques to avoid ``overfitting", users of machine learning tend to use NN that are bigger than necessary for their task \cite{lawrence1997lessons}. Intuitively, such networks thus contain redundancies.
We want to remove these redundancies to decrease the size of the network and make it more scalable for verification.

Existing approaches group together similar neurons, and then choose a representative. Instead, we propose to replace a neuron with a linear combination of other neurons.
More specifically, we want to replace a neuron $i$ of layer $\ell$, not by one single neuron $j$, but rather by a clever combination of several neurons, called the \emph{basis}, $\basis{\ell}\subset\{1,\dots,n_\ell\}\backslash\{i\}$, which is a subset of all neurons of this layer and in this case given as their indices.
We assume that the behavior of a neuron can be simulated by a linear combination of the behavior of the basis neurons, i.e. by $\sum_{j\in\basis{\ell}}\lcoeff{i}{j}{\ell}\cdot \mathbf{Z}_{j,*}^{(\ell)}$ for some $\lcoeff{i}{j}{\ell}\in\mathbb{R}$. 

\paragraph{Example}\label{par:one} Consider the neural network in \cref{fig:example}. It has an input layer with two neurons $n_1^0, n_2^0$, one hidden layer with three neurons $n_1^1,n_2^1,n_3^1$, and an output layer with two neurons $n_1^2,n_2^2$.
We assume that we are given the basis $\basis{1}=\{n_2^1,n_3^1\}$, marked with blue color in the figure, and the linear coefficients $\lcoeff{1}{1}{1},\lcoeff{1}{2}{1}$. 
That is, we assume that $n_1^1$ can be simulated by the linear combination $\lcoeff{1}{1}{1}\cdot n_2^1+\lcoeff{1}{2}{1}\cdot n_3^1$.
We can remove neuron $n_1^1$ and its outgoing weights $[1,2]^\intercal$, and add the outgoing weights scaled by the linear coefficients to the basis neurons instead. 
We add $\lcoeff{1}{1}{1}\cdot[1,2]^\intercal$ to the outgoing weights of neuron $n_2^1$, so we get $[-1,3]^\intercal+\lcoeff{1}{1}{1}\cdot[1,2]^\intercal=[-1+\lcoeff{1}{1}{1}\cdot 1, 3+\lcoeff{1}{1}{1}\cdot 2]^\intercal$, and respectively, we get $[-2+\lcoeff{1}{2}{1}\cdot 1,1+\lcoeff{1}{2}{1}\cdot 2]^\intercal$ as the outgoing weights of neuron $n_3^1$.\\

\definecolor{light-gray}{gray}{0.9}
\definecolor{light-blue}{RGB}{156,196,255}
\begin{figure}[t]
	\begin{subfigure}[b]{0.45\textwidth}
		\scalebox{0.8}{
			\begin{tikzpicture}[every node/.style=draw,minimum width=0.75,circle]
				\coordinate (i0) at (0,0);
				\coordinate (i1) at (0,2);
				\node[] (n01) at (1,0) {$n_2^0$};
				\draw[->] (i0) -- (n01);
				\node[] (n02) at (1,2) {$n_1^0$};
				\draw[->] (i1) -- (n02);
				\node[] (n11) at (3,2.5) {$n_1^1$};
				\node[fill=light-blue] (n12) at (3,1) {$n_2^1$};
				\node[fill=light-blue] (n13) at (3,-0.5) {$n_3^1$};
				\foreach \x in {1,2,3} \draw[->] (n01) -- (n1\x);
				\foreach \x in {1,2,3} \draw[->] (n02) -- (n1\x);
				\node[] (n21) at (5,2) {$n_1^2$};
				\node[] (n22) at (5,0) {$n_2^2$};
				\tikzset{every node/.style={}};
				\draw[->,color=white] (n13) -- (n22) node[midway,below,xshift=-0.1cm,yshift=0.5cm, rotate=12] {\color{white}$1+\lcoeff{1}{2}{1}\cdot2$};
				\draw[->] (n11) -- (n21) node[midway,above] {1};
				\draw[->] (n11) -- (n22) node[midway,above,xshift=-0.4cm, yshift=0.5cm] {2};
				\draw[->] (n12) -- (n21) node[midway,above,xshift=-0.5cm, yshift=-0.3cm] {-1};
				\draw[->] (n12) -- (n22) node[midway,above,xshift=-0.45cm, yshift=0.1cm] {3};
				\draw[->] (n13) -- (n21) node[midway,above,xshift=-0.7cm, yshift=-0.9cm] {-2};
				\draw[->] (n13) -- (n22) node[midway,below] {1};
				\coordinate (o0) at (6,0);
				\coordinate (o1) at (6,2);
				\draw[->] (n21) -- (o1);
				\draw[->] (n22) -- (o0);
				\node[shape=rectangle] () at (3,3.5) {$W^{(1)}=\left(\begin{matrix}1&-1&-2\\2&3&1\end{matrix}\right)$};
		\end{tikzpicture}}
	\end{subfigure}\hfil
	\begin{subfigure}[b]{0.45\textwidth}
		\scalebox{0.8}{
			\begin{tikzpicture}[every node/.style=draw,minimum width=0.75,circle]
				\coordinate (i0) at (0,0);
				\coordinate (i1) at (0,2);
				\node[] (n01) at (1,0) {$n_2^0$};
				\draw[->] (i0) -- (n01);
				\node[] (n02) at (1,2) {$n_1^0$};
				\draw[->] (i1) -- (n02);
				\node[color=light-gray] (n11) at (3,2.5) {$n_1^1$};
				\node[fill=light-blue] (n12) at (3,1) {$n_2^1$};
				\node[fill=light-blue] (n13) at (3,-0.5) {$n_3^1$};
				\foreach \x in {2,3} \draw[->] (n01) -- (n1\x);
				\foreach \x in {2,3} \draw[->] (n02) -- (n1\x);
				\draw[->,color=light-gray] (n01) -- (n11);
				\draw[->,color=light-gray] (n02) -- (n11);
				\node[] (n21) at (5,2) {$n_1^2$};
				\node[] (n22) at (5,0) {$n_2^2$};
				\tikzset{every node/.style={}};
				\draw[->,color=light-gray] (n11) -- (n21) node[midway,above] {1};
				\draw[->,color=light-gray] (n11) -- (n22) node[midway,above,xshift=-0.4cm, yshift=0.6cm] {2};
				\draw[->] (n12) -- (n21) node[midway,above,xshift=0.4cm, yshift=-0.5cm, rotate=30] {\color{red}$-1+\lcoeff{1}{1}{1}\cdot1$};
				
				\draw[->] (n12) -- (n22) node[midway,above,xshift=1cm, yshift=-0.7cm, rotate=10] {\color{red}$3+\lcoeff{1}{1}{1}\cdot2$};
				\draw[->] (n13) -- (n21) node[midway,above,xshift=0.5cm, yshift=-0.6cm, rotate=52] {\color{red}$-2+\lcoeff{1}{2}{1}\cdot1$};
				\draw[->] (n13) -- (n22) node[midway,below,xshift=-0.1cm,yshift=0.5cm, rotate=12] {\color{red}$1+\lcoeff{1}{2}{1}\cdot2$};
				\coordinate (o0) at (6,0);
				\coordinate (o1) at (6,2);
				\draw[->] (n21) -- (o1);
				\draw[->] (n22) -- (o0);
				\node[shape=rectangle] () at (3,3.5) {$\tilde{W}^{(1)}=\left(\begin{matrix}0&-1+\lcoeff{1}{1}{1}\cdot 1&-2+\lcoeff{1}{2}{1}\cdot1\\0&3+\lcoeff{1}{1}{1}\cdot 2&1+\lcoeff{1}{2}{1}\cdot 2\end{matrix}\right)$};
		\end{tikzpicture}}
	\end{subfigure}
	\caption{Linear Abstraction - On the left, the original network with the basis $B$ in blue. On the right, the abstracted network with the removed neuron $n_1^1$ and the changed output weights of the basis neurons $n_2^1,n_3^1$, where we assume that $n_1^1$ can be simulated by $\lcoeff{1}{1}{1}\cdot n_2^1+\lcoeff{1}{2}{1}\cdot n_3^1$.}
	\label{fig:example}
\end{figure}
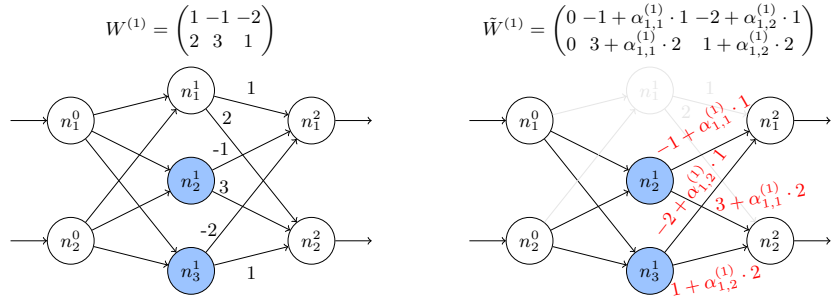

The computational overhead to compute a linear combination compared to finding a representative is negligible, as we will see in our experiments (see \cref{sec:comparison-existing-work}). 
On the other hand, 
they provide more expressive power, subsuming the aforementioned clustering-based approach \cite{deepabstract}. In particular, we can detect scaled weights that previous approaches failed to identify.

Please note that although it is possible to replace a neuron with a linear combination of any other neurons in the network, we will only use neurons from the same layer due to more efficient support by modern neural network frameworks.

In the following sections, we will answer three questions: 
How can one find a set of neurons that serves as a basis (\ref{sec:basis})?
How to find the coefficients for the linear combination (\ref{sec:coefficients})?
How to replace a neuron, once its representation as a linear combination is given (\ref{sec:replacement})?

\subsection{Finding the Basis}\label{sec:basis}
Our approach is meant to find a sufficient smaller subset of neurons in one layer, which is enough to represent the behavior of the whole layer.
We will make use of the semantic information of a layer $\ell$, given as $\mathbf{Z}^{(\ell)}=(z_i^{(\ell)}(\vecx_j))_{i,j}$ (see \ref{sec:semantic-information}).
Based on this, we try to find a basis of neurons, i.e. a set of indices for neurons in this layer $ \{j_1, \dots j_{k_\ell}\}=\basis{\ell}\subset\{1,\dots,n_\ell\}$, which can represent the whole space as well as possible.
To this end we want to find a subset of size $k=|\basis{\ell}|$ such that $\|\sum_{j\in\basis{\ell}}\lcoeff{i}{j}{\ell}\cdot \mathbf{Z}_{j,*}^{(\ell)}-\mathbf{Z}_{i,*}\|$ is minimized.
We denote with
\begin{equation}\label{eq:projection-matrix1}
	A_B = 
	\left[
	\begin{array}{cccc}
		\vert &         & \vert \\
		\mathbf{Z}^{(\ell)}_{j_1,*} & \ldots & \mathbf{Z}^{(\ell)}_{j_{k_\ell},*}    \\
		\vert &         & \vert 
	\end{array}
	\right]
\end{equation}
the matrix containing the activations $\mathbf{Z}^{(\ell)}_{j,*}$ of the neurons in the basis as columns.

\subsubsection{Greedy Algorithm}
The problem of finding an optimal basis of size $k$ w.r.t. L$_2$ distance can be seen as a variation of the \emph{column subset selection problem} which is known to be NP-complete \cite{shitov2021column}. As a consequence, we use a variant of a greedy algorithm \cite{greedyCSSP}. While it does not always yield the optimal solution, it has been observed to work reasonably well in practice \cite{farahat2011efficient, farahat2013fast}.

It has already been observed that layers closer to the output usually contain more condensed information and more redundancies, and can, thus, be compressed more aggressively \cite{deepabstract}.  
We present a greedy algorithm that chooses which layer contains more information and needs a larger basis instead of decreasing the basis sizes equally fast in each layer.
\begin{algorithm}[t]
	\caption{Greedy algorithm over all layers}
	\begin{algorithmic}[1]
		\State \text{Given:} $k$ neurons to be removed
		\State $\forall l\in\{1,\dots,L\}\,:\,B^{(\ell)} \gets \{1,\dots,n_l\}$
		\State $error_{min}\gets\infty$, 
		$l_{best} \gets -1$,
		$n_{best} \gets -1$
		\For{$i \in {1, \dots, k}$}
			\For{$l \in {1,\dots, L}$}
				\For{$j \in {0,\dots, n_l}$}
					\State Compute the projection error $error_j$ of $A_{B^{(\ell)} \setminus \{j\}}$
					\If{$error_j<error_{min}$} 
					\State $l_{best} \gets l$
					\State $n_{best} \gets j$
					\State $error_{min} \gets error_j$ 
					\EndIf
					\EndFor
			\EndFor
			\State $B^{l_{best}} \gets B^{l_{best}} \setminus \{n_{best}\}$
		\EndFor
		\State \Return $B^1,\dots,B^L$
	\end{algorithmic}
	\label{alg:greedy-all-layers}
\end{algorithm}

In \cref{alg:greedy-all-layers}, we see that the procedure iteratively removes neurons from the basis.
To this end, it iterates over all layers $l\in\{1,\dots,L\}$ in the network.
It tries to remove one neuron at a time from the basis. 
Then it computes the projection error of the smaller basis, which is defined as $\|\mathbf{Z}^{(\ell)^\intercal}-\Pi_{A_B} \mathbf{Z}^{(\ell)^\intercal}\|$, where $\Pi_{A_B}$ is the matrix that projects the columns of $\mathbf{Z}^{(\ell)^\intercal}$ onto the column space of $A_B$. The columns of $A_B$ are the rows of $\mathbf{Z^{(\ell)}}$ whose neurons belong to $B$.
It greedily evaluates all neurons in all layers and selects the best neuron of the best layer to be removed.
After checking every layer, the algorithm decides on the best layer and neuron to be removed, i.e. the one with the smallest error.

Since the approach thoroughly evaluates all possibilities, its runtime depends on both the number of layers and neurons. 
A natural alternative would be a heuristic that guides us similarly well through the search space. We provide our choice of heuristic below.

\subsubsection{Variance-based Heuristic} Instead of a step-wise decision that takes a lot of computation time, we propose to use a variance-based heuristic. 
We define the variance of a vector $\mathbf{v} \in \R^n$ in the usual way by
$\Var(\mathbf{v}) = \sum_{i=0}^n (v_i - \Mean(\mathbf{v}))^2$
where $\Mean(\mathbf{v})$ is the mean of the vector values. 
W.l.o.g. let the neurons be numbered in such a way that $\Var(\vecz^{(\ell)}_1) \geq \dots \geq \Var(\vecz^{(\ell)}_{n_\ell})$. 
We then choose the basis to contain the neurons with the $k_\ell$ largest variances, i.e. $B=\{1,\dots,k\}$.
We assume that neurons with a higher variance in their output values carry more information, and are, therefore, more relevant.
Indeed, we can see in our experiments, i.e. \cref{fig:basis-finding}, that the heuristic-based approach can achieve similar results, but in far less time.

\subsection{Finding the Coefficients}\label{sec:coefficients}
Given a basis  $\basis{\ell}$ for some layer $\ell$, computed with the before-mentioned approach, we want to find the coefficients that can be used to replace the remaining neurons which are not part of the basis. 
We fix a neuron $i$ in layer $\ell$ that we want to replace and whose values are stored in $\mathbf{Z}^{(\ell)}_{i,*}$, and we want to minimize $\|\sum_{j\in\basis{\ell}}\lcoeff{i}{j}{\ell}\cdot \mathbf{Z}_{j,*}^{(\ell)}-\mathbf{Z}_{i,*}\|$ for $\alpha_{i,j}^{(\ell)}$.

Since we want to find a linear combination of vectors, a natural choice is \textbf{linear programming}.
The linear program is straightforward and can be found in \cref{sec:appendix-linear-program}. Note that with the linear program, we are minimizing the L$_1$-distance between the neuron's values and its replacement, i.e. $\|\sum_{j\in\basis{\ell}}\lcoeff{i}{j}{\ell}\cdot \mathbf{Z}_{j,*}^{(\ell)}-\mathbf{Z}_{i,*}\|_1$.

In a different way, we can also consider the vectors $\mathbf{Z}^{(\ell)}_{j,*}$ for $j\in B^{(\ell)}$ to span a vector space.
If we are given a subset $\{\mathbf{Z}^{(\ell)}_{j,*} | j\in B^{(\ell)}\subset\{1,\dots,n_\ell\}\}$
that forms a basis for this space, i.e. $\text{span}((\mathbf{Z}^{(\ell)}_{j,*})_{j\in \basis{\ell}})=\text{span}((\mathbf{Z}^{(\ell)}_{j,*})_{j\in\{1,\dots,n_\ell\}})$, we can represent any other vector $\vecz^{(\ell)}_i$ in terms of this basis.
However, we usually cannot represent one neuron perfectly by a linear combination of other neurons. 
\textbf{Orthogonal projection} gives us the closest point in the subspace $\text{span}((\mathbf{Z}^{(\ell)}_{j,*})_{j\in \basis{\ell}})$ for any vector, in terms of L$_2$-distance.
Then, $\vecalpha = [\lcoeff{i}{j_1}{\ell}, \ldots, \lcoeff{i}{j_{k_\ell}}{\ell}]^\intercal \coloneqq (A_B^\top A_B)^{-1} A_B^\top \mathbf{Z}^{(\ell)}_{i,*}$ gives us the coefficients for the orthogonal projection of $\mathbf{Z}^{(\ell)}_{i,*}$ on the linear space spanned by the columns of $A_B$.
For a more detailed description of orthogonal projection see e.g. \cite[Chapter~6.8]{kirkwood2017elementary}.
Note that we assume that the columns of $A_B$ are linearly independent. If not we can simply replace the respective neurons directly.

\subsection{Replacement}\label{sec:replacement}

Assuming, we have a basis $\basis{\ell}$ of this layer and we already know the coefficients $\lcoeff{i}{j}{\ell}\in\mathbb{R}$ for $j\in\basis{\ell}$ that we need to simulate the behavior of neuron $i$. 
This means, we have a linear combination $\sum_{j\in\basis{\ell}}\lcoeff{i}{j}{\ell}\cdot \mathbf{Z}_{j,*}^{(\ell)}$, which we want to use instead of neuron $i$ itself.
We will replace the outgoing weights $\matr{W}{\ell}$ of this layer, such that for all $j\in\basis{\ell}$
\begin{align}\label{eq:repl}
	\matr{\tilde{W}}{\ell}_{*,j} &= [\weight{\ell}{1}{j}+\lcoeff{i}{j}{\ell}\weight{\ell}{1}{i}, \dots, \weight{\ell}{n_{\ell+1}}{j}+\lcoeff{i}{j}{\ell}\weight{\ell}{n_{\ell+1}}{i}]^\intercal\\
	&=\matr{W}{\ell}_{*,j}+\lcoeff{i}{j}{\ell}\matr{W}{\ell}_{*,i}\label{eq:repl2}
\end{align}
Furthermore, we set
	$\matr{\tilde{W}}{\ell}_{*,i}=[0,\dots,0]^\intercal$, and $\matr{\tilde{W}}{\ell}_{i,*}=[0,\dots,0]^\intercal$.
This means that we will not use the output of neuron $i$ anymore, but rather a weighted sum of the outputs of neurons in $\basis{\ell}$, and that we will not even compute the value of $i$.
Additionally, we keep track of the changes we apply to the different neurons with a matrix $D^{(\ell)} = (d^{(\ell)}_{j,i}) \in \R^{n_{\ell} \times n_{\ell+1}}$.
Initially, $D^{(\ell)}$ is $0$ and after each replacement, we add $\lcoeff{i}{j}{\ell} \cdot \weight{\ell}{i}{i'}$ to  $d^{(\ell)}_{j,i'}$ for $j \in \basis{\ell}$ and $i' \in\{1,\dots,n_{\ell+1}\}$. 
This is necessary for restoring neurons at a later point.
\vspace*{2pt}

In the optimal case, the replacement will not change the overall behavior of the neural network.
We can derive a the same semantic equivalence from \cite{Prabhakar22} incorporated into our setting:\begin{proposition}[Semantic Equivalence]\label{prop:sem-equi}
	Let $N$ be a neural network with $L$ layers, $\ell$ a layer of $N$, $i$ a neuron of this layer, and $\basis{\ell}\subset\{1,\dots,n_\ell\}\backslash\{i\}$ a chosen basis. 
	Let $\tilde{\NN}$ be the NN after replacing neuron $i$ by a linear combination of basis vectors with coefficients $\lcoeff{i}{j}{\ell}$, with the procedure as described above.
	
	If for all inputs $\vecx\in X\subset \mathcal{X}$, $\z^{(\ell)}_i(\vecx)=\sum_{j\in\basis{\ell}}\lcoeff{i}{j}{\ell}\z^{(\ell)}_j(\vecx)$, then $\NN$ and $\tilde{\NN}$ are semantically equal, i.e. for all inputs $\vecx\in X$, $\tilde{\NN}(\vecx)=\NN(\vecx)$.
\end{proposition}
It is easy to see that this proposition is true, for a full proof see \cref{sec:proof-of-prop1}.
However, the proposition assumes equality of $\z^{(\ell)}_i(\vecx)$ and $\sum_{j\in\basis{\ell}}\lcoeff{i}{j}{\ell}\z^{(\ell)}_j(\vecx)$ for $\vecx\in X$, which virtually never holds for real-world neural networks. 
Therefore, we want to minimize the difference $|\z^{(\ell)}_i(\vecx)-\sum_{j\in\basis{\ell}}\lcoeff{i}{j}{\ell}\z^{(\ell)}_j(\vecx)|$, which will not yield a semantically equivalent abstraction, but an abstraction with very similar behavior. 
We can then \textbf{quantify the difference} between the output of the original network and the abstraction, i.e. the \emph{induced error} with the following Theorem.
\newpage
\begin{theorem}[Over-approximation Guarantee]\label{th:bisim-error}
	Let $N$ be an NN with $L$ layers. For each layer $\ell$, we have a basis of neurons $\basis{\ell}$, and a set of replaced neurons $\repl{\ell}$. Then, let $\tilde{N}$ be the network after replacing neurons in $\repl{\ell}$ as described above.
	
	We can \emph{over-approximate the error} between the output of the original network $N^L$ and the output of the abstraction $\tilde{N}^L$ for $\vecx\in X\subset \mathcal{X}$ by 
	\vspace{5pt}
	$$\|\tilde{N}^{L}(\vecx)-N^L(\vecx)\|\leq b(1-a^{L-1})/(1-a)$$
	\vspace{-4pt}
	
	\noindent with $a=\lambda(\|W\|+\eta)$, $b=\lambda\|W\|\epsilon$, with $\lambda^{(\ell)}$ being the Lipschitz-constant of the activation function in layer $\ell$, $\lambda=\max_\ell \lambda^{(\ell)}$, $\|W\|=\max_\ell \|W^{(\ell)}\|_1$, $\eta=\max_\ell \eta^{(\ell)}$, and $\epsilon=\max_\ell \epsilon^{(\ell)}$,
	assuming that for all layers $\ell\in\{1,\dots,L\}$ and for all inputs $\vecx\in X$, we have
	\begin{itemize}[topsep=0pt]
		\item for $i\in\repl{\ell}\; :\;|\z^{(\ell)}_i(\vecx)-\sum_{j\in\basis{\ell}}\lcoeff{i}{j}{\ell}\z^{(\ell)}_j(\vecx)|\leq\epsilon^{(\ell)}$
		\item $|\sum_{i\in I^{(\ell)}}\smash{\matr{W}{\ell}_{*,i}}\sum_{t\in\basis{\ell}}\lcoeff{i}{t}{\ell}|\leq\eta^{(\ell)}$
	\end{itemize}	
\end{theorem}
In other words, we can over-approximate the difference in the output of the original and the abstracted network by a value that depends on the weight matrices, the activation function and the tightness of the abstracted neurons to their replacements.
The full proof can be found in \cref{sec:proof-of-th1}. This Theorem provides us with the \textbf{theoretical guarantee} that, given our abstraction, we can provide a valid over-approximation of the output of the original network.\\

\paragraph{Comparison to the $\delta$-bisimulation}
Let us recap the error definition from \cite{Prabhakar22}.
The difference of the bisimulation and the original network is bounded by $[(2a)^k-1]b/(2a-1)$, where $a=\lambda|S|\|W\|$ and $b=\lambda(|P|L(\mathcal{N})\|x\|+1)\delta$\footnote{Please note that this statement is slightly different from the paper ($(2a)^k$ instead of $(2/a)^k$), which we believe to be a typo in the paper.}.
In this notation, $|S|$ is the maximum number of neurons per layer in the whole network, $|P|$ the maximum number of neurons in the bisimulation (can be understood as the number of neurons in an abstraction), $L(\mathcal{N})$ is the maximum Lipschitz-constant of all layers, and $\delta$ is the maximum absolute difference of the bias and sum of the incoming weights.

The drawbacks of that approach are twofold: (i) the error is based on one specific input, and (ii) it makes use of the Lipschitz-constant of the whole network. Calculating the Lipschitz constant of an NN is still part of ongoing research \cite{fazlyab2019efficient} and not a trivial problem. 
In contrast, we improve on both. Our error calculation generalizes over a set of inputs. Additionally, we use local information, stored in the weight-matrices, to circumvent using the Lipschitz-constant of the NN.

\section{Refinement}\label{sec:refinement}For certain inputs the abstraction might not reflect the behavior of the original network. For these inputs, so-called \emph{counterexamples}, we may want to \emph{refine} the abstraction, as opposed to starting the abstraction from the original network again.
We consider an input to be a counterexample whenever the abstraction assigns it a different label than the original network. 
However, a counterexample can be any input that does not align with the specifications. 

We propose to refine the abstraction by restoring some of the replaced neurons. 
To do this, we need to know which neurons should be replaced and how.
We first briefly mention three heuristics to choose a neuron for restoration. 
Afterward, we explain how to restore a neuron.
Note that the refinement offers more than a ``roll-back'' of the most recent step of the abstraction since it picks the step-to-be-rolled-back in retrospect reflecting all other steps, leading to a more informed choice. This could in principle be done directly in the abstraction phase, but at an infeasible cost of a huge look-ahead.

\subsubsection{Refinement Heuristics}
We propose three different heuristics: difference-guided, gradient-guided, and look-ahead.\vspace{-5pt}
\begin{itemize}
\item The \emph{difference-guided} refinement looks at the difference of a neuron in the original and its representation as a linear combination in the abstraction. It replaces the neuron with the largest difference.
\item The \emph{gradient-guided} refinement additionally takes the gradient of the NN into account, that is computed as in the training phase of the NN. This takes into account how the whole network would need to change to fix the counterexample.
\item The \emph{look-ahead} is the most greedy method and would try out every replaced neuron. It would check how much the network would improve if the neuron was replaced and then chooses the neuron with the highest improvement.
\end{itemize}
More details on the approaches can be found in \cref{sec:refinement-heuristics}.

\subsubsection{Restoration of a Neuron}
The restoration principle can be seen as the counterpart of the replacement. 
Let $\dbtilde{\NN}$ be the network obtained by replacing several neurons in the original network $\NN$, where we want to restore a deleted neuron $i$ of layer $\ell$.
To do this, we need not only to get the original neuron back, including its incoming and outgoing weights but also to remove the additional outgoing weights from the basis neurons.
Intuitively, the restoration removes the linear combination, ensures that the original outgoing weights for the neuron are used, and adjusts the incoming weights of the neuron.
We may have changed layer $\ell-1$, and thus we cannot restore the original incoming weights of neuron $i$, but we have to adapt it to changes in the basis $\basis{\ell-1}$.
This can be done with the following changes:
\begin{itemize}[topsep=0pt]
	\item $\forall j \in \basis{\ell}$: $\matr{\tilde{W}}{\ell}_{*,j}=\matr{\dbtilde{W}}{\ell}_{*,j}-\alpha_j\matr{W}{\ell}_{*,j}$
	\item $\matr{\tilde{W}}{\ell}_{*,i}=\matr{W}{\ell}_{*,i}$
	\item $\forall j \in \basis{\ell-1}$: $\weighta{\ell-1}{i}{j} = \weight{\ell-1}{i}{j} + d^{(\ell-1)}_{j, i}$
\end{itemize}
Afterward, we subtract $\alpha_j \cdot \weight{\ell}{i}{i'}$ from $d^{(\ell)}_{j,i'}$ for $i' \in \{1,\dots,n_{\ell+1}\}$ and $j \in \basis{\ell}$.

\newpage 
\section{Experimental Results}\label{sec:experiments}

Our experimental section is divided into several parts:
The first one covers how the different methods for finding a basis and the coefficients compare, as described in \cref{sec:coefficients} and \cref{sec:basis}.
The second part shows experiments on our approach in comparison to existing works, namely \emph{DeepAbstract} \cite{deepabstract} and our implementation of bisimulation \cite{Prabhakar22} (which was not implemented before).
The third part contains the comparison between the abstraction based on syntactic and semantic information.
The fourth part describes our experiments on abstraction refinement. Finally, the last part contains experiments on the error induced by our abstraction.
Note that supplemental experiments can be found in the Appendix.

Lastly, the work of Katz et al.~\cite{katz_abstraction} tightly couples the abstraction with the subsequent particular verification, by integrating the specification as layers into the network. It is, thus, not clear how an abstraction from \cite{katz_abstraction} could be extracted from the tool and reused for another purpose.
Additionally, our abstraction would have to be connected with some verification algorithm (DeepPoly, as done by DeepAbstract, or some other) to compare. \emph{Any comparison of the two works would then mostly compare the different verification tools, not really the abstractions.} Although a comparison of different verifiers linked to our LiNNA is an interesting next step into one of the possible applications, it is out of the scope of this paper, which examines the abstraction itself (see Introduction).

\paragraph{Implementation}
We implemented the approach in our tool \emph{LiNNA} (\emph{Li}near \emph{N}eural \emph{N}etwork \emph{A}bstraction)\footnote{https://github.com/cxlvinchau/LiNNA}.
We used networks that were trained on MNIST \cite{lecun1998mnist}, CIFAR-10 \cite{krizhevsky2009learning}, and FashionMNIST \cite{xiao2017fashion} for our experiments.
In the following, we refer to the corresponding trained networks with ``$L \times n$", where $L$ denotes the number of hidden layers and $n$ is the number of neurons in these hidden layers. 
All experiments were conducted on a computer with Ubuntu 22.04 LTS with 2.6 GHz Intel© Core™ i7 processors, and 32GB of RAM. 

\paragraph{Performance Measures}
We will compare 
the approaches mostly on (i) the reduction rate and (ii) the accuracy on a test set.
Intuitively, the \emph{reduction rate} describes how much the NN was reduced by abstraction. If an NN $\NN$ has in total $n$ neurons, but after reduction, there are $m$ neurons left, then the reduction rate is then defined as $RR(\NN) = 1-\frac{m}{n}$.
The \emph{accuracy} of a NN on a test set is defined as the ratio of 
how many inputs are predicted with the correct label. This is the key performance indicator in machine learning and shows how well a network generalizes to unseen data. 
In evaluating our abstraction, we follow the same principle since we want to know how well the NN generalizes after abstraction. Note that this test set was not used for training or computing the abstraction.

\subsection{Abstraction}
\subsubsection{Finding the Basis}
We have given two different methods in \cref{sec:basis} to find a good basis $B$.
While the orthogonal projection yields an equally good abstraction  compared to linear programming, it outperforms the latter in terms of runtime by magnitudes.
Hence, we conducted the rest of the experiments with orthogonal projection. 
The full comparison between orthogonal projection and linear programming can be found in \cref{fig:lp-op-comp} in \cref{sec:basis-finding}. 
\begin{figure}[!t]
	\centering
	\begin{subfigure}[t]{0.32\textwidth}
	\centering
	\includegraphics[width=\textwidth]{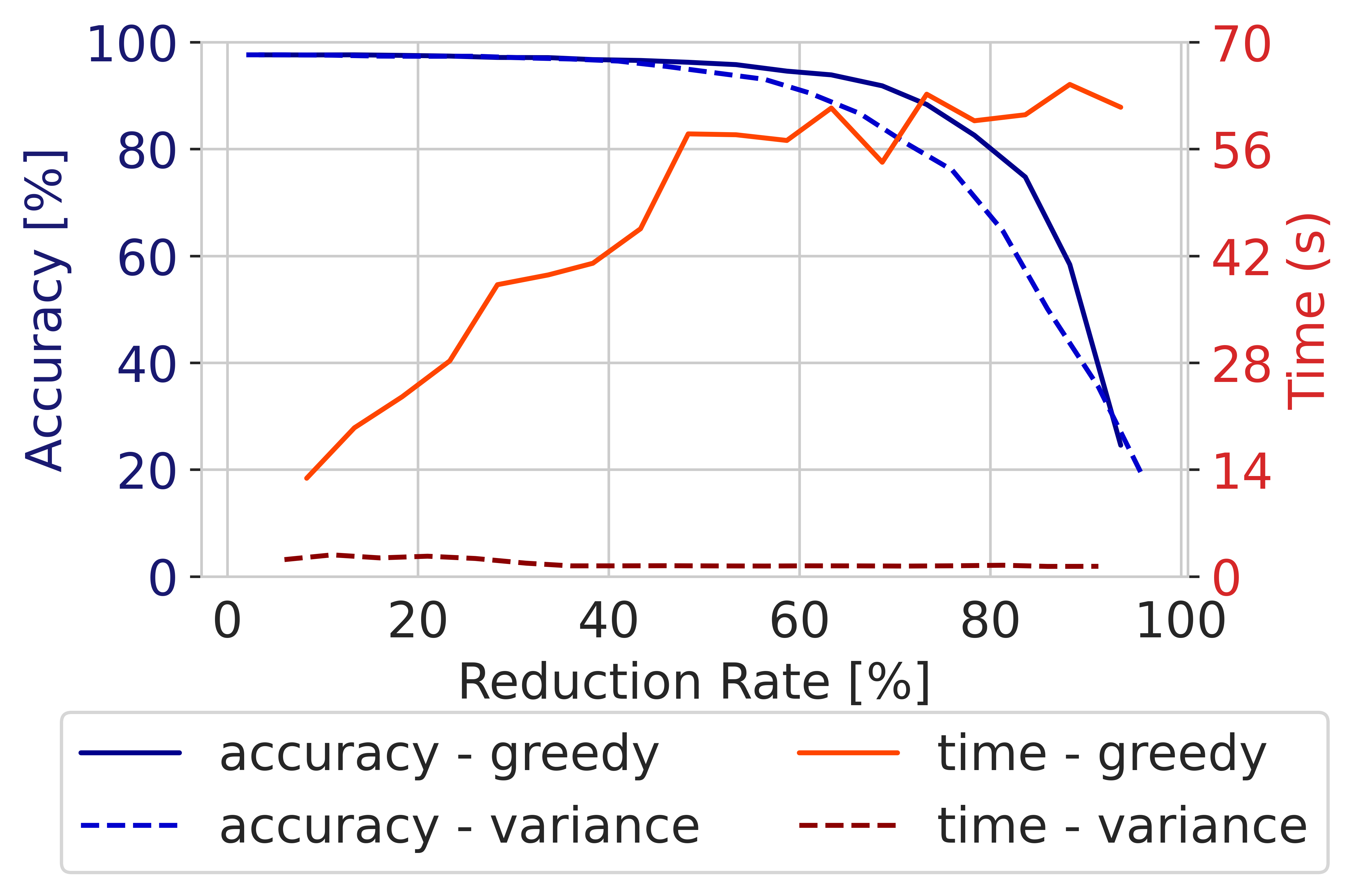}
	 \caption{MNIST 3x100}
	\label{fig:basis-finding-mnist}
	\end{subfigure}\hfill
	\begin{subfigure}[t]{0.32\textwidth}
	\centering
	\includegraphics[width=\textwidth]{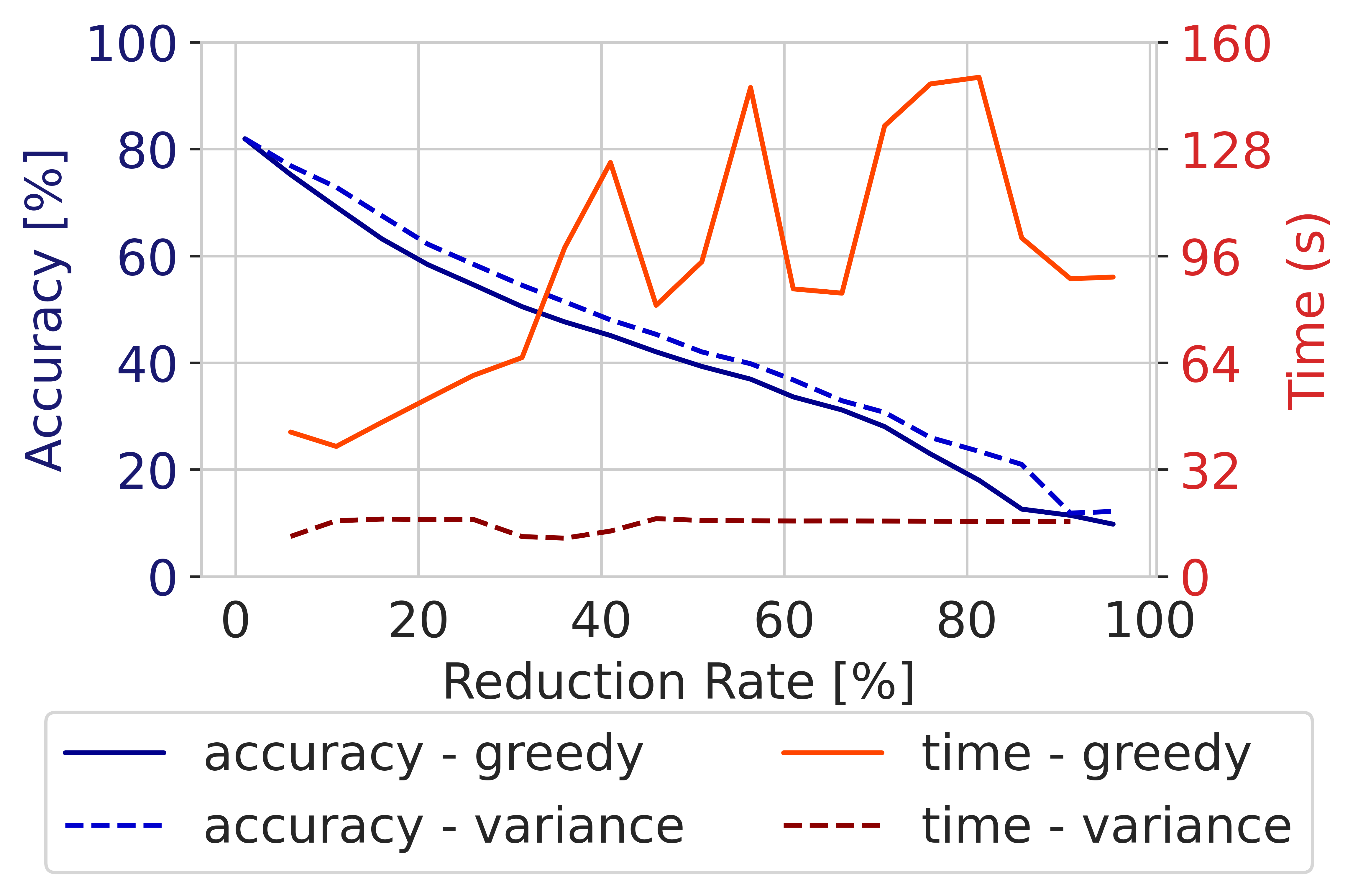}
	\caption{CIFAR-10 3x2500}
	\label{fig:basis-finding-cifar}
	\end{subfigure}\hfill
	\begin{subfigure}[t]{0.32\textwidth}
	\centering
	\includegraphics[width=\textwidth]{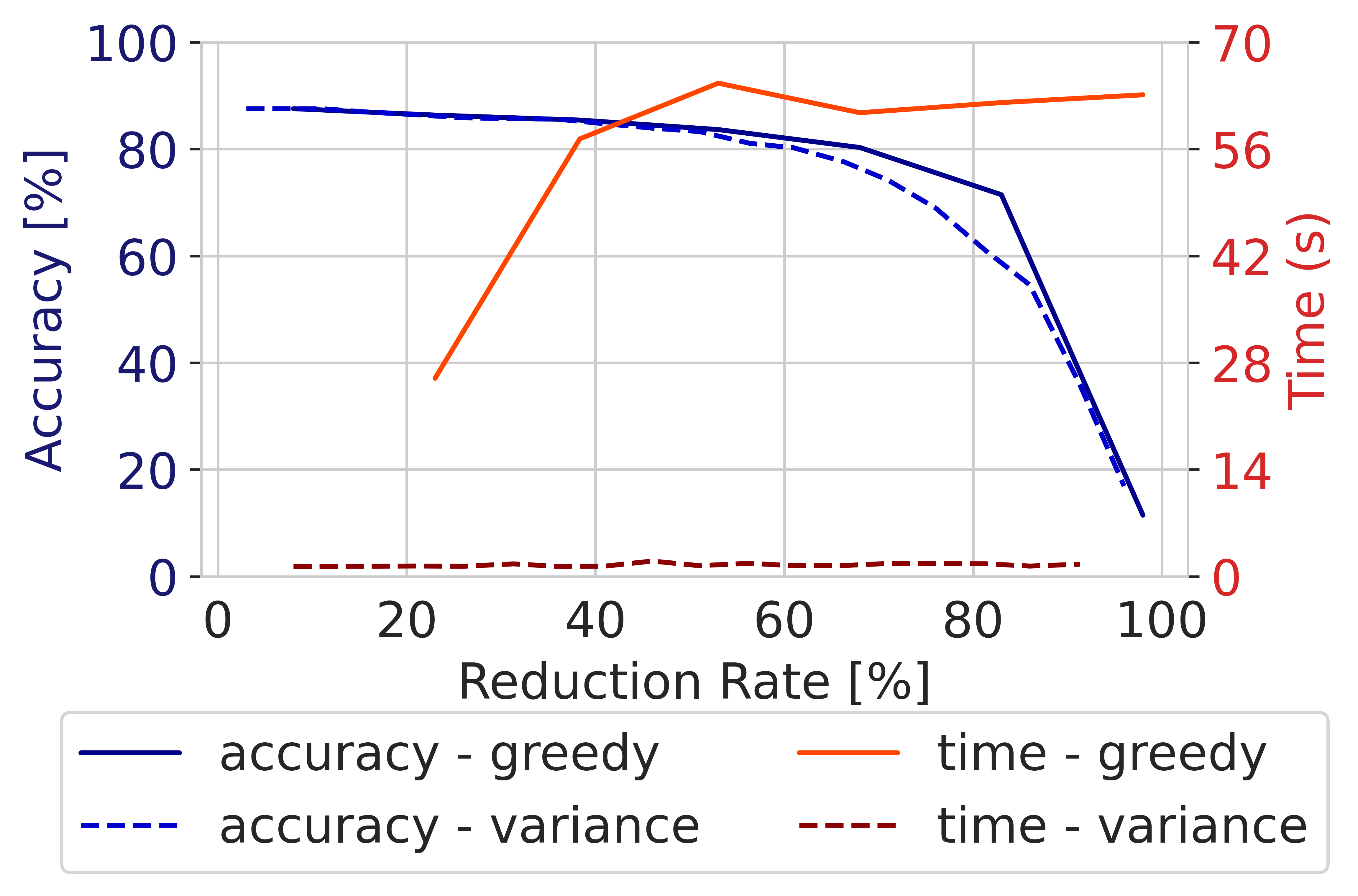}
	\caption{FashionMNIST 3x100}
	\label{fig:basis-finding-fmnist}
	\end{subfigure}
\caption{\emph{Finding the basis for replacement} - Evaluation on different datasets. The plots contain a comparison of LiNNA while using the greedy variant (solid) and the variance-based heuristic (dashed) for finding a basis with orthogonal projection. Comparison of accuracy (blue) in percent and computation time (red) in seconds.}
\label{fig:basis-finding}
\end{figure}

When we compare the greedy and the heuristic-based approach, shown in \cref{fig:basis-finding}, we see that the former outperforms the latter in terms of accuracy on MNIST and FashionMNIST.
On CIFAR-10, the variance-based approach is slightly better.
However, the variance-based approach is always faster than the greedy approach and scales better, as can be seen for all datasets. 
Unsurprisingly, the greedy approach takes more time for higher reduction rates, because it needs to evaluate many candidates.
The variance-based approach just takes the best neurons according to their variance, which has to be calculated only once.
Therefore, the calculation is constant in terms of removed neurons.

The plots show one more difference in the behavior: On MNIST and FashionMNIST, we see a quite stable accuracy until a reduction rate of 60\%. We cannot see the same behavior on CIFAR-10. 
We believe this is due to the accuracy and size of the networks. Whereas it is fairly easy to train a feedforward network for MNIST and FashionMNIST on a regular computer, this is more challenging for CIFAR-10. 
We plan to include more extensive experiments including more involved NN architectures in future work. Finally, our abstraction relies on the assumption that NNs contain a lot of redundant information. 

We want to emphasize, that in machine learning, it is common to train a huge network that contains many more neurons than necessary to solve the task \cite{ZhangBHRV16}. 
After the introduction of regularization techniques (e.g. \cite{Schmidhuber_2015}), the problem of over-fitting (e.g. \cite{NIPS2000_059fdcd9}) has become often negligible. 
Therefore, the automatic response to a bad neural network is often to increase its size, either in depth or in width.
Our approach can detect these cases and abstract away the redundant information.

\subsubsection{Finding the Coefficients}
We have in total four different approaches to finding the coefficients:
greedy or heuristic-based linear programming, and greedy or heuristic-based orthogonal projection. 
All four have similar accuracies for the same reduction rate, whereas the heuristic ones are mostly just slightly worse than the greedy ones. For a more detailed evaluation, please refer to 
\cref{sec:lp-op-comparison}. 
The runtimes of the four approaches, however, differ a lot. Take for example an MNIST 3x100 network. 
We assume that the abstraction is performed by starting with the full network and reducing up to a certain reduction rate. Thus, we have runtimes for each of the approaches for each reduction rate.
We take the average over all the reductions and get 47s for the greedy orthogonal projection, 5130s for the greedy linear programming, 1s for the heuristic orthogonal projection, and 2s for the heuristic linear programming. 
Linear programming takes a lot more time than orthogonal projection, and, as already seen before, the heuristic approaches are much faster than the greedy ones. 
Please refer to \cref{sec:time-tables} for more experiments on the runtime. Therefore, we propose to use the heuristic approach and the orthogonal projection.

\subsubsection{Scalability}
\begin{figure}[!t]
	\begin{minipage}[t]{0.49\textwidth}
		\centering
		\includegraphics[width=\textwidth]{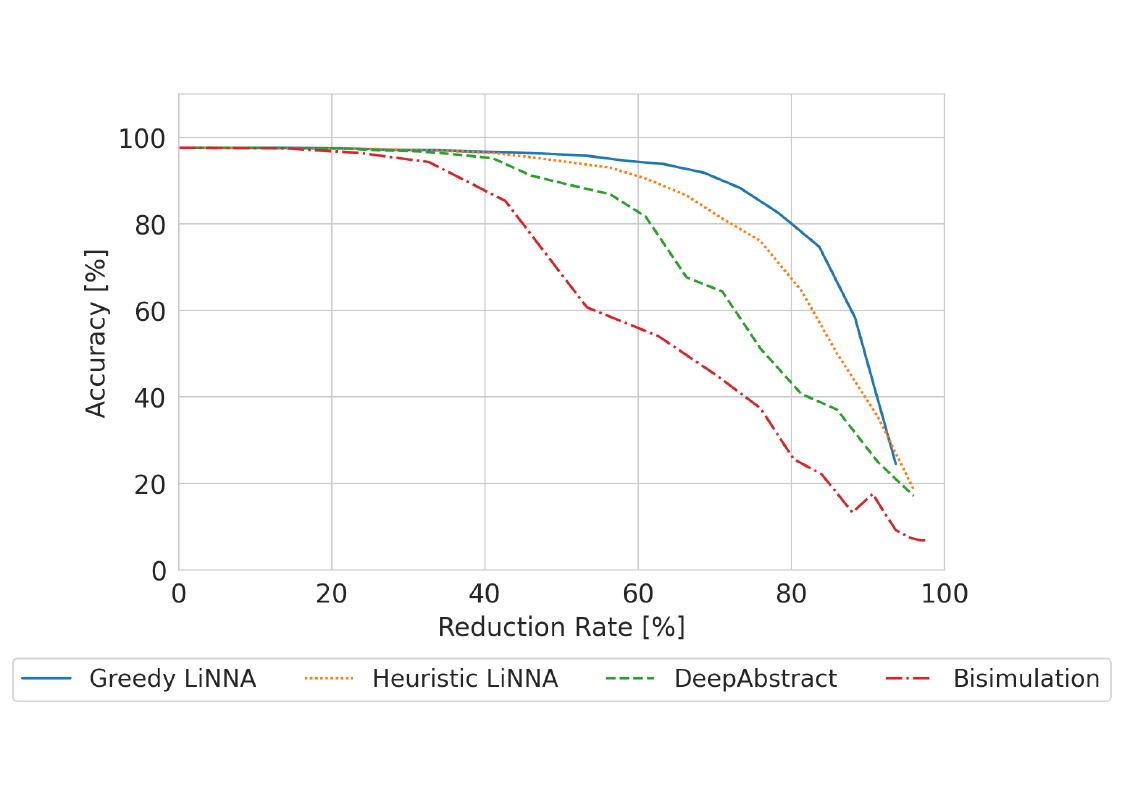}
		\caption{\emph{Comparison of LiNNA to related work - } LiNNA (greedy and heuristic-based variant), \emph{DeepAbstract} \cite{deepabstract}, and our implementation of the bisimulation \cite{Prabhakar22} is evaluated in terms of accuracy on the test set for a certain reduction rate. The experiment was conducted on an MNIST 3x100 network.}
		\label{fig:rw-comp}
	\end{minipage}\hfill
	\label{fig:finding}
	\begin{minipage}[t]{0.49\textwidth}
	\centering
	\includegraphics[width=\textwidth]{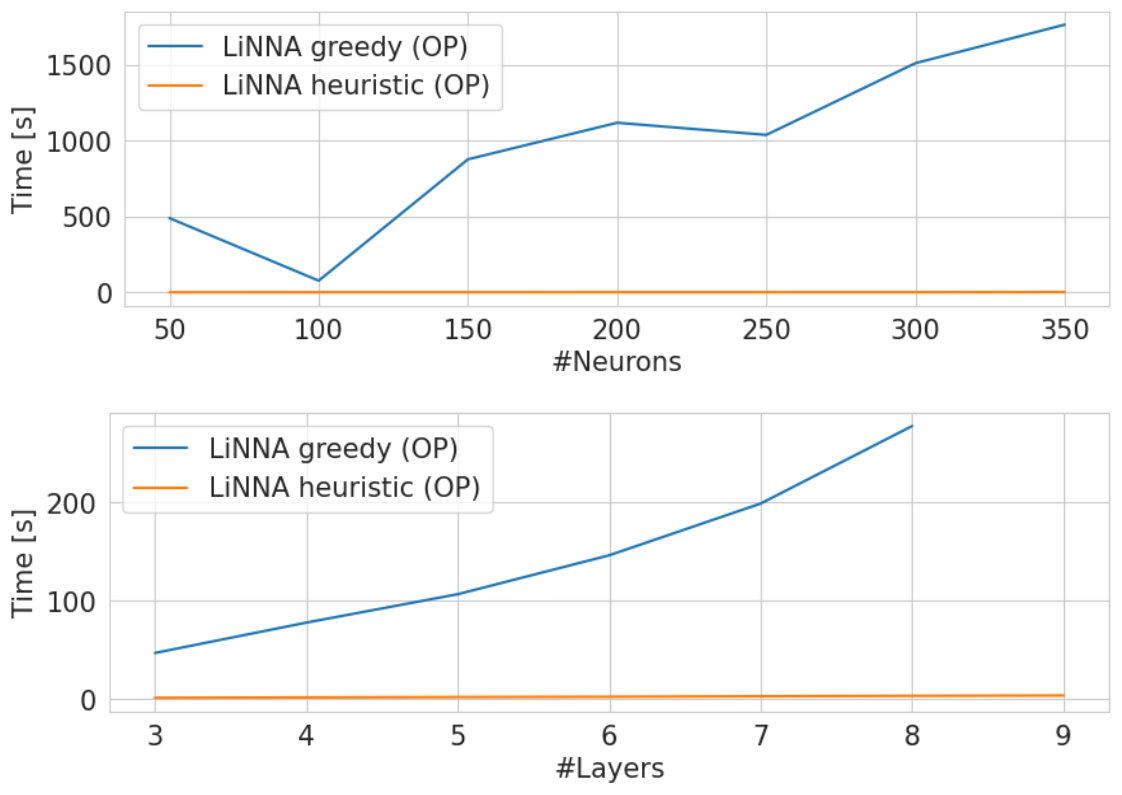}
	\caption{\emph{Scalability of LiNNA} -  Average runtime for 20 different reduction rates on one network. The plot at the top depicts the runtime for MNIST networks with 4 layers, w.r.t. number of neurons. The plot at the bottom shows the runtime for MNIST networks with 100 neurons per layer, w.r.t. number of layers.
	}
	\label{fig:scalability}
	\end{minipage}
     
\end{figure}
We evaluate how our approach scales to networks of different sizes. We evaluate (1) how our approach scales with an increasing number of layers, and (2) how it scales with a fixed number of layers but an increasing number of neurons.
We show our experiments in \cref{fig:scalability}.
The runtime is the average runtime over 20 different reduction rates on the same network. 
One can imagine this as averaging the runtimes shown in \cref{fig:basis-finding}.
We can see that the variance-based approach has almost constant runtime, whereas the runtime of the greedy approach is increasing for both a higher number of layers and neurons.
\subsubsection{Final Assessment}
We have four possibilities on how to abstract an NN: greedy orthogonal projection, greedy linear programming, heuristic-based orthogonal projection, and heuristic-based linear programming.
Given that the orthogonal projection outperforms linear programming in terms of accuracy and computation time, we propose to use orthogonal projection. 
We believe that it is sufficient to use the heuristic-based approach, thereby gaining faster runtimes and only barely sacrificing any accuracy.
Whenever we refer to \emph{LiNNA} from now on without any additions, it will be the heuristic-based orthogonal projection.

\subsection{Comparison to Existing Work}\label{sec:comparison-existing-work}
We want to show how our approach compares to existing works, i.e. \emph{DeepAbstract} and the \emph{bisimulation}.
Since there is no implementation available for the latter, we implemented it ourselves.
Please refer to \cref{sec:app-bisimulation} for the details.
The results of the comparison are shown in \cref{fig:rw-comp}. 
It is evident that DeepAbstract achieves higher accuracies than the bisimulation, but LiNNA outperforms DeepAbstract and the bisimulation in terms of accuracy for all reduction rates.

Concerning the runtime, we measure the runtime of each approach for a certain reduction rate, starting from the full network. 
We find that (in the median) LiNNA (greedy) needs 55s up to 199s, LiNNA (heuristic) 2s up to 3s, DeepAbstract 187s up to 2420s, and the bisimulation 1s up to 2s, on MNIST networks of different sizes (starting from 4x50 up to 11x100).
The details can be found in \cref{sec:time-tables}.
The bisimulation performs best, however just slightly ahead of the heuristic-based LiNNA.
The greedy LiNNA, as well as DeepAbstract both have a much higher computation time.

However, in terms of accuracy, greedy LiNNA seems to be the best-performing approach, given sufficient time. 
Due to efficiency, we suggest using heuristic-based LiNNA, as it is as fast as the bisimulation, but its accuracy is a lot better and even close to greedy LiNNA.

\begin{figure}[!t]
		\begin{subfigure}[t]{0.45\textwidth}
		\centering
		\includegraphics[width=\textwidth]{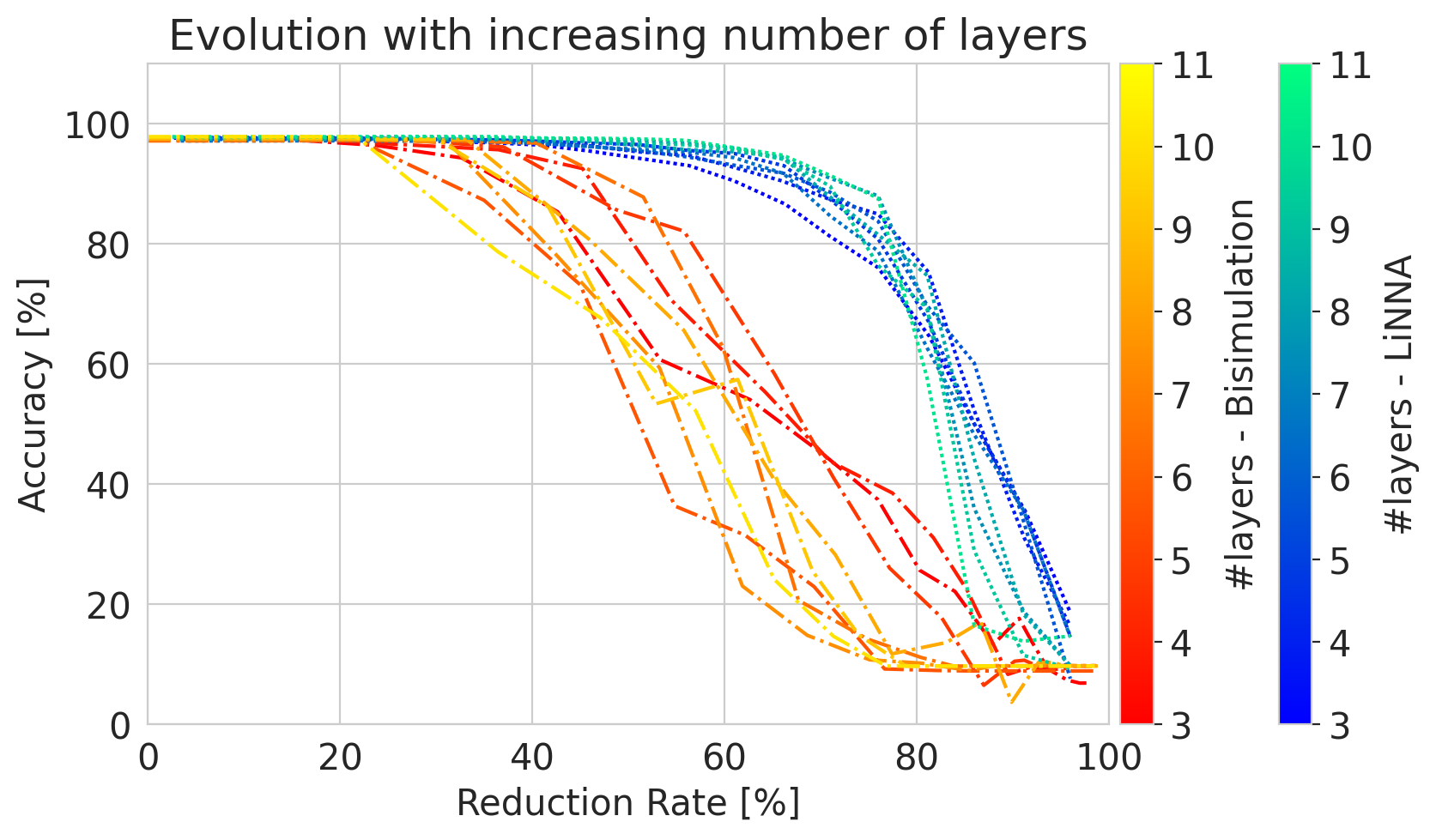}
		\label{fig:evolution-depth}
	\end{subfigure}\hfill
	\begin{subfigure}[t]{0.45\textwidth}
		\centering
		\includegraphics[width=\textwidth]{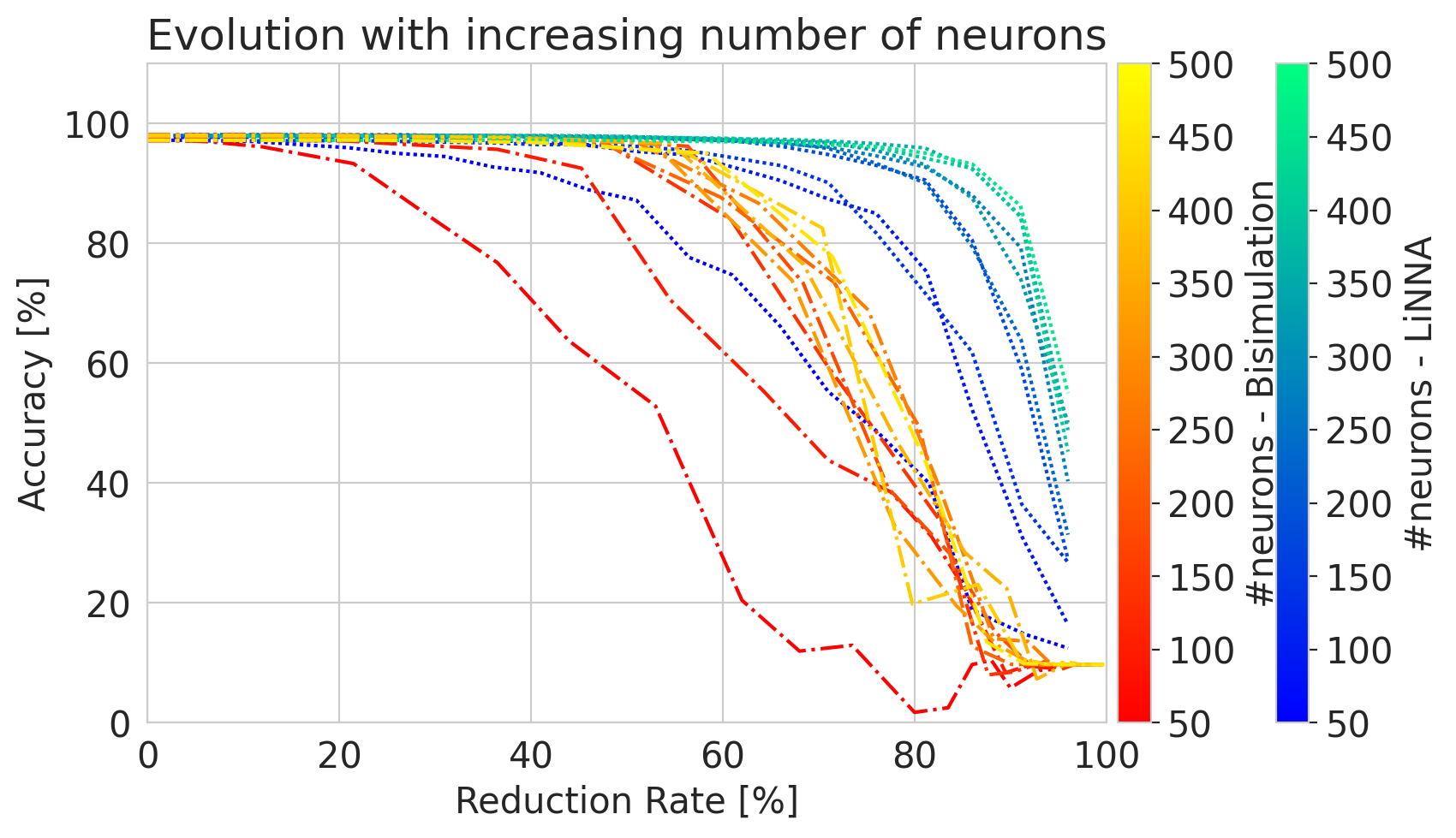}
		\label{fig:evolution-width}
	\end{subfigure}
	\caption{\emph{Evolution of the accuracy} on the test set for different reduction rates, for an increasing number of layers, or neurons. We show LiNNA (blue-green) for semantic abstraction, and for syntactic abstraction, bisimulation (red-yellow). The networks were trained on MNIST and have a fixed number of neurons (100) on the left, and a fixed number of layers (4) on the right.}
	\label{fig:evolution}
\end{figure}

Since we are interested in the general behavior of the abstraction, we want to see how the methods work for varying sizes of networks, but not only in terms of scalability. In \cref{fig:evolution}, we show the trend for bisimulation and LiNNA for an increasing number of layers resp. neurons per layer.
On the left, we fix the number of neurons per layer to 100 and incrementally increase the number of layers. 
On the right, we fix the number of layers to four and increase the number of neurons. 

We can see that the performance of the networks from the bisimulation varies a lot and gets slightly worse when there are more layers, whereas LiNNA has a very small variation and the performance of the abstractions increases slightly for more layers.
Both approaches compute abstractions that perform better the more neurons are in a layer, but LiNNA converges to a much steeper curve at high reduction rates. 

For NNs with 400 or more neurons, LiNNA can reduce 80\% of the neurons without a significant loss in accuracy, whereas the bisimulation can do the same only for up to a reduction rate of 55\%.

\subsection{Semantic vs Syntactic}
In the following, we want to show the differences between 
semantic and syntactic abstractions.
Recall that syntactic abstraction makes use of the weights of the network, the syntactic information, with no consideration of the actual behavior of the NN on the inputs.
Semantic abstraction, on the other hand, focuses on the values of the neurons on an input dataset, which also incorporates information about the weights.
DeepAbstract and LiNNA, both use semantic information, whereas bisimulation uses syntactic information.
We additionally evaluate the performance of LiNNA on syntactic information.

\emph{Which type of information is better for abstraction: semantic or syntactic?}
Note that both DeepAbstract and the bisimulation represent a group of neurons by one single representative, whereas LiNNA makes use of a linear combination.
\begin{figure}[!t]
	\centering
	\begin{minipage}[t]{0.48\textwidth}
		\includegraphics[width=\textwidth]{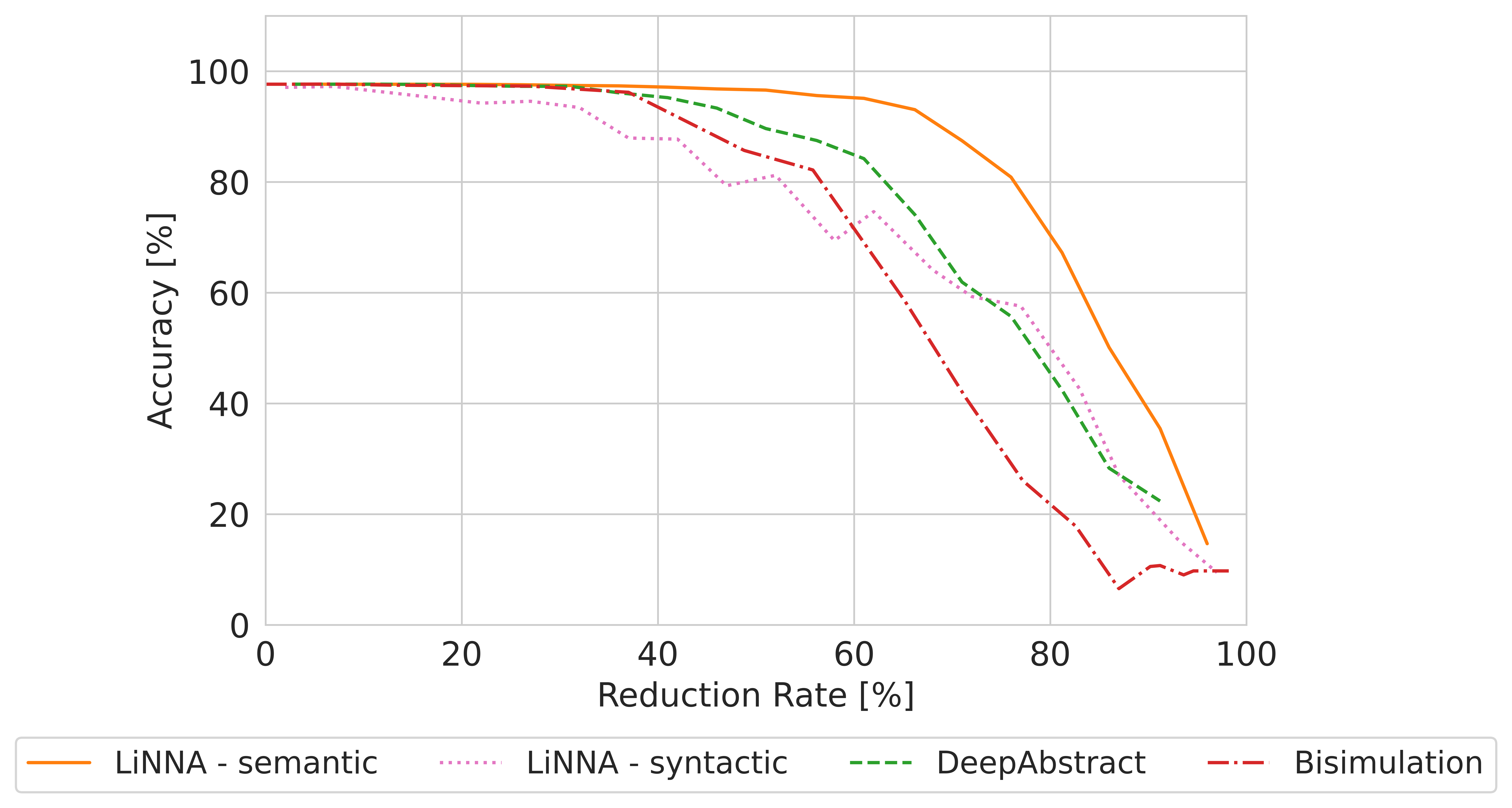}
		\caption{\emph{Syntactic VS. Semantic} - This plot shows the difference between using semantic resp. syntactic information for the abstraction on an MNIST 5x100 network. Semantic: LiNNA (semantic) and DeepAbstract. Syntactic: LiNNA (syntactic) and the bisimulation.}
		\label{fig:comp-syn-sem}
	\end{minipage}\hfill
	\begin{minipage}[t]{0.48\textwidth}
		\includegraphics[width=\textwidth]{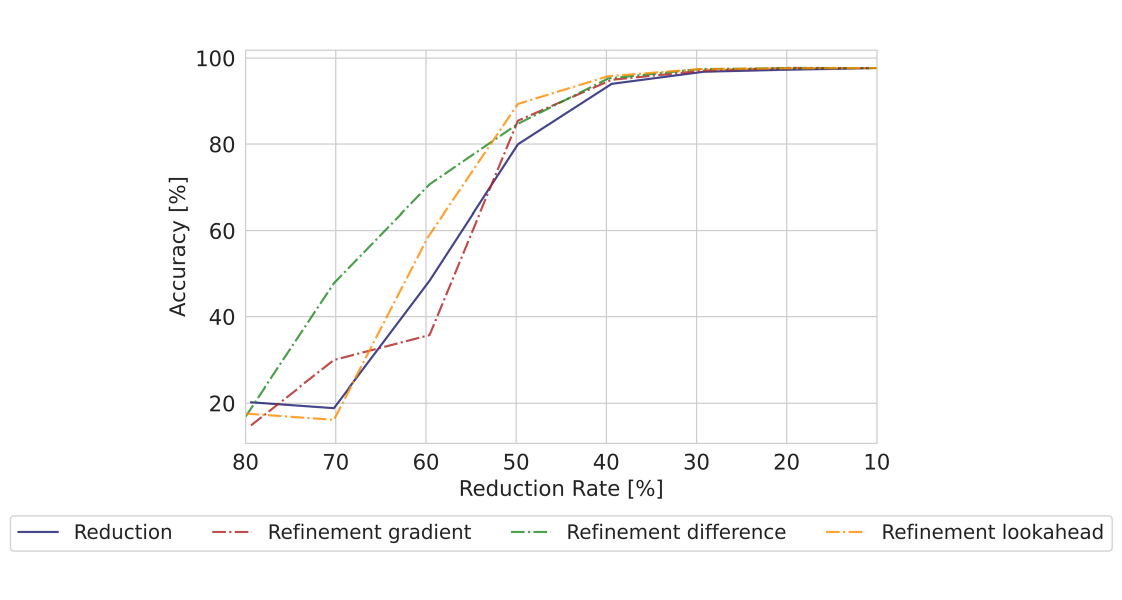}
		\caption{\emph{Refinement} - This plot shows the accuracy of an MNIST 5x100 network that was abstracted and refined to a certain reduction rate $R$. 			There is also a plot for an abstraction to the same reduction rate as after the refinement but without refining.}
		\label{fig:refinement-argument}	
	\end{minipage}
\end{figure}

We summarize our results in \cref{fig:comp-syn-sem}. 
For smaller reduction rates, the bisimulation performs better than LiNNA on syntactic information; for higher reduction rates it is reversed.
In general, the approaches based on semantics (DeepAbstract and LiNNA - semantic) outperform the other two approaches w.r.t. accuracy. While abstraction based on syntactic information can provide global guarantees for any input, abstraction based on semantic information relies on the fact that its inputs during abstraction are similar to the ones it will be evaluated on later. 
However, we see that still the semantic information is more appropriate for preserving accuracy because it combines the knowledge about possible inputs with the knowledge about the weights.

\FloatBarrier
\subsection{Refining the Network}
We propose refinement of the abstraction in cases where it does not capture all the behavior anymore, instead of restarting the abstraction process. We consider networks that are abstracted up to certain reduction rates, i.e. 20\%, 30\%, $\dots$, 90\%, and use the refinement to regain 10\% of the neurons. 
For example, we reduce the network by 90\% and then use refinement to get back to a reduction rate of 80\%. 
We evaluate this refined network on the test dataset and plot its accuracy.
Additionally, we show the accuracy of the same NN which is directly reduced to an 80\% reduction rate, without refinement.
This plot is shown in \cref{fig:refinement-argument} for a 5x100 network, trained on MNIST.

The gradient and look-ahead refinement have a similar performance.
However, the difference-based approach even outperforms the direct reduction itself. 
This behavior can be explained by the fact that the refinement and the abstraction look at different metrics for removing/restoring neurons. 
The refinement can focus directly on optimizing for the inputs at hand, whereas the abstraction was generated on the training set.
In conclusion, the refinement can even improve the abstraction and it is beneficial to abstract slightly more than required, and refine for the relevant inputs, rather than having a finer abstraction directly.

\begin{figure}[t!]
	\centering
	\begin{minipage}[t]{0.4\textwidth}
		\includegraphics[width=0.9\textwidth]{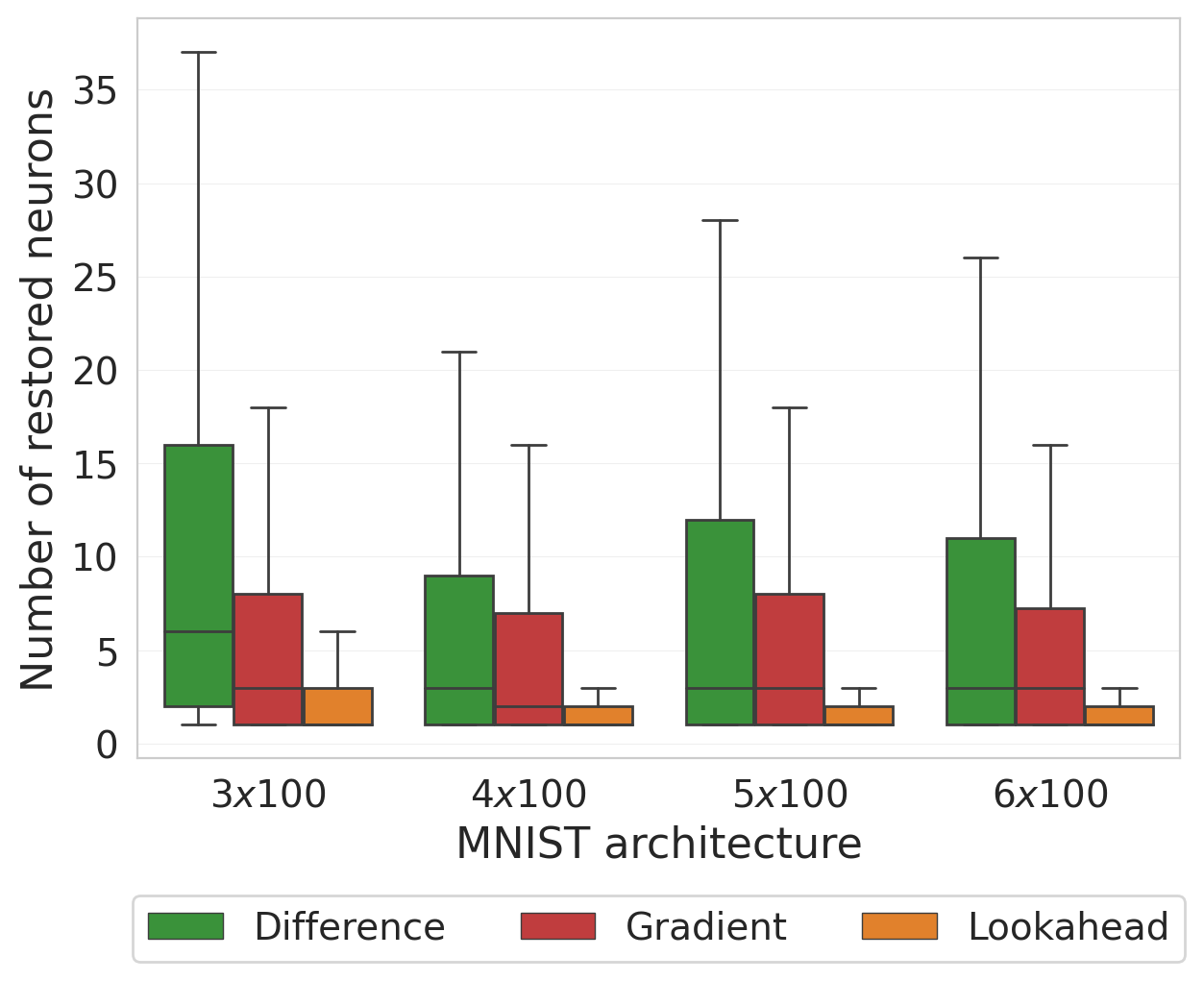}
		\caption{\emph{Comparison of refinement techniques} on different architectures for MNIST. The respective networks were abstracted with a reduction rate of $50\%$. The lines show the variance, the box represents 50\% of the data, the line in the box shows the median.}
		\label{fig:boxplot-refinement}
	\end{minipage}\hfill
	\begin{minipage}[t]{0.55\textwidth}
		\centering
		\includegraphics[width=0.9\textwidth]{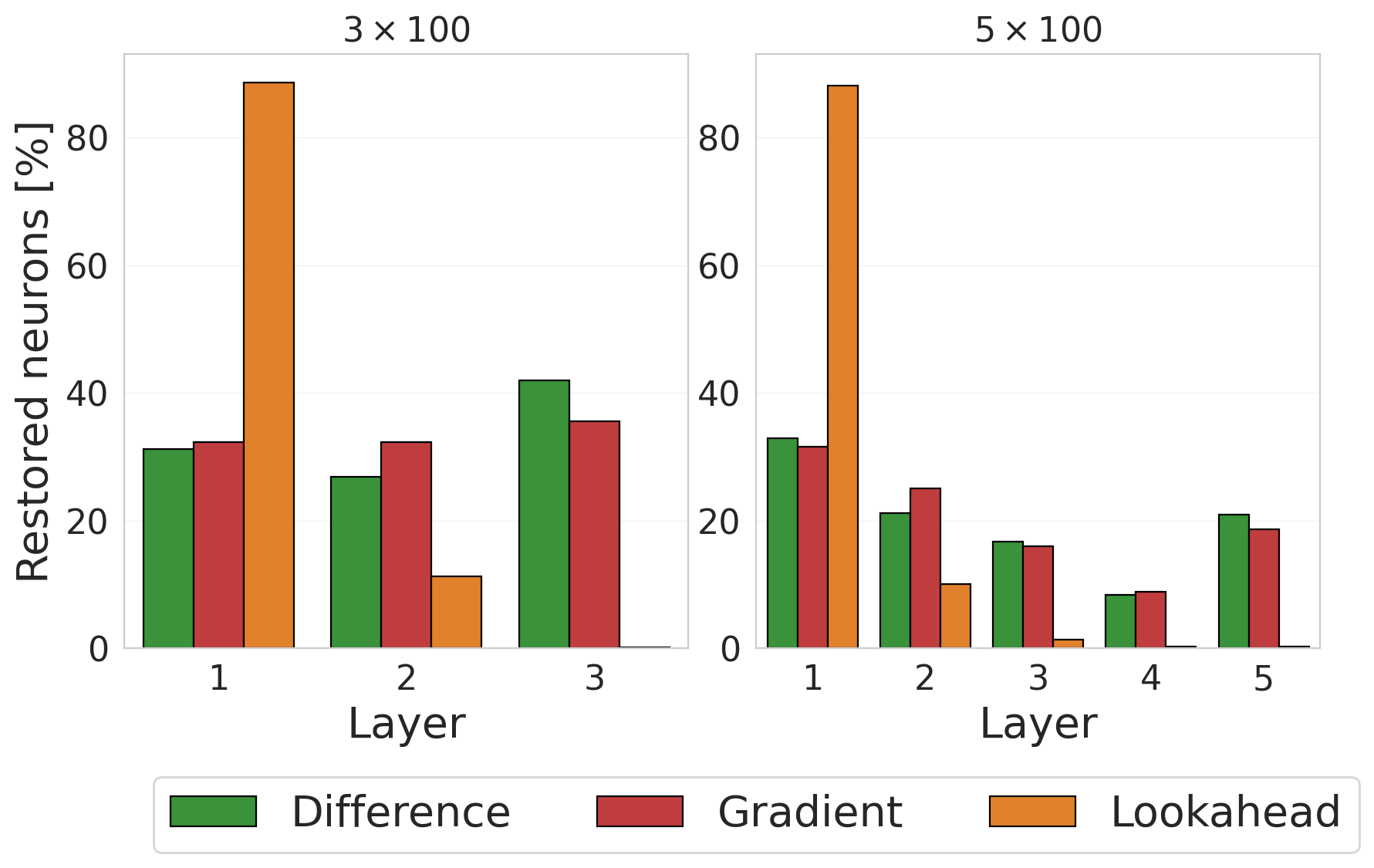}
		\caption{\emph{Refinement on different layers} - We considered abstractions that were obtained with a $50\%$ reduction rate and fixed $1000$ counterexamples. The plots depict the percentage of restored neurons in the layers of the different MNIST networks.}
		\label{fig:barplot-refinement}
	\end{minipage}\label{fig:refinement}
\end{figure}
\subsubsection{Comparison of the Different Approaches}
We collect images that are labeled differently by the abstraction and observe the number of neurons that are restored in order to fix the classification of each image. 
We ran the experiment on different networks that were abstracted with a $50\%$ reduction rate and considered $1000$ counterexamples for each network. 
The results are summarized in \cref{fig:boxplot-refinement}, where we have boxplots for each refinement method on four different network architectures. 
The look-ahead approach is the most effective technique since it requires the smallest number of restored neurons. 
In the median, it only requires $1$ to $2$ operations. 
The gradient-based approach performs noticeably worse but outperforms the difference-based approach on all networks.
The computation time, however, gives a different perspective: Repairing one counterexample takes on average $<$1s for the difference-based approach, 1s for the gradient-based, but the look-ahead approach takes on average 4s. 
Interestingly, the look-ahead approach restores fewer neurons but performs worse in accuracy. The difference-based performs better in terms of accuracy while restoring more neurons. 
\subsubsection{Insight on the Relevance of Layers}
We also investigated in which layers the different refinement techniques tend to restore the neurons. 
The plots in \cref{fig:barplot-refinement} illustrate the percentage of restored neurons in each layer. Notably, the look-ahead approach restores most neurons in the first layer, and very few or none in the later layers, whereas the other approaches have a more uniform behavior.
However, the more layers the network has, the more the gradient- and difference-based approaches tend to restore more neurons in the first layer.
As reported already by \cite{deepabstract}, the first layers seem to have a larger influence on the network's output and hence should be focused on during refinement.
It is even more interesting that the difference-based approach does not focus on the first layers as much as the look-ahead approach, but it is better in terms of accuracy. \FloatBarrier
\subsection{Error Calculation}
We want to show how the abstraction simulates the original network on unseen data not only w.r.t.~the output but on every single neuron.
In other words, is the discrepancy between the concrete and abstract network higher on the \emph{test} data than on the \emph{training} data that generate the abstraction, or does the link between the neuron and its linear abstraction generalize well?

\begin{figure}[t!]
	\centering
	\begin{minipage}[t]{0.3\textwidth}
		\centering
		\includegraphics[width=0.9\textwidth]{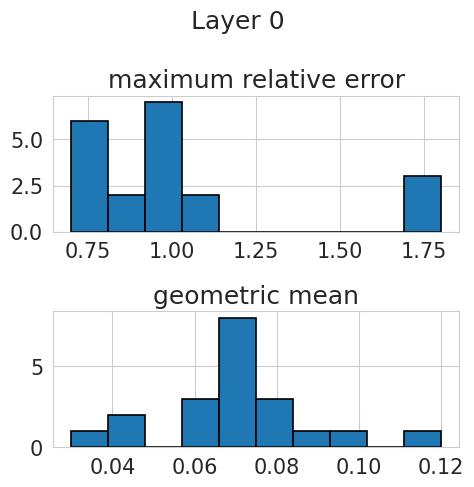}
	\end{minipage}
	\begin{minipage}[t]{0.3\textwidth}
		\centering
		\includegraphics[width=0.935\textwidth]{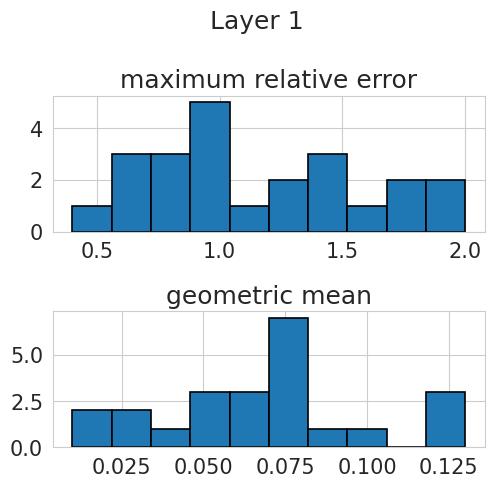}
	\end{minipage}
	\begin{minipage}[t]{0.3\textwidth}
		\centering
		\includegraphics[width=0.9\textwidth]{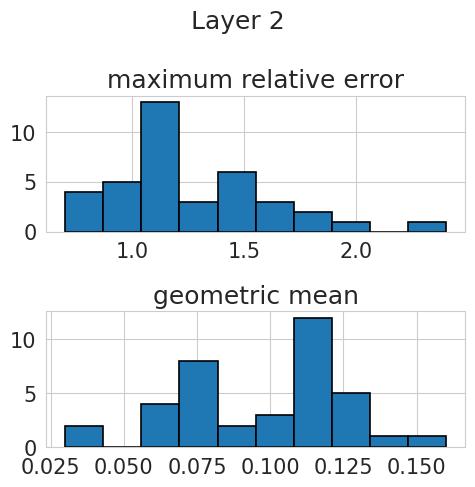}
	\end{minipage}
	\caption{\emph{Histograms of the relative error} of the values of the neurons in an MNIST 3x100 network and its abstraction (reduced by 30\%). 		The first row shows the maximum relative error of each neuron in the NN, that occurred for some input from the test set. The second row shows the geometric mean of the relative error of each neuron over 100 images of the test set.}
	\label{fig:error}
\end{figure}

In \cref{fig:error}, we look at this ratio (``relative error of the abstraction''), i.e. the absolute difference of (activation values of) a simulated abstract neuron to the original neuron, once on the test dataset divided by the maximum value on the training dataset. 
We can see that there are cases where the error can be greater than one (meaning ``larger than on the training set''), see the first row of the plot. However, the geometric mean, defined as $\left(\Pi_{i=1}^Na_i \right)^{\frac{1}{N}}$, calculated over all images is very small. 
Note that more experiments can be found in \cref{sec:appendix-error}.
In conclusion, we can say that our abstraction is close to the original also on the test dataset, although the theoretical error calculation does not guarantee so tight a simulation.
Future work should reveal how to further utilize the empirical proximity in transferring the reasoning from the abstraction to the original.

\section{Conclusions}
The focus of this work was to examine abstraction not as a part of a verification procedure, but rather as a stand-alone transformation, which can later be used in different ways: as a preprocessing step for verification, as means of obtaining an equivalent smaller network, or to gain insights about the network and its training, such as identifying where redundancies arise in trained neural networks. 
(This is analogous to the situation of bisimulation, which has been largely investigated on its own not necessarily as a part of a verification procedure, and its use in verification is only one of the applications.)

We have introduced \emph{\linna}, which abstracts a network by replacing neurons with \emph{linear combinations} of other neurons and also equip it with a \emph{refinement} method. We bound the error and thus the difference between the abstraction and the original network in Theorem \ref{th:bisim-error}. The theorem yields a lower and an upper bound on the network’s output, thereby providing its over-approximation. 

We showed that the linear extension provides better performance than existing work on abstraction for classification networks, both DeepAbstract, and the bisimulation-based approach. 
We focused our experimental evaluation on \emph{accuracy}, since the aim of the abstraction is to faithfully mimic the \emph{whole classification process} in the smaller, abstract network, not just one concrete property to be verified, which describes only a very specific aspect of the network.
Interestingly, the practical error is dramatically smaller than the worst-case bounds. 
We hope this first, experimental step will stimulate interest in research that could utilize this actual advantage, which is currently not supported by any respective theory.

Furthermore, we show that the use of \emph{semantic information should be preferred} over syntactic information because it allows for higher reductions while preserving similar behavior and being cheap since the I/O sets  can be quite small.
Bringing back semantics could take us closer to the efficiency of classical software abstraction, where the semantics of states is the very key, going way beyond bisimulation.

\newpage

\printbibliography

\newpage
\appendix
\section*{Appendix}
\section{Proof of \cref{prop:sem-equi}}\label{sec:proof-of-prop1}
\begin{proof}
	We show that $\vecz^{(\ell+1)}(\vecx)=\tilde{\vecz}^{(\ell+1)}(\vecx)$ for all $\vecx\in X$. After this, it follows immediately that $\tilde{\NN}(\vecx)=\NN(\vecx)$.
	
	Let $\vecx\in X$ be fixed. We write $\vecz^{(\ell+1)}$ instead of $\vecz^{(\ell+1)}(\vecx)$ to abbreviate. We have $\vecz^{(\ell+1)}=\phi(\matr{W}{\ell}\vecz^{(\ell)}+\vecb^{(\ell+1)})$ and $\tilde{\vecz}^{(\ell+1)}=\phi(\matr{\tilde{W}}{\ell}\vecz^{(\ell)}+\vecb^{(\ell+1)})$, since $\vecz^{(\ell)}$ has not changed.
	
	For each neuron $s$ of layer $\ell+1$, we have:
	\begin{align}
\nonumber\tilde{z}_s^{(\ell+1)}&=\phi\left(\matr{\tilde{W}}{\ell}_{s,*}\vecz^{(\ell)}+b_s^{(\ell+1)}\right)\\
\label{eq:prop1-l1}&=\phi\Bigg(\sum_{t\in\{1,\dots,n_\ell\}}\weighta{\ell}{s}{t}z_t^{(\ell)} + b_s^{(\ell+1)}\Bigg)\\
\label{eq:prop1-l2}&=\phi\Bigg(\sum_{t\notin\basis{\ell},t\neq i}\weight{\ell}{s}{t}z_t^{(\ell)}\\
\nonumber&\hspace{9mm}+\sum_{t\in\basis{\ell}}\weighta{\ell}{s}{t}z_t^{(\ell)} + \weighta{\ell}{s}{i}z_i^{(\ell)} + b_s^{(\ell+1)}\Bigg)\\
\label{eq:prop1-l3}&=\phi\Bigg(\sum_{t\notin\basis{\ell},t\neq i}\weight{\ell}{s}{t}z_t^{(\ell)}\\
\nonumber&\hspace{9mm}+\sum_{t\in\basis{\ell}}(\weight{\ell}{s}{t}+\alpha_t\weight{\ell}{s}{i})z_t^{(\ell)} + \weighta{\ell}{s}{i}z_i^{(\ell)} + b_s^{(\ell+1)}\Bigg)\\
\label{eq:prop1-l4}&=\phi\Bigg(\sum_{t\notin\basis{\ell},t\neq i}\weight{\ell}{s}{t}z_t^{(\ell)}+\sum_{t\in\basis{\ell}}\weight{\ell}{s}{t}z_t^{(\ell)}\\
\nonumber&\hspace{9mm}+ \sum_{t\in\basis{\ell}}\alpha_t\weight{\ell}{s}{i}z_t^{(\ell)} + b_s^{(\ell+1)}\Bigg)\\
\label{eq:prop1-l5}&=\phi\Bigg(\sum_{t\notin\basis{\ell},t\neq i}\weight{\ell}{s}{t}z_t^{(\ell)}+\sum_{t\in\basis{\ell}}\weight{\ell}{s}{t}z_t^{(\ell)} \\
\nonumber&\hspace{9mm}+ \weight{\ell}{s}{i}z_i^{(\ell)} + b_s^{(\ell+1)}\Bigg)\\
\label{eq:prop1-l6}&=\phi\Bigg(\sum_{t\in\{1,\dots,n_\ell\}}\weight{\ell}{s}{t}z_t^{(\ell)} + b_s^{(\ell+1)}\Bigg)\\
\nonumber&=z_s^{(\ell+1)}
	\end{align}
	where
	\begin{itemize}
		\item\cref{eq:prop1-l1} is taking apart the matrix multiplication.
		\item\cref{eq:prop1-l2} splits the neurons of layer $\ell$ into three classes: neuron $i$ itself, neurons within the basis $t\in\basis{\ell}$, and neurons not in the basis. Neurons that are not part of the basis don't change their weights in the replacement process (\ref{sec:replacement}) (usually they would be removed in a later stage). 
		\item\cref{eq:prop1-l3} replaces $\weighta{\ell}{s}{t}$ by the original weights $\weight{\ell}{s}{t}$ plus the additional weights from neuron $i$, i.e. $\alpha_t\weight{\ell}{s}{i}$.
		\item\cref{eq:prop1-l4} is just sorting the weights in anoter way
		\item\cref{eq:prop1-l5} uses the equality $\z^{(\ell)}_i(\vecx)=\sum_{j\in\basis{\ell}}\lcoeff{i}{j}{\ell}\z^{(\ell)}_j(\vecx)$.
		\item\cref{eq:prop1-l6} gathers all sums in one sum again. This works now because we have a sum over all three classes (neuron $i$ itself, neurons within the basis $t\in\basis{\ell}$, and neurons not in the basis) only over the original weights.
	\end{itemize}
	\qed
	 
\end{proof}
\section{Proof of \cref{th:bisim-error}}\label{sec:proof-of-th1}
Following the idea of \cite{Prabhakar22}, we first introduce a lemma that shows the difference of the abstraction to the original for one layer and then show how the theorem follows from that.
\begin{lemma}[Replacement Error of One Layer]\label{lem:bisim-error}
	Let $N$ be a neural network with $L$ layers. For one layer $\ell$, we have a basis of neurons $\basis{\ell}$, and a set of replaced neurons $\repl{\ell}$. Then, let $\tilde{N}$ be the network after replacing neurons in $\repl{\ell}$ as described in \cref{sec:replacement}. Furthermore, let $\lambda^{(\ell)}$ be the Lipschitz constant of the activation function of that layer.
	If for all inputs $\vecx\in X\subset \mathcal{X}$, we have
	\begin{enumerate}
		\item for $i\in\repl{\ell}\; :\;|\z^{(\ell)}_i(\vecx)-\sum_{j\in\basis{\ell}}\lcoeff{i}{j}{\ell}\z^{(\ell)}_j(\vecx)|\leq\epsilon^{(\ell)}$
		\item $|\sum_{i\in I^{(\ell)}}\smash{\matr{W}{\ell}_{*,i}}\sum_{t\in\basis{\ell}}\lcoeff{i}{j}{\ell}|\leq\eta^{(\ell)}$
		\item $\|\tilde{\vecz}^{(\ell)}(\vecx)-\vecz^{(\ell)}(\vecx)\|\leq\delta^{(\ell)}$
	\end{enumerate}
	then, we get $\|\tilde{\vecz}^{(\ell+1)}-\vecz^{(\ell+1)}\|\leq \lambda^{(\ell)}(\|\matr{W}{\ell}\|_1\delta^{(\ell)}+\|\matr{W}{\ell}\|_1\epsilon^{(\ell)}+\eta^{(\ell)}\delta^{(\ell)})$
\end{lemma}
Note that the error value $\epsilon$ talks about the differences in the values of the neurons before and after replacement, but without any change to any other layer. 
In constrast, the term $\delta$ references the difference between the values of the neurons before and after changing layers before layer $\ell$.
\begin{proof}[of \cref{lem:bisim-error}]
	Let $\vecx\in X$ be fixed. We write $\vecz^{(\ell+1)}$ instead of $\vecz^{(\ell+1)}(\vecx)$ to abbreviate. 
	We have $\vecz^{(\ell+1)}=\phi(\vech^{(\ell+1)})$ and similarly $\tilde{\vecz}^{(\ell+1)}=\phi(\tilde{\vech}^{(\ell+1)})$.
	It is 
	\begin{align*}
		\|\tilde{\vech}^{(\ell+1)}-\vech^{(\ell+1)}\|&=\|\matr{\tilde{W}}{\ell}\tilde{\vecz}^{(\ell)}+\vecb^{(\ell+1)} - (\matr{W}{\ell}\vecz^{(\ell)}+\vecb^{(\ell+1)})\|\\
		&=\|\sum_{t\in\basis{\ell}} \left( \smash{\matr{\tilde{W}}{\ell}_{*,t}}\tilde{z}^{(\ell)}_t\right) +\vecb^{(\ell+1)} \\
		&\hspace{5mm}-\sum_{t\in\basis{\ell}\cup\repl{\ell}} \left( \smash{\matr{W}{\ell}_{*,t}}z^{(\ell)}_t\right) - \vecb^{(\ell+1)}\|\\
		&=\|\sum_{t\in\basis{\ell}}\left(\smash{\matr{\tilde{W}}{\ell}_{*,t}}\tilde{z}^{(\ell)}_t-\smash{\matr{W}{\ell}_{*,t}}z^{(\ell)}_t\right)-\sum_{t\in\repl{\ell}} \left( \smash{\matr{W}{\ell}_{*,t}}z^{(\ell)}_t\right)\|
		\intertext{Using the replacement $\matr{\tilde{W}}{\ell}_{*,j}=\matr{W}{\ell}_{*,j}+\lcoeff{i}{j}{\ell}\matr{W}{\ell}_{*,i}$ from (\ref{eq:repl2})}\\
		&=\|\sum_{t\in\basis{\ell}}\left(\left(\matr{W}{\ell}_{*,t}+\sum_{i\in\repl{\ell}}\lcoeff{i}{t}{\ell}\matr{W}{\ell}_{*,i}\right)\tilde{z}^{(\ell)}_t-\smash{\matr{W}{\ell}_{*,t}}z^{(\ell)}_t\right)\\
		&\hspace{5mm}-\sum_{t\in\repl{\ell}} \left( \smash{\matr{W}{\ell}_{*,t}}z^{(\ell)}_t\right)\|\\
		&=\|\sum_{t\in\basis{\ell}} \left( \matr{W}{\ell}_{*,t}(\tilde{z}^{(\ell)}_t-z^{(\ell)}_t)+\sum_{i\in\repl{\ell}} \lcoeff{i}{t}{\ell}\matr{W}{\ell}_{*,i}\tilde{z}^{(\ell)}_t\right) \\
		&\hspace{5mm}-\sum_{t\in\repl{\ell}}\left( \smash{\matr{W}{\ell}_{*,t}}z^{(\ell)}_t\right) \|\\
		\intertext{with 3., we get}
		&\leq\|\sum_{t\in\basis{\ell}}\left(\matr{W}{\ell}_{*,t}\delta^{(\ell)}\right)\\
		&\hspace{5mm}+\sum_{i\in\repl{\ell}}\sum_{t\in\basis{\ell}}\left(\lcoeff{i}{t}{\ell}\matr{W}{\ell}_{*,i}\tilde{z}^{(\ell)}_t-\smash{\matr{W}{\ell}_{*,i}}z^{(\ell)}_i\right)\|\\
		&\leq\|\sum_{t\in\basis{\ell}}\left(\matr{W}{\ell}_{*,t}\delta^{(\ell)}\right)\\
		&\hspace{5mm}+\sum_{i\in\repl{\ell}}\sum_{t\in\basis{\ell}}\left(\lcoeff{i}{t}{\ell}\matr{W}{\ell}_{*,i}(z_t^{(\ell)}+\delta^{(\ell)})-\smash{\matr{W}{\ell}_{*,i}}z^{(\ell)}_i\right)\|\\
		&\leq\|\sum_{t\in\basis{\ell}}\left(\matr{W}{\ell}_{*,t}\delta^{(\ell)}\right)+\sum_{i\in\repl{\ell}}\Bigg(\matr{W}{\ell}_{*,i}\sum_{t\in\basis{\ell}}\left(\lcoeff{i}{t}{\ell}z_t^{(\ell)}\right)-z^{(\ell)}_i\\
		&\hspace{5mm}+\sum_{t\in\basis{\ell}}\left(\lcoeff{i}{t}{\ell}\matr{W}{\ell}_{*,i}\delta^{(\ell)}\right)\Bigg)\|\\
		\intertext{with 1., we get}
		&\leq\|\delta^{(\ell)}\sum_{t\in\basis{\ell}}\left(\matr{W}{\ell}_{*,t}\right)
		+\epsilon^{(\ell)}\sum_{i\in\repl{\ell}}\left(\matr{W}{\ell}_{*,i}\right)\\
		&\hspace{5mm}+\delta^{(\ell)}\sum_{i\in\repl{\ell}}\sum_{t\in\basis{\ell}}\left(\lcoeff{i}{t}{\ell}\matr{W}{\ell}_{*,i}\right)\|\\
		\intertext{with 2., we get}
		&\leq\|\delta^{(\ell)}\sum_{t\in\basis{\ell}}\left(\matr{W}{\ell}_{*,t}\right)
		+\epsilon^{(\ell)}\sum_{i\in\repl{\ell}}\left(\matr{W}{\ell}_{*,i}\right)
		+\eta^{(\ell)}\delta^{(\ell)}\|\\
		&\leq\|\matr{W}{\ell}\|_1\delta^{(\ell)}+\|\matr{W}{\ell}\|_1\epsilon^{(\ell)}+\eta^{(\ell)}\delta^{(\ell)}
		\end{align*}
	From the Lipschitz-continuity of $\phi$, we get
	$$\|\tilde{\vecz}^{(\ell+1)}-\vecz^{(\ell+1)}\|\leq \lambda^{(\ell)}(\|\matr{W}{\ell}\|_1\delta^{(\ell)}+\|\matr{W}{\ell}\|_1\epsilon^{(\ell)}+\eta^{(\ell)}\delta^{(\ell)})$$
\qed
\end{proof}

Given the proof of \cref{lem:bisim-error}, we can now easily give the proof for \cref{th:bisim-error}.

\begin{proof}[of \cref{th:bisim-error}]
	We define $\lambda=\max_\ell \lambda^{(\ell)}$, $\|W\|=\max_\ell \|W^{(\ell)}\|_1$, $\eta=\max_\ell \eta^{(\ell)}$, and $\epsilon=\max_\ell \epsilon^{(\ell)}$.
	The first layer, the input layer, cannot be changed, thus we have $\delta^{(1)}=0$.
	
    We set $a=\lambda(\|W\|+\eta)$ and $b=\lambda\|W\|\epsilon$.
	
	By \cref{lem:bisim-error}, we have for the last layer $L$ that 
	$$\|\tilde{\vecz}^{(L)}-\vecz^{(L)}\|\leq \lambda^{(L-1)}(\|\matr{W}{L-1}\|_1\delta^{(L-1)}+\|\matr{W}{L-1}\|_1\epsilon^{(L-1)}+\eta^{(L-1)}\delta^{(L-1)})$$
	
	It holds that $\|\tilde{\vecz}^{(L)}-\vecz^{(L)}\|\leq a\delta^{(L-1)}+b$, since $a$ and $b$ consist of the maxima over all layers.
	Unrolling this leads to
	$$ a\delta^{(L-1)}+b \leq a(a\delta^{(L-2)}+b)+b \leq ... \leq a^L\delta^{(1)} + \sum_{i=0}^{L-2}ba^i$$
	and from $\delta^{(1)}=0$, we remain with $\sum_{i=0}^{L-2}ba^i=b(1-a^{L-1})/(1-a)$.
	\qed
\end{proof}

\section{Linear Program for Finding the Coefficients}\label{sec:appendix-linear-program}
We want to minimize $\|\sum_{j\in\basis{\ell}}\lcoeff{i}{j}{\ell}\cdot \mathbf{Z}_{j,*}^{(\ell)}-\mathbf{Z}_{i,*}\|$ for $\alpha_{i,j}^{(\ell)}$.
In the following, we describe the linear program for computing optimal coefficients $\alpha_{i,j}^{(\ell)}$.

\begin{equation}
	\begin{array}{lr@{}c@{}r@{}l}
		\text{minimize}   & \vecone^\top \vecbeta^+ + \vecone^\top \vecbeta^-  \\
		\text{subject to } & \sum_{j \in B} \vecz_j^{(\ell)} \lcoeff{i}{j}{\ell} - \id_d \vecbeta^+ + \id_d \vecbeta^- = \vecz^{(\ell)}_i\\
		&\lcoeff{i}{j}{\ell} \in \R \text{ for } j \in B,\;\;
		\vecbeta^+, \vecbeta^- \in {\R}^d,\;\;
		\vecbeta^+, \vecbeta^- \geq 0
	\end{array}
\end{equation}

We use $\vecone$ to denote a vector containing only ones and $\id_d$ to denote the identity matrix with $d$ rows and columns. 
Since the equation $\|\sum_{j\in\basis{\ell}}\lcoeff{i}{j}{\ell}\cdot \mathbf{Z}_{j,*}^{(\ell)}-\mathbf{Z}_{i,*}\|$ typically does not have an exact solution, we introduce the \emph{slack variables} $\vecbeta^+$ and $\vecbeta^-$. 
This is a common method to create a relaxed optimization problem for which solutions exist.
Note that the sum of the two slack variables gives us the L1 distance between the obtained linear combination and the actual activation. 
Particularly, this means that the optimal coefficients $\lcoeff{i}{j}{\ell}$ give us the best solution w.r.t. L$_1$ distance.
\newpage
\section{Refinement Heuristics}\label{sec:refinement-heuristics}
\subsubsection{Difference-guided Refinement}
Let $\vecx$ be a counterexample.
The difference guided refinement considers the difference between a neuron $\vecz^{(\ell)}_i(\vecx)$ in the original network and the linear combination $\sum_{j \in \basis{\ell}} \alpha_j \vecz^{(\ell)}_j(\vecx)$ with which it was replaced in the abstracted network.
We evaluate this for all neurons and restore the neuron with the largest difference. 
Intuitively, a large difference suggests that the linear combination does not accurately capture the behavior of the neuron on the given counterexample.

\subsubsection{Gradient-guided Refinement}
The gradient-guided refinement follows a similar approach, but additionally, we take the gradient into account. 
Suppose the original network outputs $\vecy$ and the reduced network $\vec{\tilde y}$. 
We can quantify the error with a typical loss function $g(\vecy, \tilde{\vecy})$ for NN outputs
, e.g. mean-squared error or cross-entropy loss \cite{Wang2020}. 
For the neurons in the basis $j \in {B}^{(\ell)}$ and all neurons of the layers before, $j' \in \{1,\dots,n_{\ell-1}\}$, we then compute the gradient of the loss function w.r.t. their incoming weights.
That is we compute the gradient $\diff{g}{\weight{\ell}{j}{j'}}$, which can be done with the \emph{back-propagation} algorithm as generally used for training a NN \cite{DBLP:books/lib/Bishop07}. 
Afterward, we simulate a single optimization step with gradient descent. Consequently, the values of the neurons in the basis change and, thus, also the value of the linear combination of a replaced neuron $i$. 
Let us denote the new value of the linear combination for neuron $i$ by $\bar{\z}_i$. 
We evaluate $(\bar{\z}_i - \z^{\ell}_{i}) \cdot (\sum_{j \in \basis{\ell}} \alpha_j \z^{\ell}_{j} - \z^{\ell}_{i})$ on the counterexample. 
The first factor $\bar{\z}_i - \z^{\ell}_{i}$ describes the difference between the updated linear combination and the actual output and it indicates whether a neuron's output value needs to be decreased or increased to minimize the loss function. 
The second factor $\sum_{j \in \basis{\ell}} \alpha_j \z^{\ell}_{j} - \z^{\ell}_{i}$ shows how the value would change if we were to restore the neuron.
Therefore, we choose the neuron with the largest value and restore it. 

While the difference-guided refinement only considers the \emph{local behavior} of a neuron, the gradient-guided refinement also takes the influence of a neuron on the network's output into account.

\subsubsection{Look-ahead Refinement}
The look-ahead refinement is the most greedy approach for refinement, where we simulate the restoration of each replaced neuron and observe how it changes the difference between the output $\tilde{\vecy}$ of the abstraction and $\vecy$ of the original network.
We then choose to restore the neuron that minimizes this difference. 
Again, we can quantify the difference with an appropriate loss function, as we have done in the gradient-guided refinement.

\newpage
\section{Supplemental Experiments on Finding a Basis}\label{sec:basis-finding}
In this section, we provide more experiments on how to find a basis, similar to \cref{fig:basis-finding}. 
\begin{figure}[!h]
	\centering
	\begin{subfigure}[]{0.48\textwidth}
		\includegraphics[width=\textwidth]{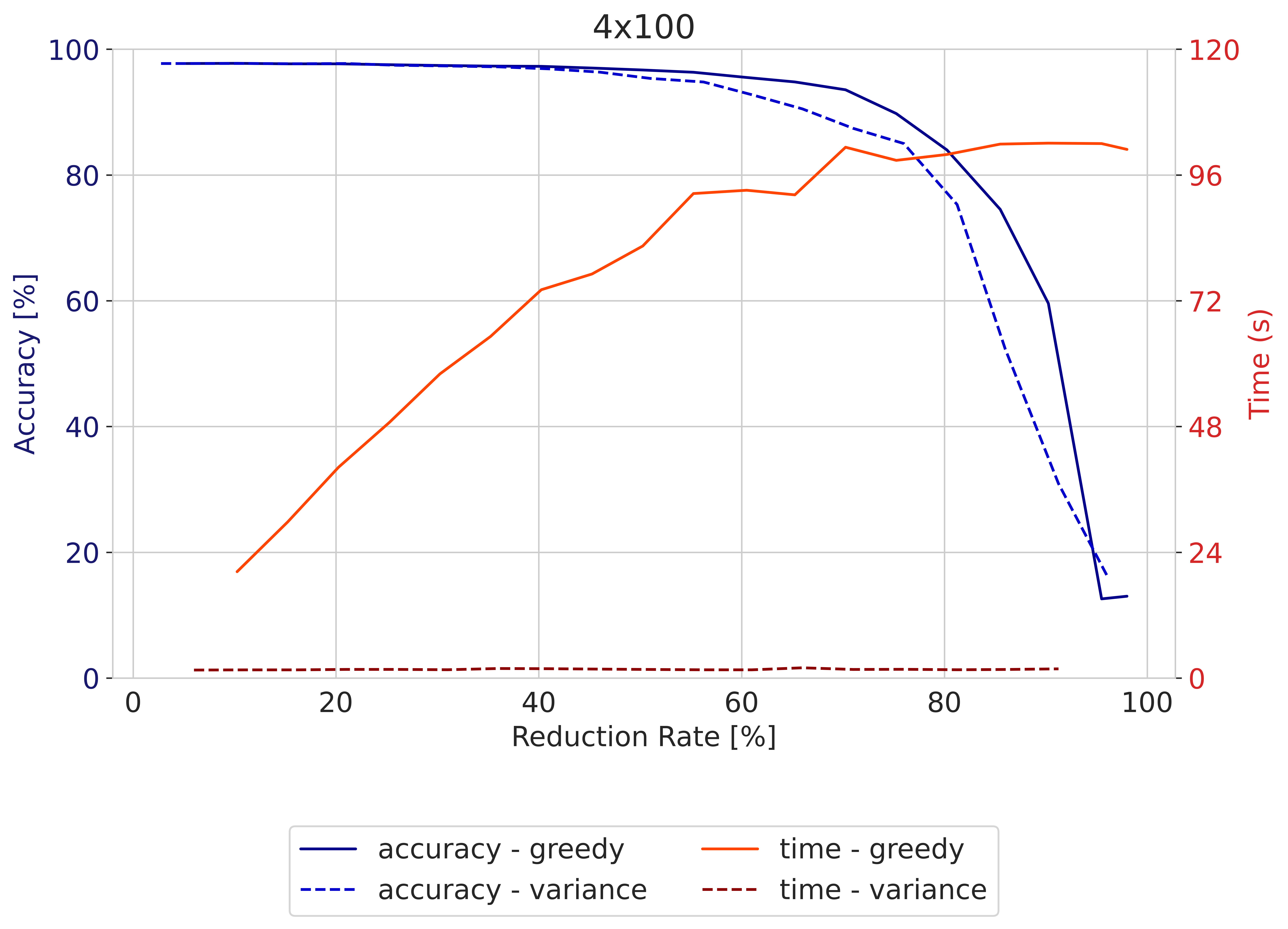}
	\end{subfigure}\hfill
	\begin{subfigure}[]{0.48\textwidth}
		\includegraphics[width=\textwidth]{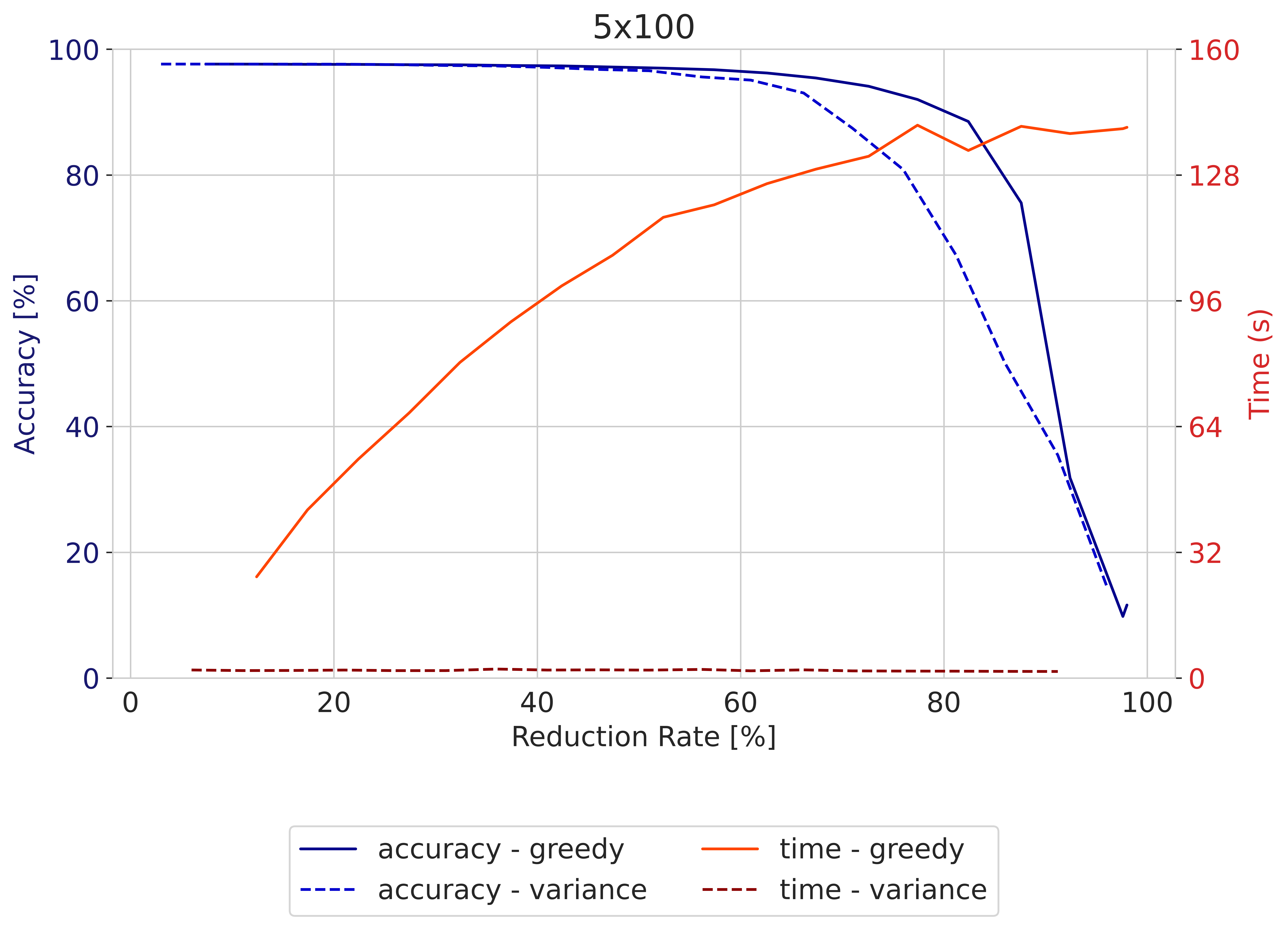}
	\end{subfigure}

	\begin{subfigure}[]{0.48\textwidth}
		\includegraphics[width=\textwidth]{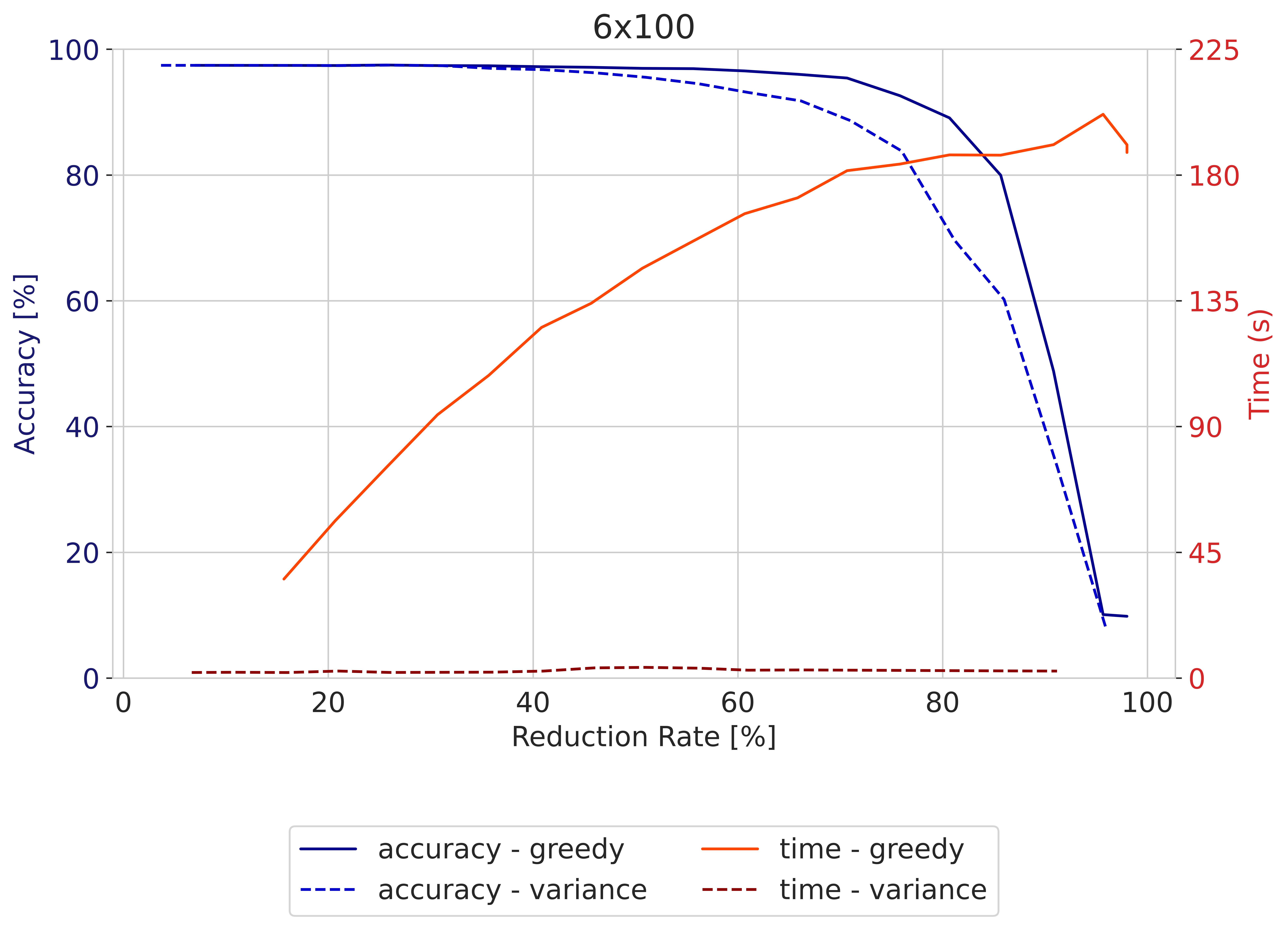}
	\end{subfigure}\hfill
	\begin{subfigure}[]{0.48\textwidth}
		\includegraphics[width=\textwidth]{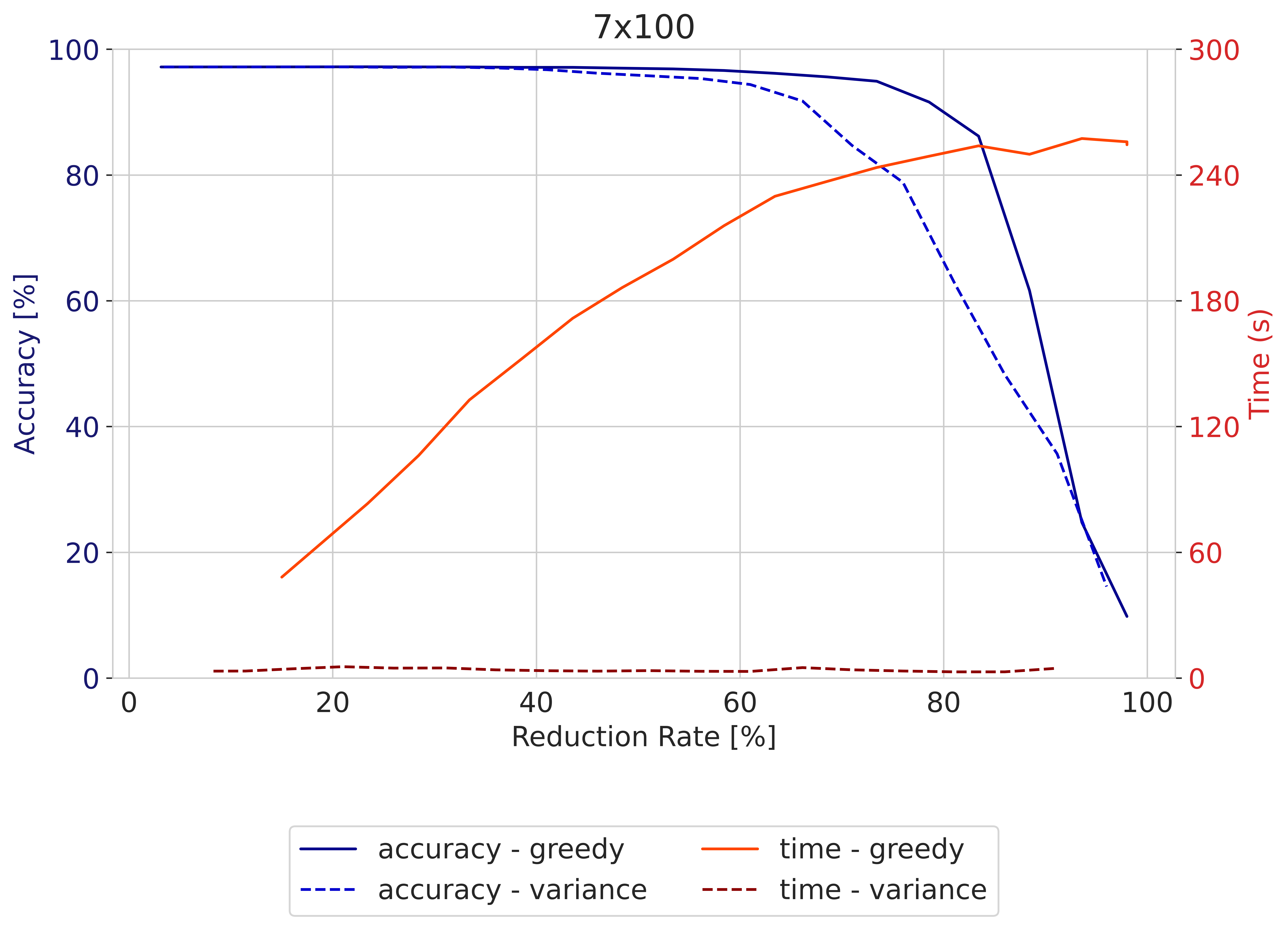}
	\end{subfigure}

	\begin{subfigure}[]{0.48\textwidth}
		\includegraphics[width=\textwidth]{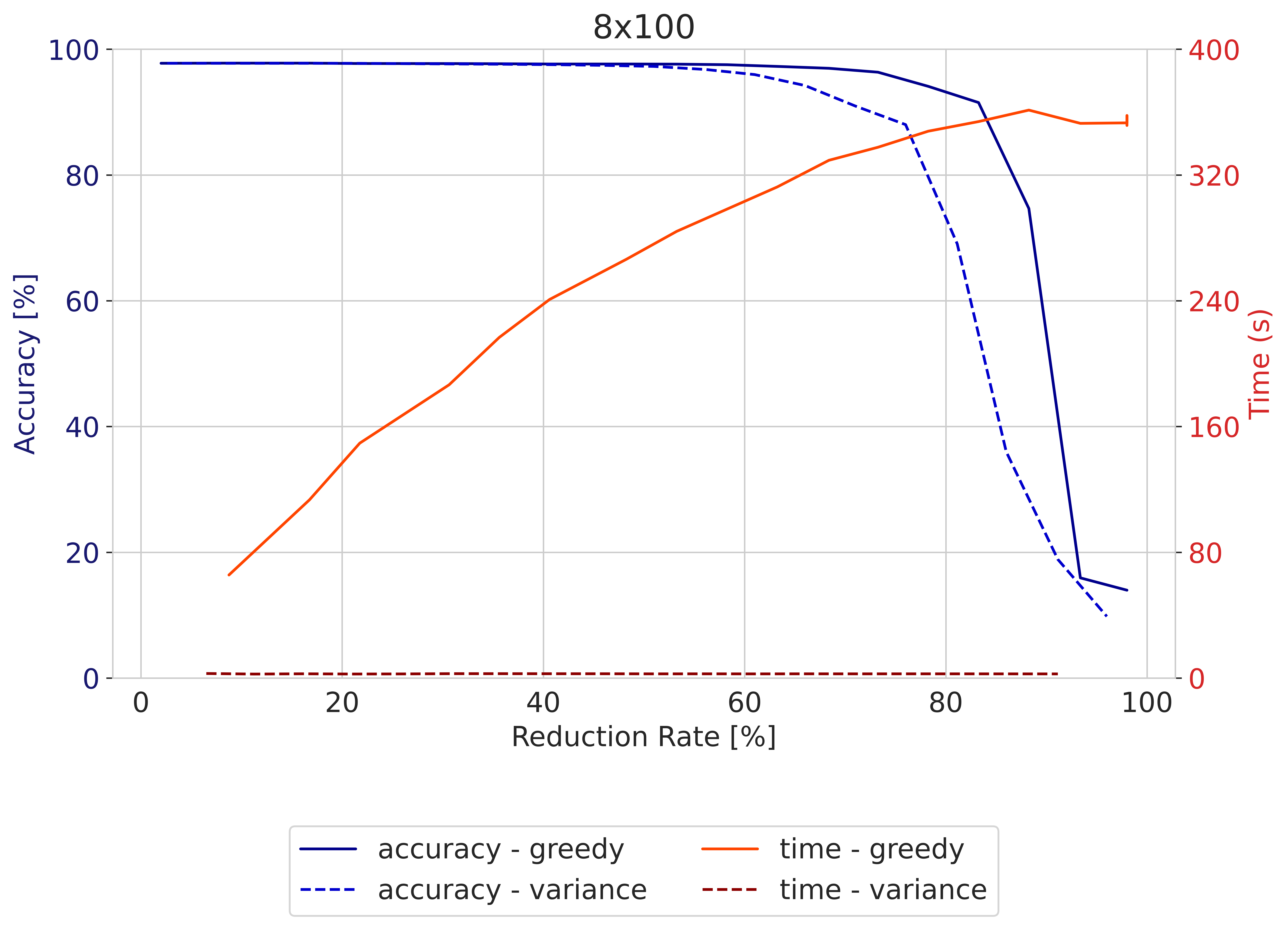}
	\end{subfigure}\hfill
	\caption{\emph{Comparison of how to find a basis on MNIST} - These plots show the accuracy and runtime for the greedy and variance-based LiNNA with orthogonal projection.}
	\label{fig:basis-finding2}
\end{figure}
\begin{figure}[!h]
	\centering
	\begin{subfigure}[]{0.48\textwidth}
		\includegraphics[width=\textwidth]{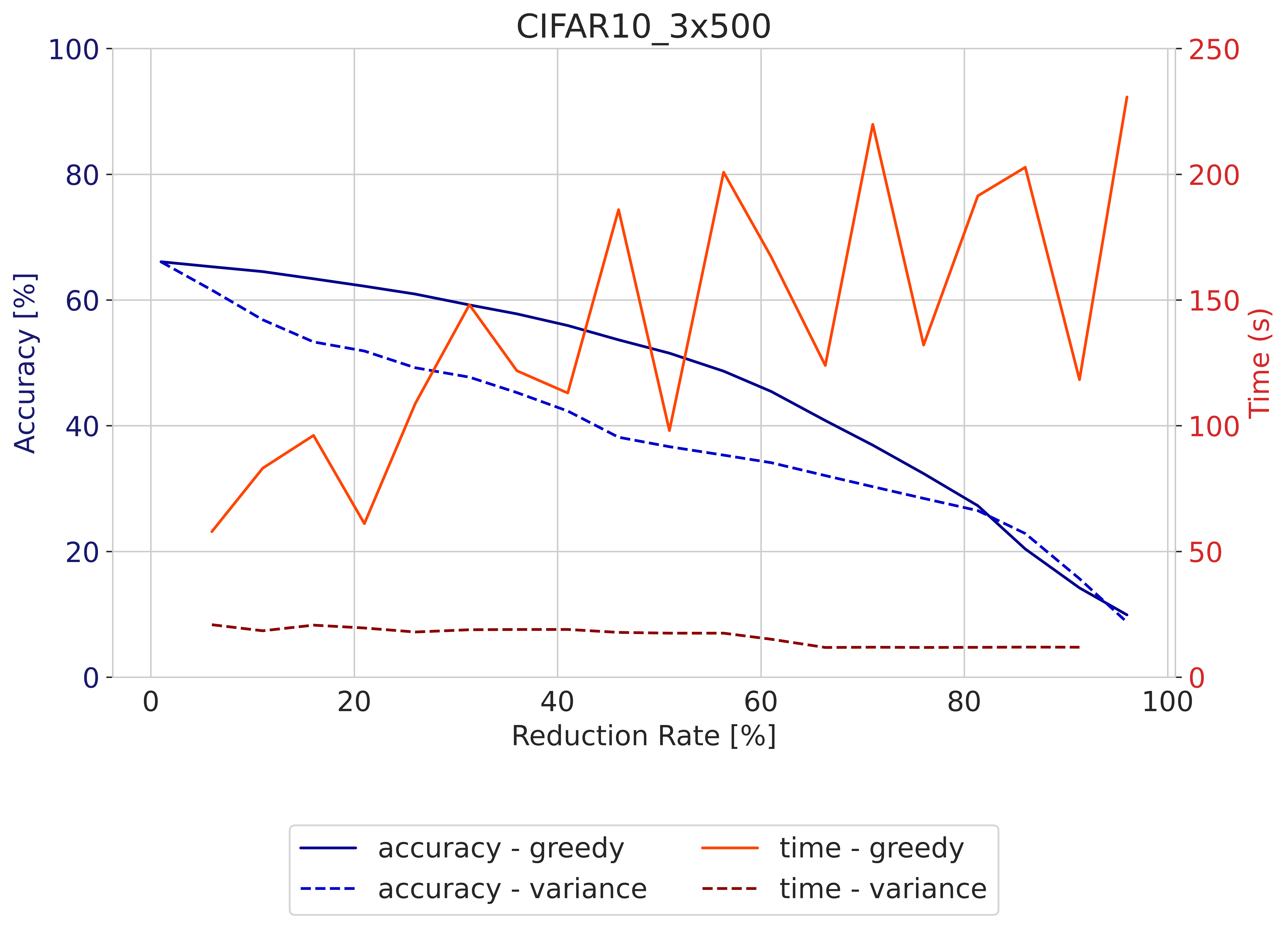}
	\end{subfigure}\hfill
	\begin{subfigure}[]{0.48\textwidth}
		\includegraphics[width=\textwidth]{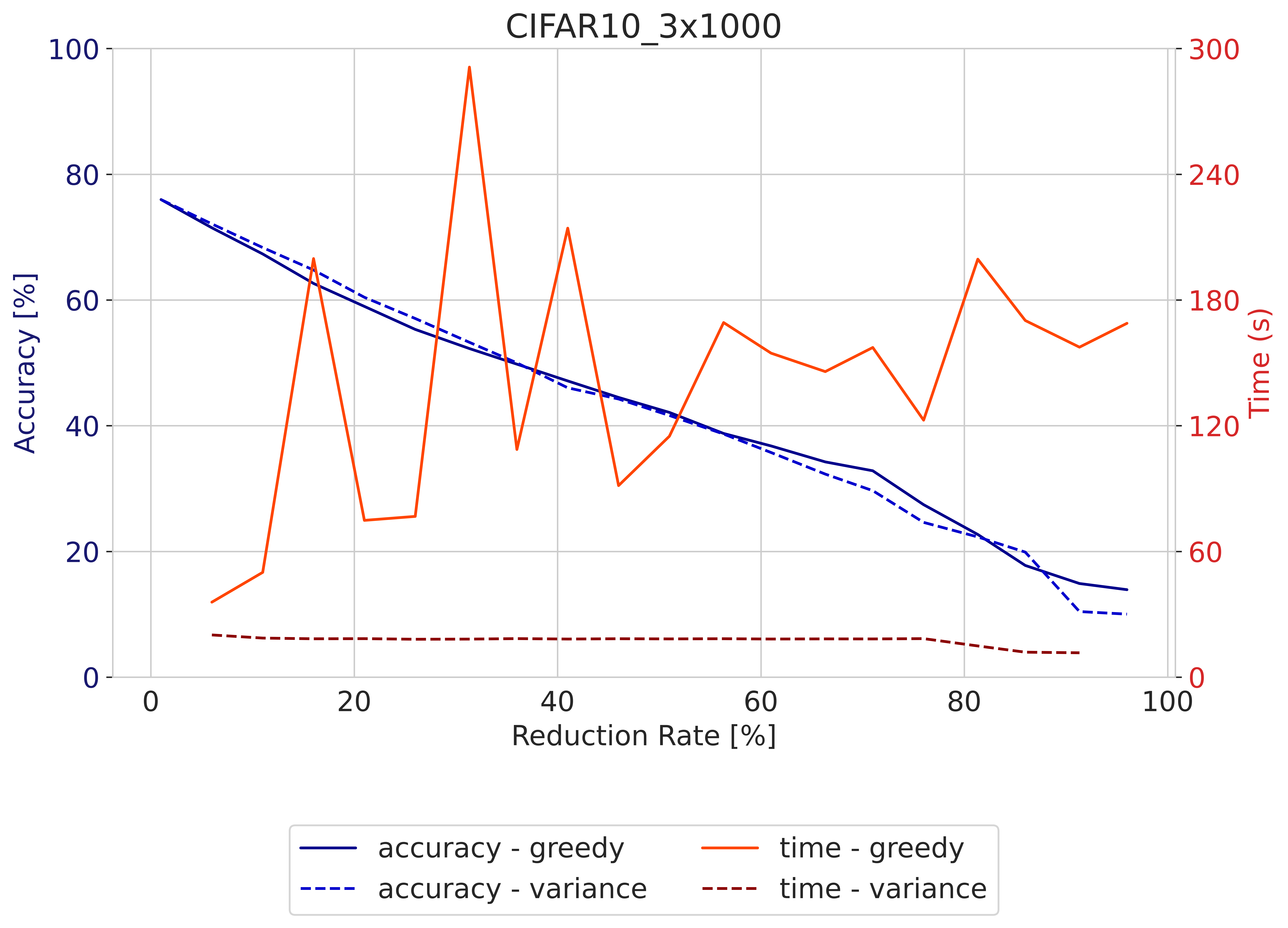}
	\end{subfigure}\hfill
	
	\begin{subfigure}[]{0.48\textwidth}
		\includegraphics[width=\textwidth]{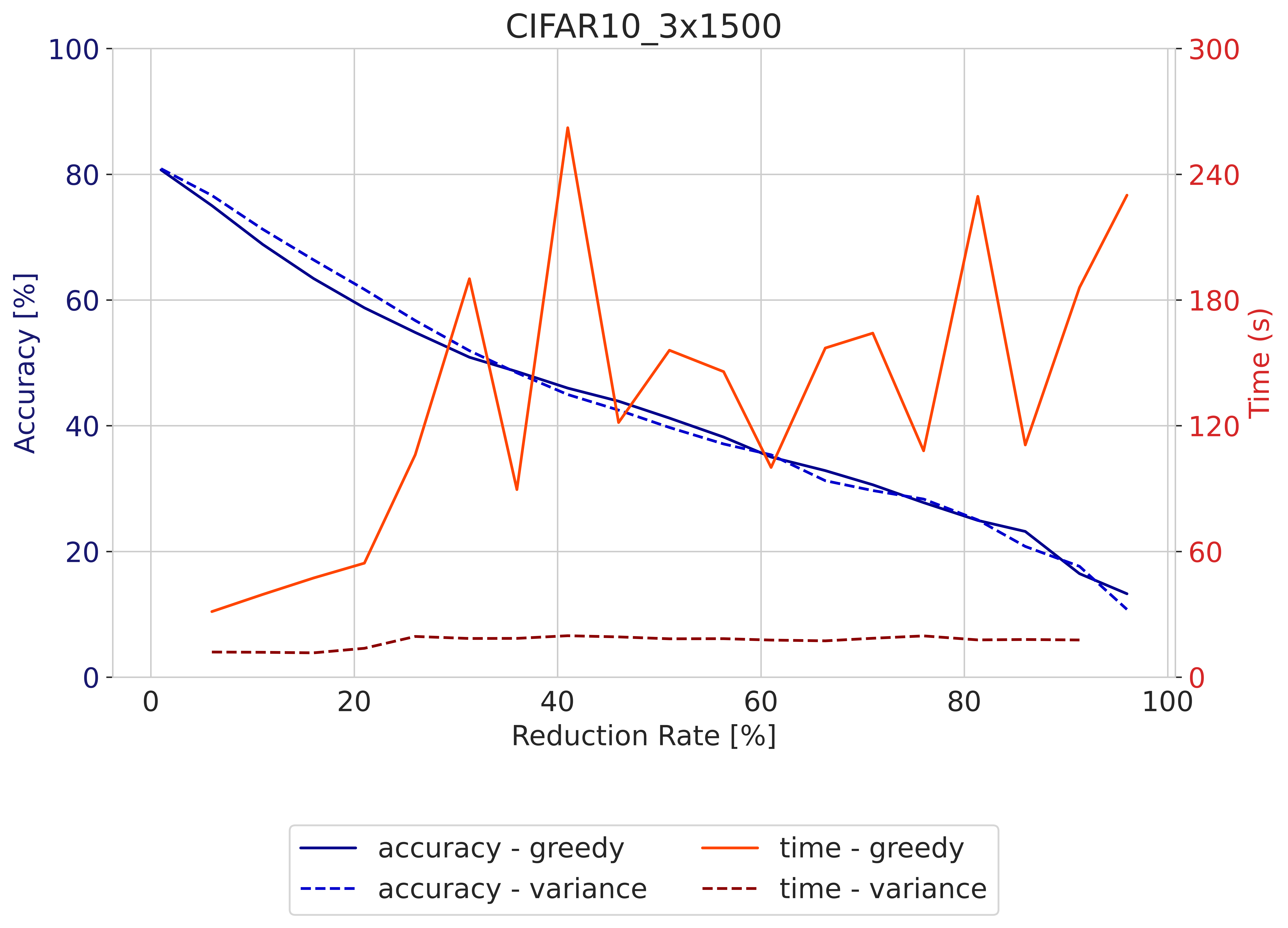}
	\end{subfigure}\hfill
	\begin{subfigure}[]{0.48\textwidth}
		\includegraphics[width=\textwidth]{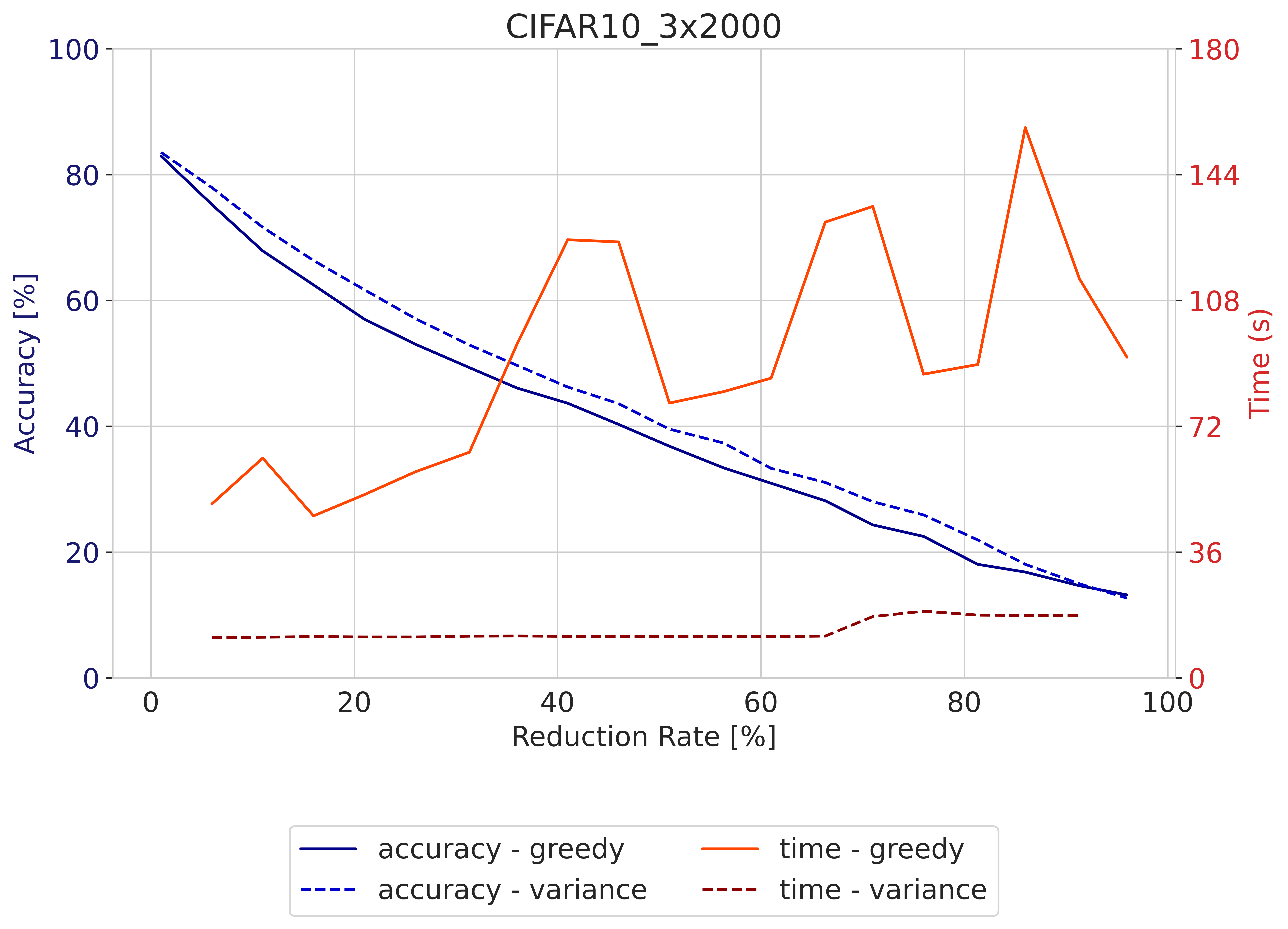}
	\end{subfigure}\hfill
	
	\begin{subfigure}[]{0.48\textwidth}
		\includegraphics[width=\textwidth]{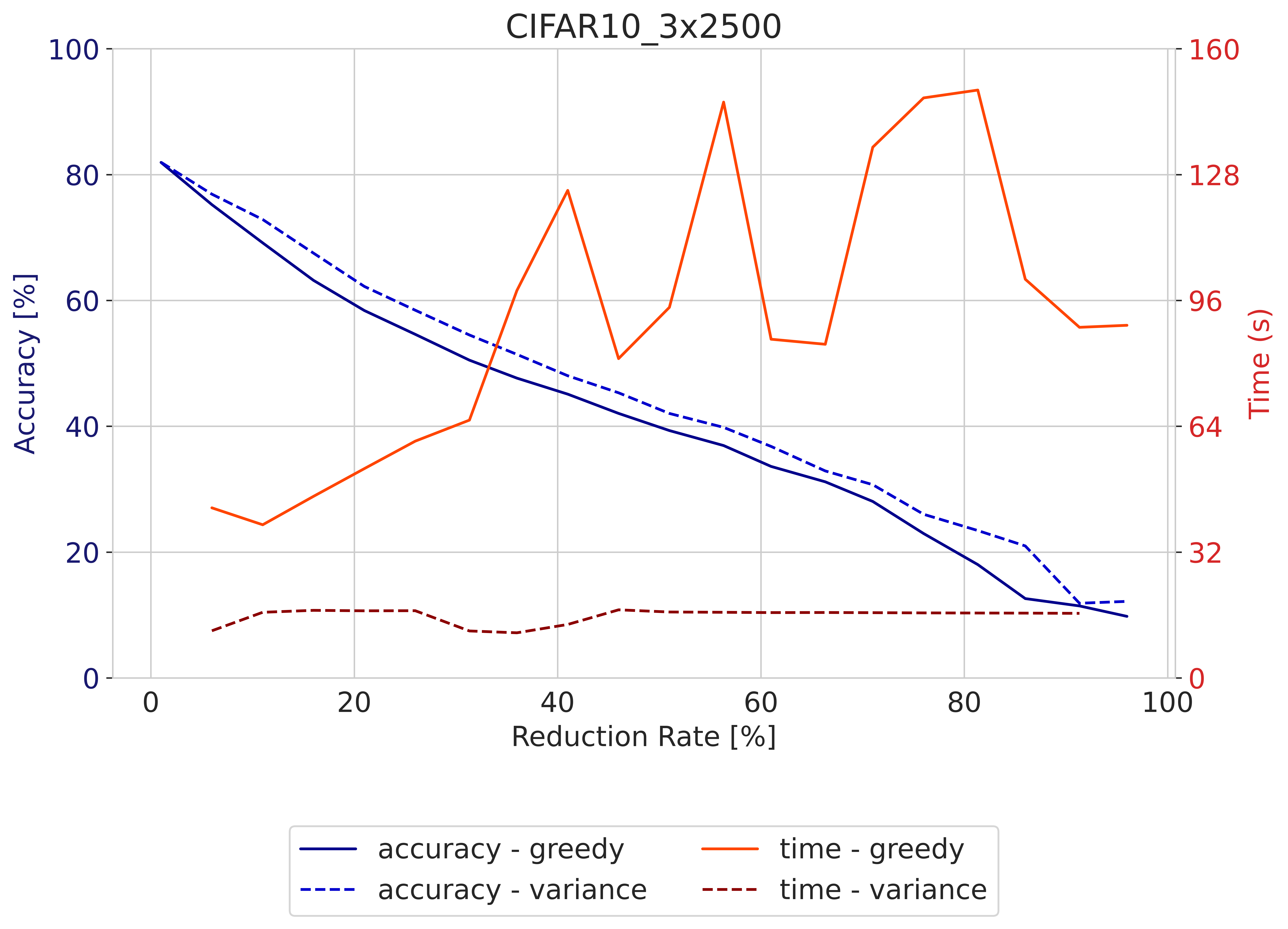}
	\end{subfigure}\hfill
	\begin{subfigure}[]{0.48\textwidth}
		\includegraphics[width=\textwidth]{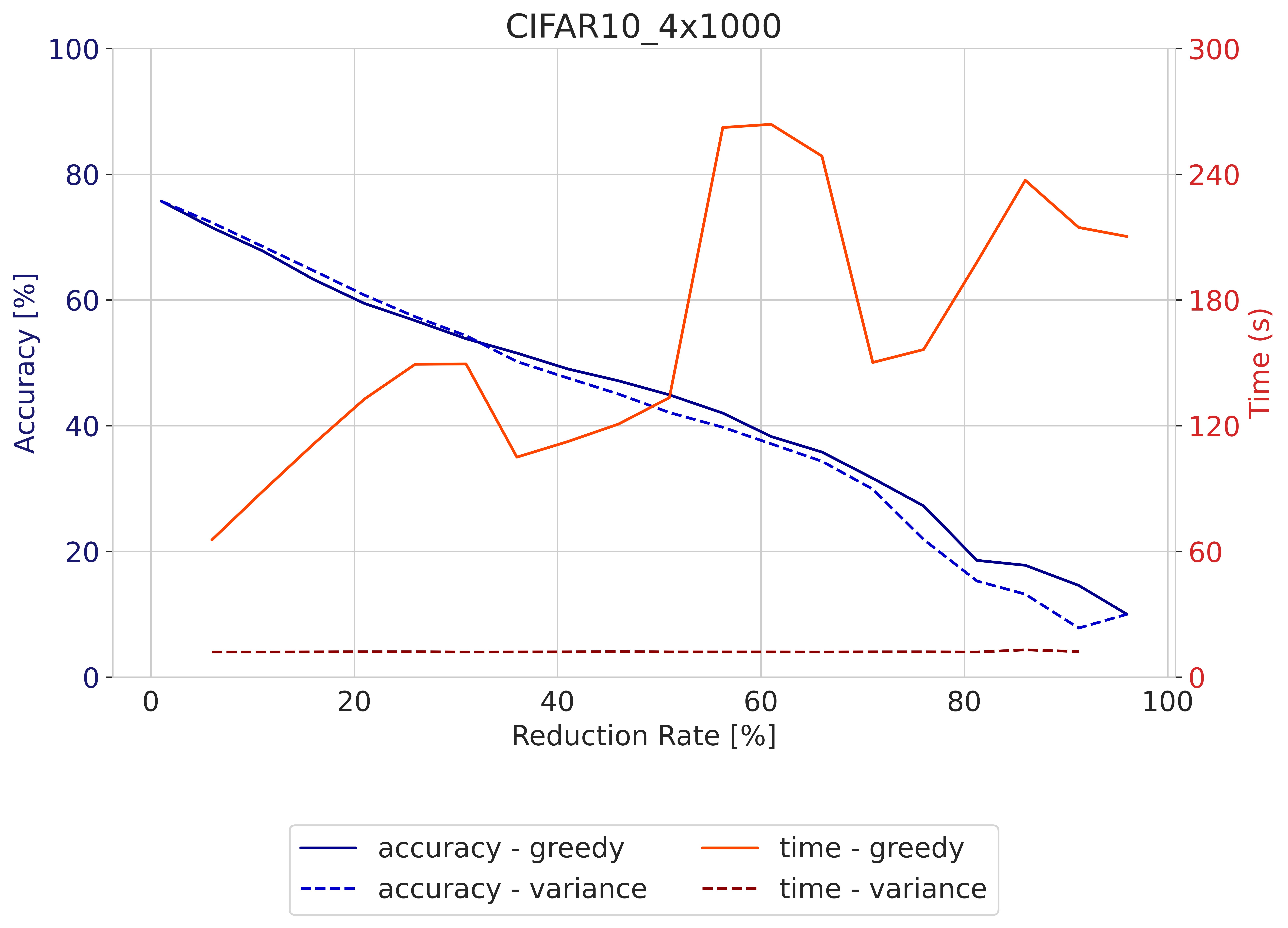}
	\end{subfigure}\hfill
	
	\begin{subfigure}[]{0.48\textwidth}
		\includegraphics[width=\textwidth]{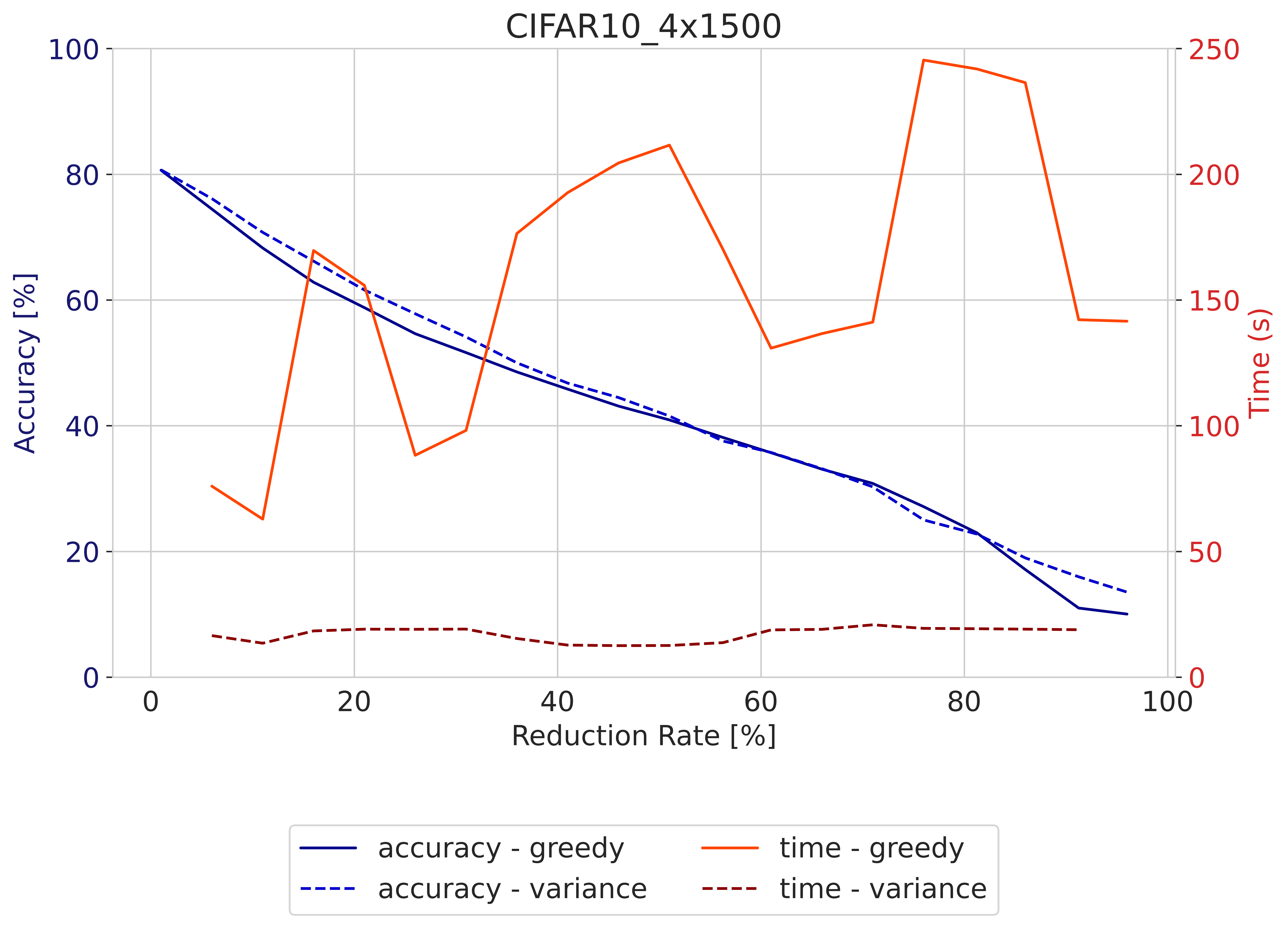}
	\end{subfigure}\hfill
	\caption{\emph{Comparison of how to find a basis on CIFAR-10} - These plots show the accuracy and runtime for the greedy and variance-based LiNNA with orthogonal projection.}
	\label{fig:basis-finding3}
\end{figure}
We can see the results in \cref{fig:basis-finding2,fig:basis-finding3}.
The x-axis shows the reduction rate. 
On the left (blue), the y-axis of the plots shows the accuracy of the network on the test data set, which was not used for generating the abstraction, and on the right (red) the time for the reduction, where the time is measured in seconds.
The title of each plot indicates the architecture.
On MNIST, \cref{fig:basis-finding2}, we can see that the greedy approach outperforms the variance-based approach in terms of accuracy, but it performs much worse in terms of runtime. 
The latter can also be seen on CIFAR-10. However, the accuracies of the greedy and the variance-based approach are much more similar on this benchmark.

\section{Supplemental Details on the Implementation of the Bisimulation}\label{sec:app-bisimulation}
The bisimulation chooses a random neuron as representative of a group, where all neurons of the group agree on their incoming weights up to a deviation of $\delta$. 
We start by calculating the distance of two neurons $i,j$ as follows:
$d(i,j)=\max_k\{|\weight{\ell-1}{i}{k}-\weight{\ell-1}{j}{k}|,|\bias{\ell}{i}-\bias{\ell}{j}|\}$.
Note that the neurons $i,j$ are $\delta$-bisimilar if and only if $d(i,j)\leq\delta$.
Afterward, we use agglomerative clustering \cite{Zepeda-Mendoza2013} to find groups of neurons that have $\delta$-similar incoming weights. 
Since agglomerative clustering belongs to the methods of hierarchical clustering, it is somewhat similar to the minimization approach as described in \cite{Prabhakar22} for the exact bisimulation without $\delta$-deviation, and we hope that it captures the same intended behavior.

\section{Supplemental Experiments on Comparing Our Approaches}\label{sec:lp-op-comparison}
\begin{figure}[!h]
	\centering
	\includegraphics[scale=0.26]{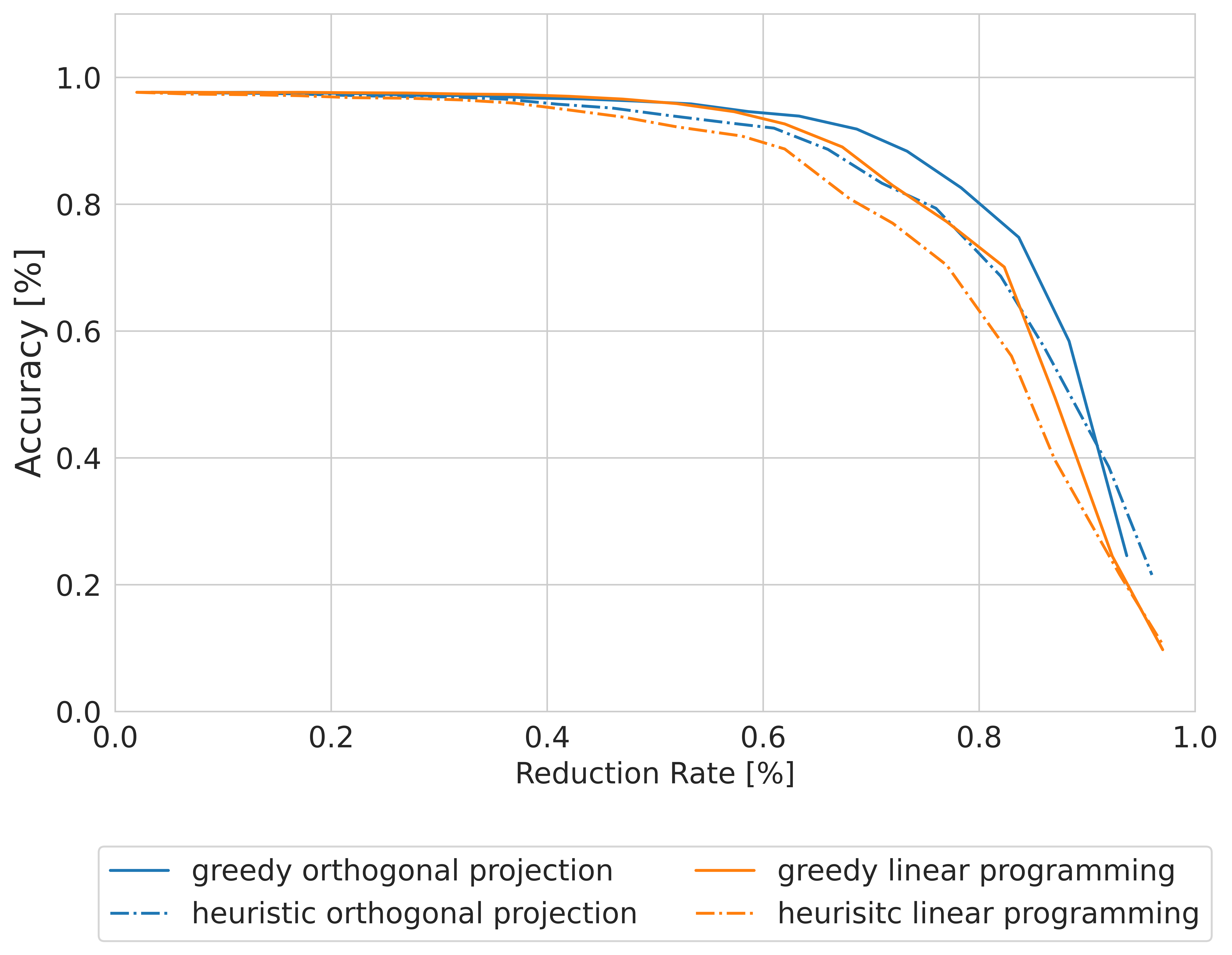}
	\caption{\emph{Comparison of Our Approaches} - On MNIST 3x100. This plot contains the comparison of all four possible approaches: orthogonal projection (blue) and linear programming (orange), each either greedy (solid) or based on a heuristic (dashed). 	}
	\label{fig:coefficient-finding}
\end{figure}
In \cref{fig:coefficient-finding}, we have four plots in total. Each two for linear programming and orthogonal projection, and the greedy and the heuristic-based approach.
The results for the linear programming are shown in green, and the results for the orthogonal projection are in blue. 
The greedy approaches are shown with a solid line and the heuristic-based approaches with a dashed line. 
As already seen, the greedy approach outperforms the heuristic-based approach. This also holds for linear programming. 
However, we can also see, that the greedy orthogonal projection always outperforms all other approaches, except for reduction rates close to 90\%.
Additionally, and more importantly, the heuristic-based orthogonal projection is almost as good as the greedy linear programming; for high reduction rates, it is even better.
\begin{figure}[!h]
	\centering
	\begin{subfigure}[]{0.48\textwidth}
		\includegraphics[width=\textwidth]{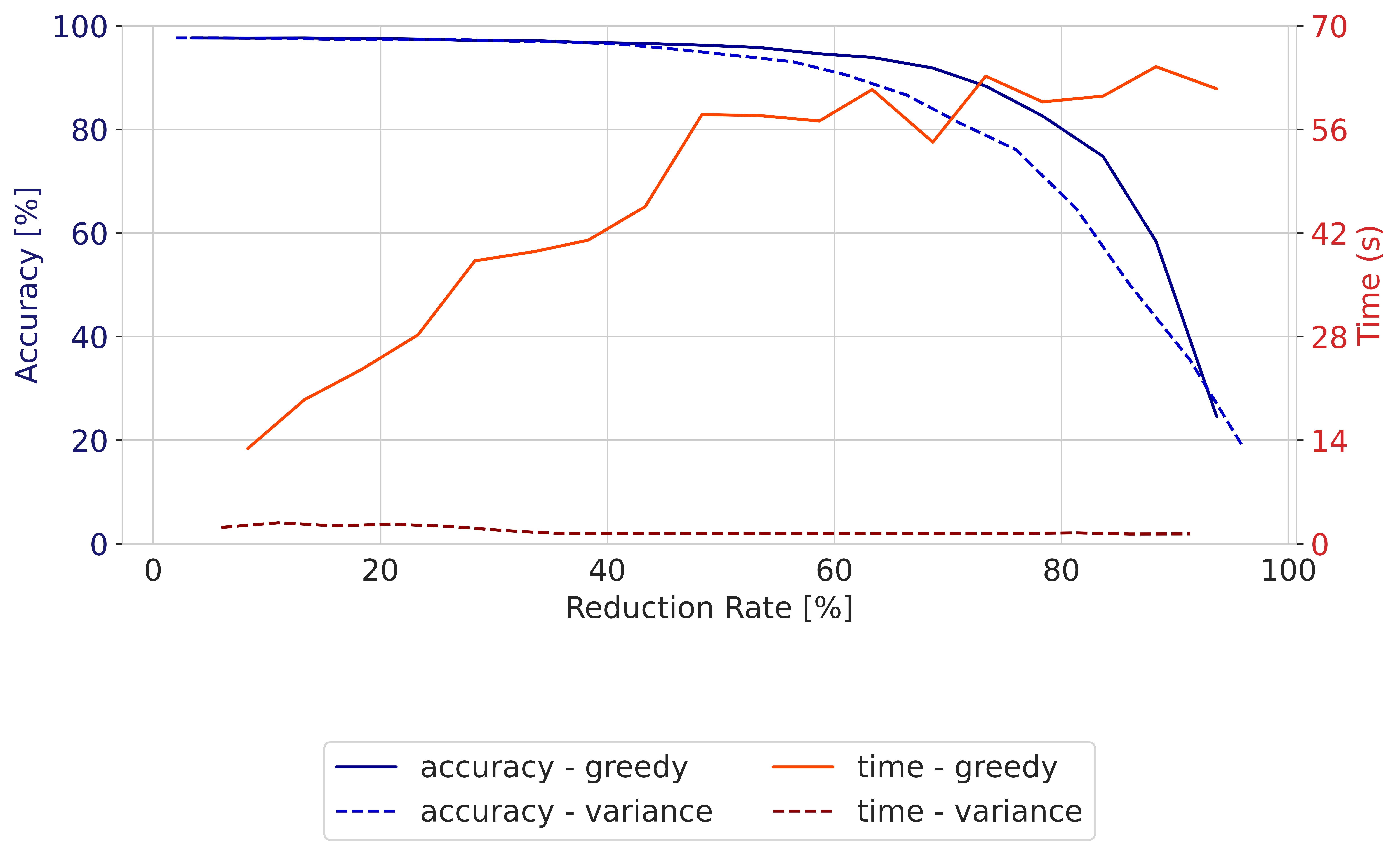}
	\end{subfigure}\hfill
	\begin{subfigure}[]{0.48\textwidth}
		\includegraphics[width=\textwidth]{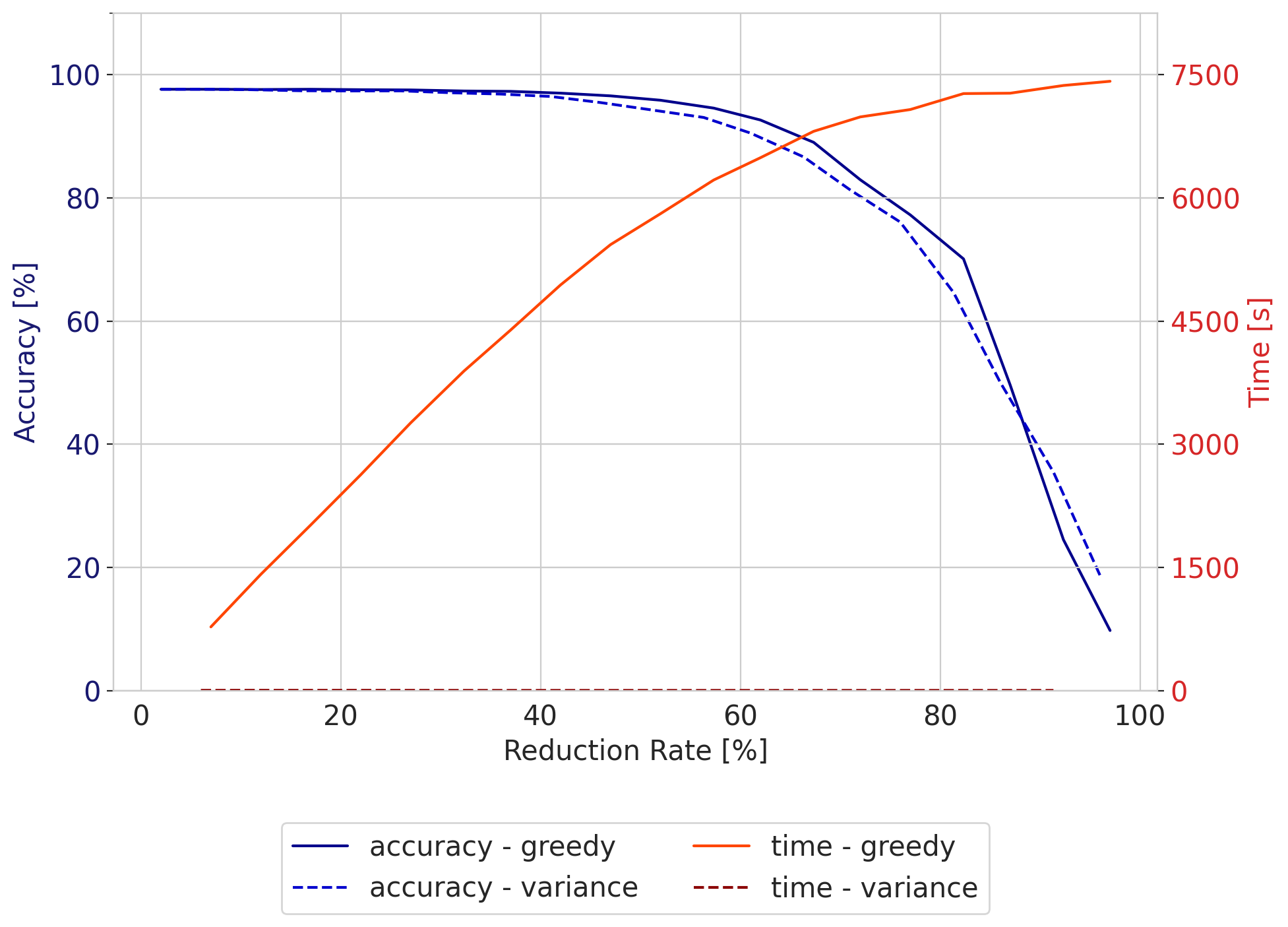}
	\end{subfigure}
	\caption{\emph{Comparison of Linear Programming to Orthogonal Projection} - Plots for the orthogonal projection (left) and linear programming (right) on MNIST 3x100. Each is either greedy (solid) or based on a heuristic (dashed). The accuracy is shown in blue (left y-axis), and the computation time is in red (right y-axis).}
	\label{fig:lp-op-comp}
\end{figure}
\cref{fig:lp-op-comp} shows on the left a plot for the orthogonal projection. 
It shows the accuracy (blue) and the time (red) of the greedy (solid) against the variance-based (dashed) method.
On the right, we have the same plot for the linear programming. 
Note here that the time for the linear programming is 100 times as much as for the orthogonal projection.
\newpage
\section{Supplemental Experiments on Comparison to Related Works}\label{sec:comparison}
We have some more experiments in this section on the comparison of LiNNA to related works, i.e. DeepAbstract and bisimulations. 
We performed the experiments from \cref{fig:rw-comp} on some more architectures of the MNIST.
\begin{figure}[!h]
	\centering
	\begin{subfigure}[]{0.45\textwidth}
		\includegraphics[width=\textwidth]{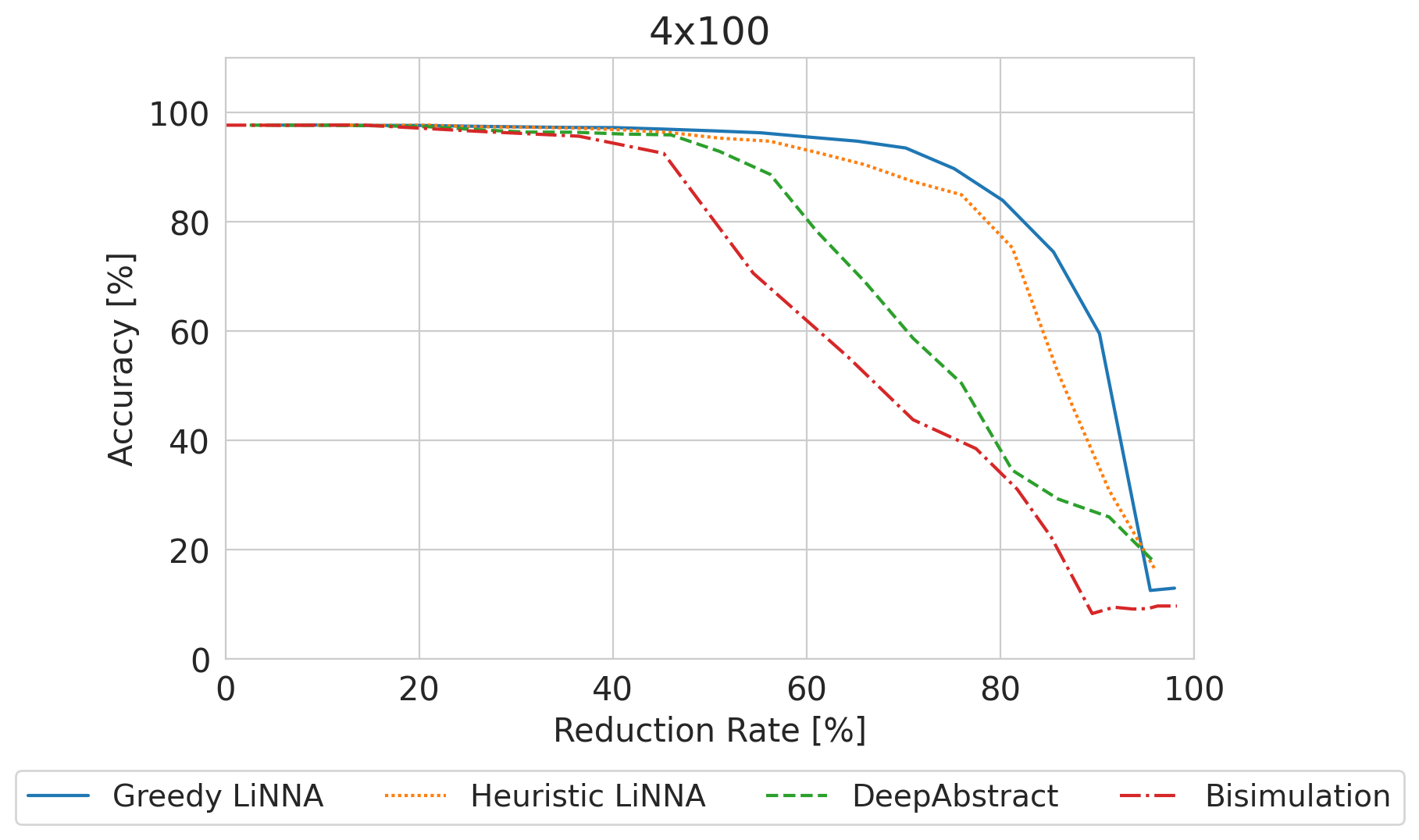}
	\end{subfigure}
	\begin{subfigure}[]{0.45\textwidth}
		\includegraphics[width=\textwidth]{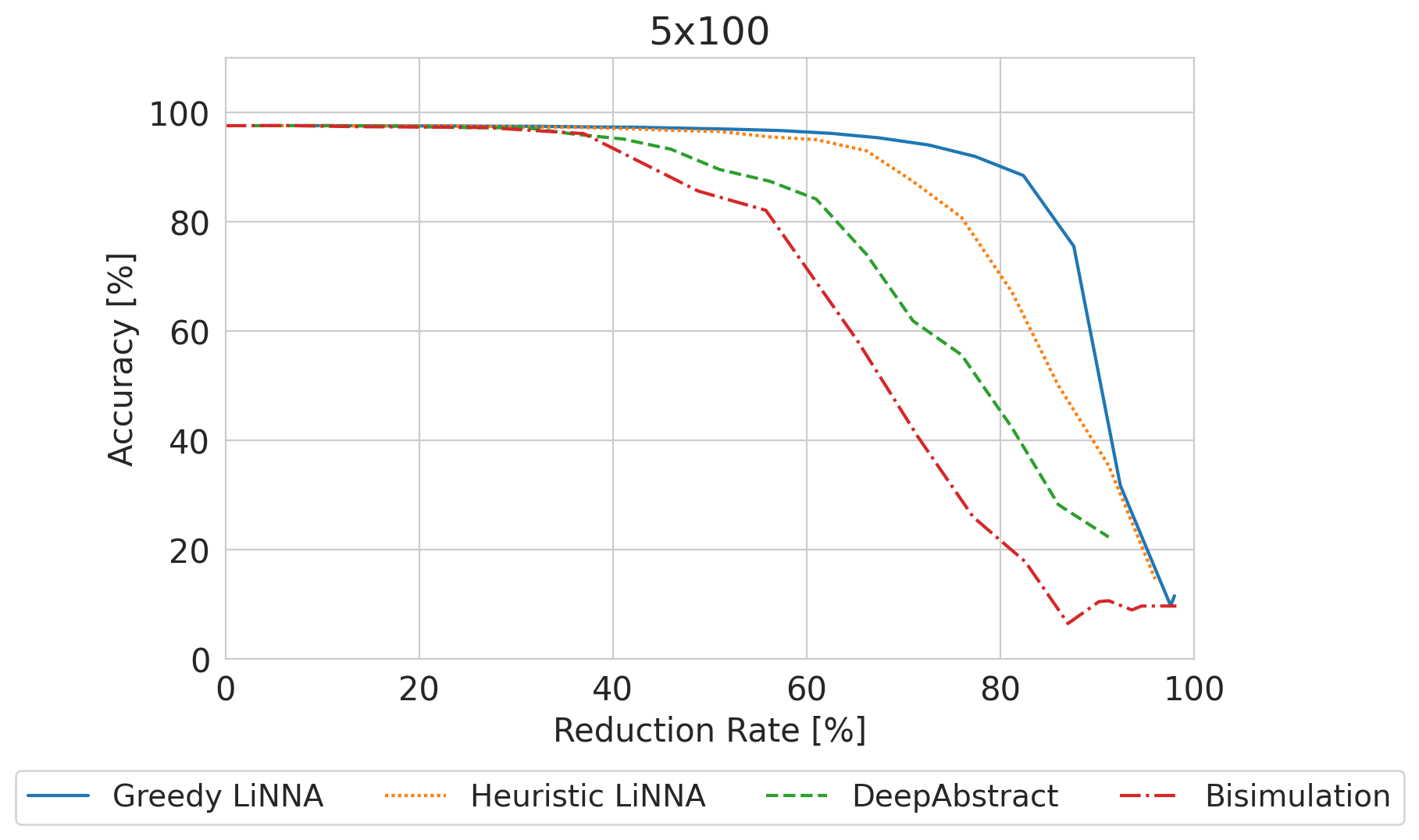}
	\end{subfigure}\hfill
	
	\begin{subfigure}[]{0.45\textwidth}
		\includegraphics[width=\textwidth]{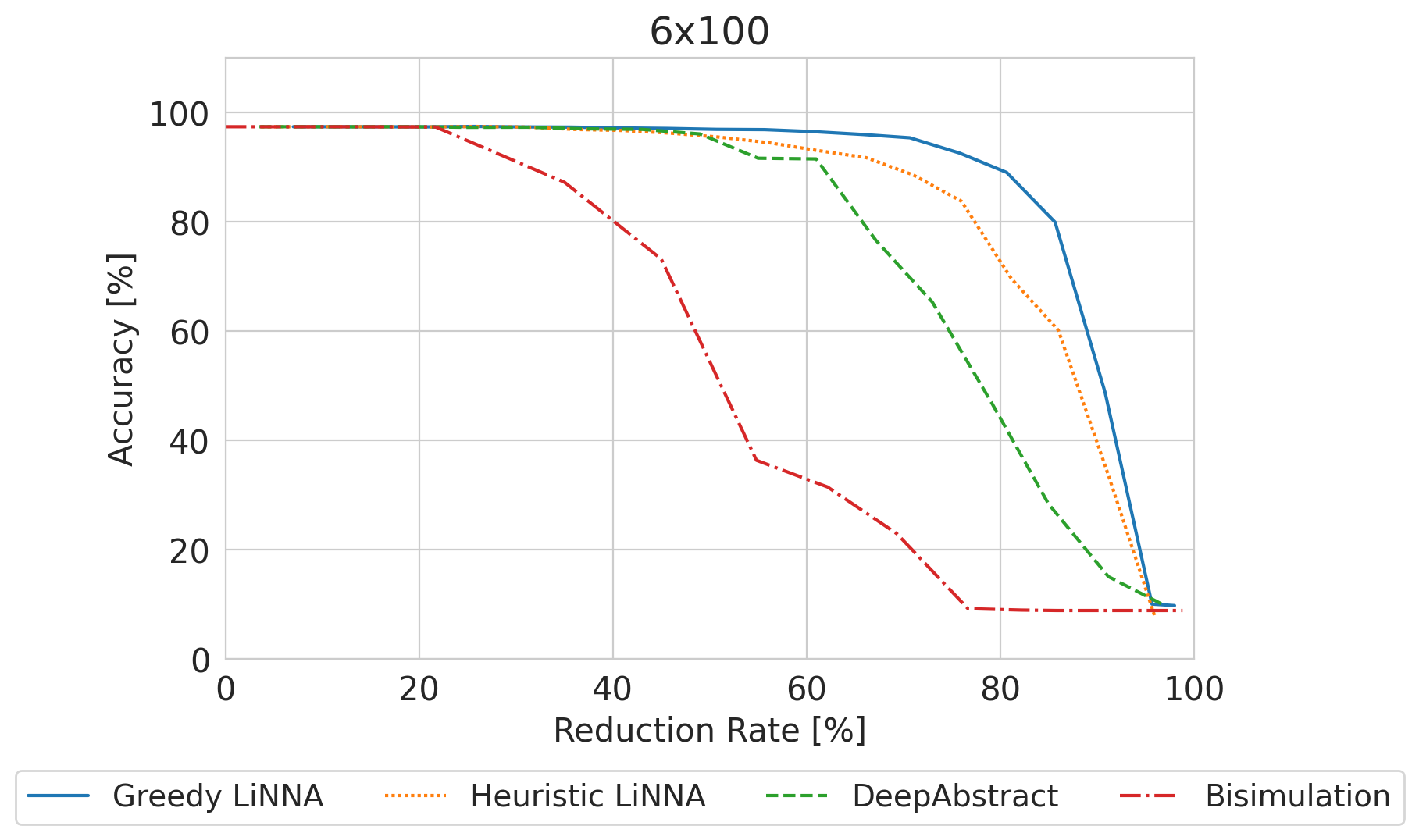}
	\end{subfigure}
	\begin{subfigure}[]{0.45\textwidth}
		\includegraphics[width=\textwidth]{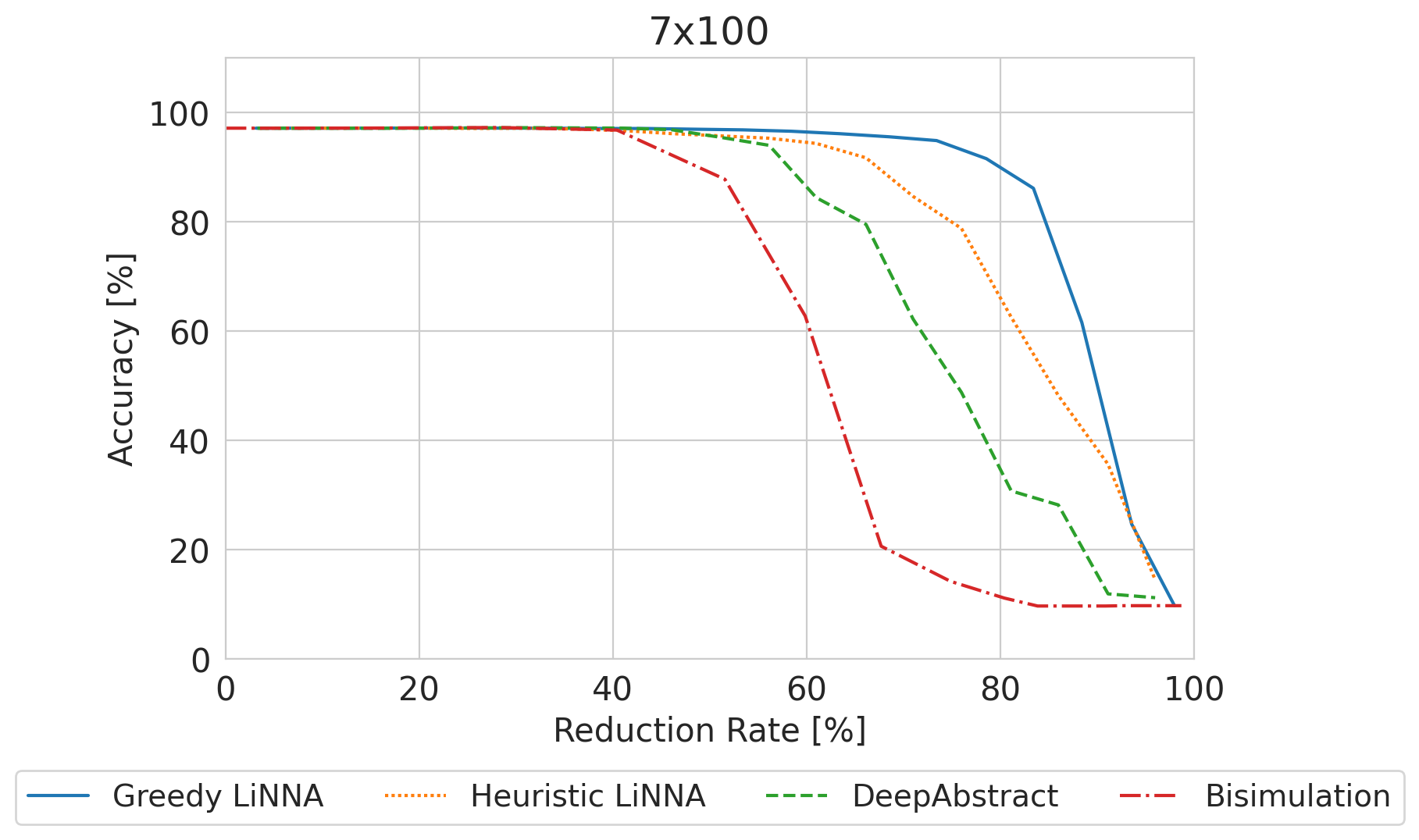}
	\end{subfigure}\hfill
	
	\begin{subfigure}[]{0.45\textwidth}
		\includegraphics[width=\textwidth]{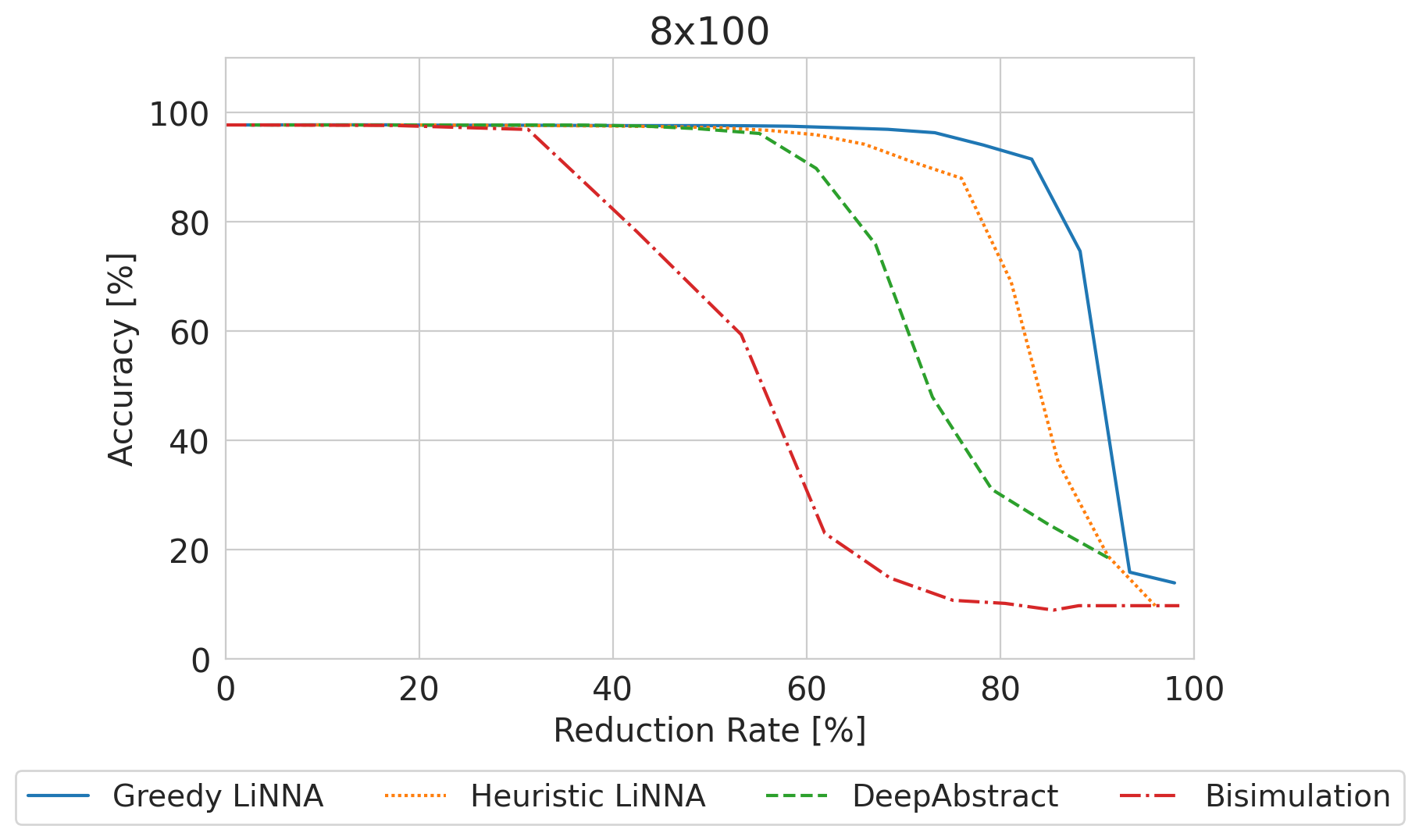}
	\end{subfigure}
	\caption{\emph{Comparison on MNIST} - There are plots for different architectures. Each plot contains four graphs: Greedy LiNNA, Heuristic LiNNA, DeepAbstract \cite{deepabstract}, and the bisimulation \cite{Prabhakar22}.}
	\label{fig:rw-comp2}
\end{figure}
We have also conducted some experiments on another dataset: FashionMNIST. 
In contrast to MNIST, which contains digits from zero to 10, FashionMNIST contains images of different clothes.
Note that the results are very similar to the ones that we generated on MNIST.
\begin{figure}[!h]
	\centering
	\begin{subfigure}[]{\textwidth}
		\includegraphics[width=\textwidth]{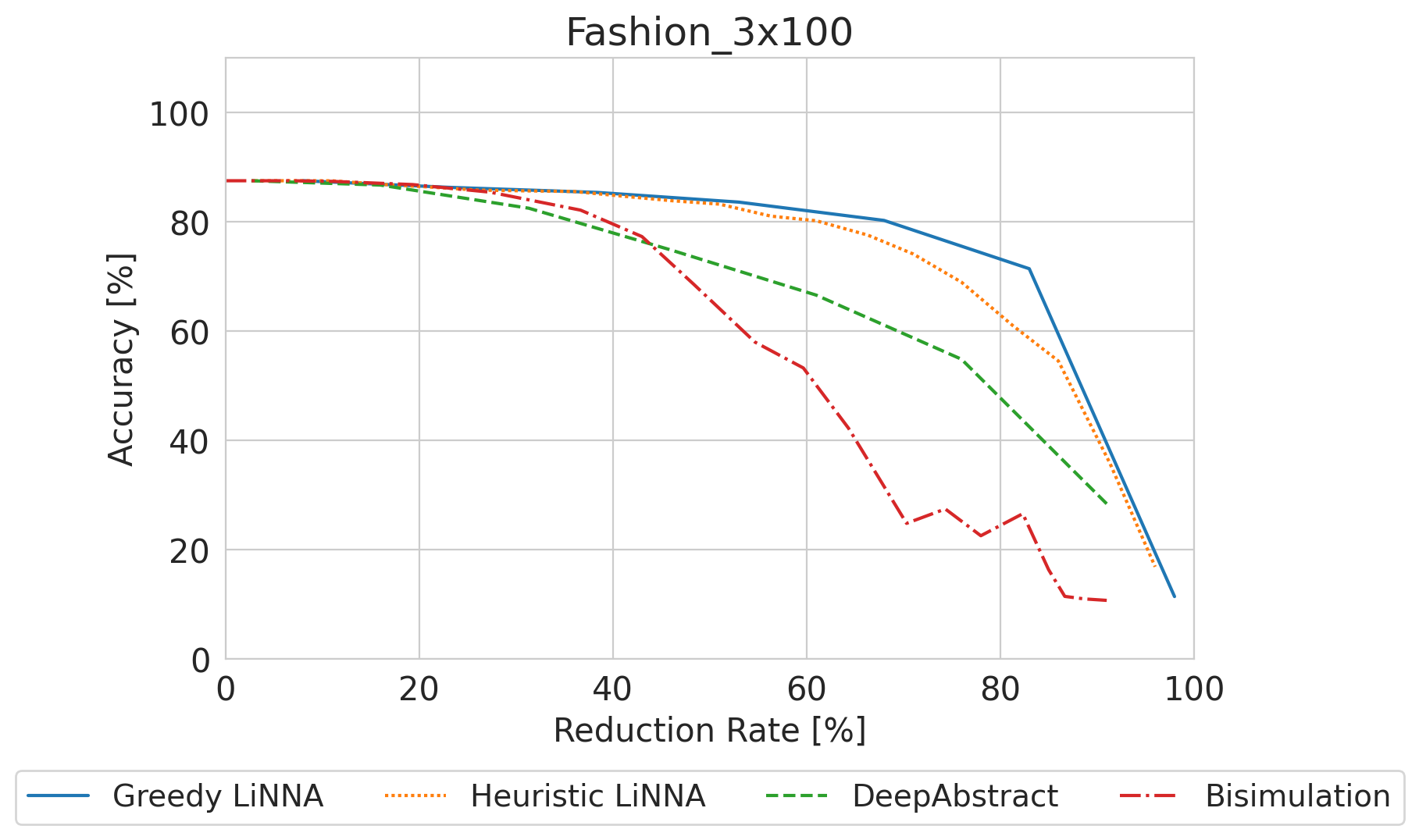}
	\end{subfigure}\hfill
	\caption{Comparison of all related work on FashionMNIST}
	\label{fig:fmnist-related-work}
\end{figure}

\begin{figure}[!h]
	\centering
	\begin{subfigure}[]{0.48\textwidth}
		\includegraphics[width=\textwidth]{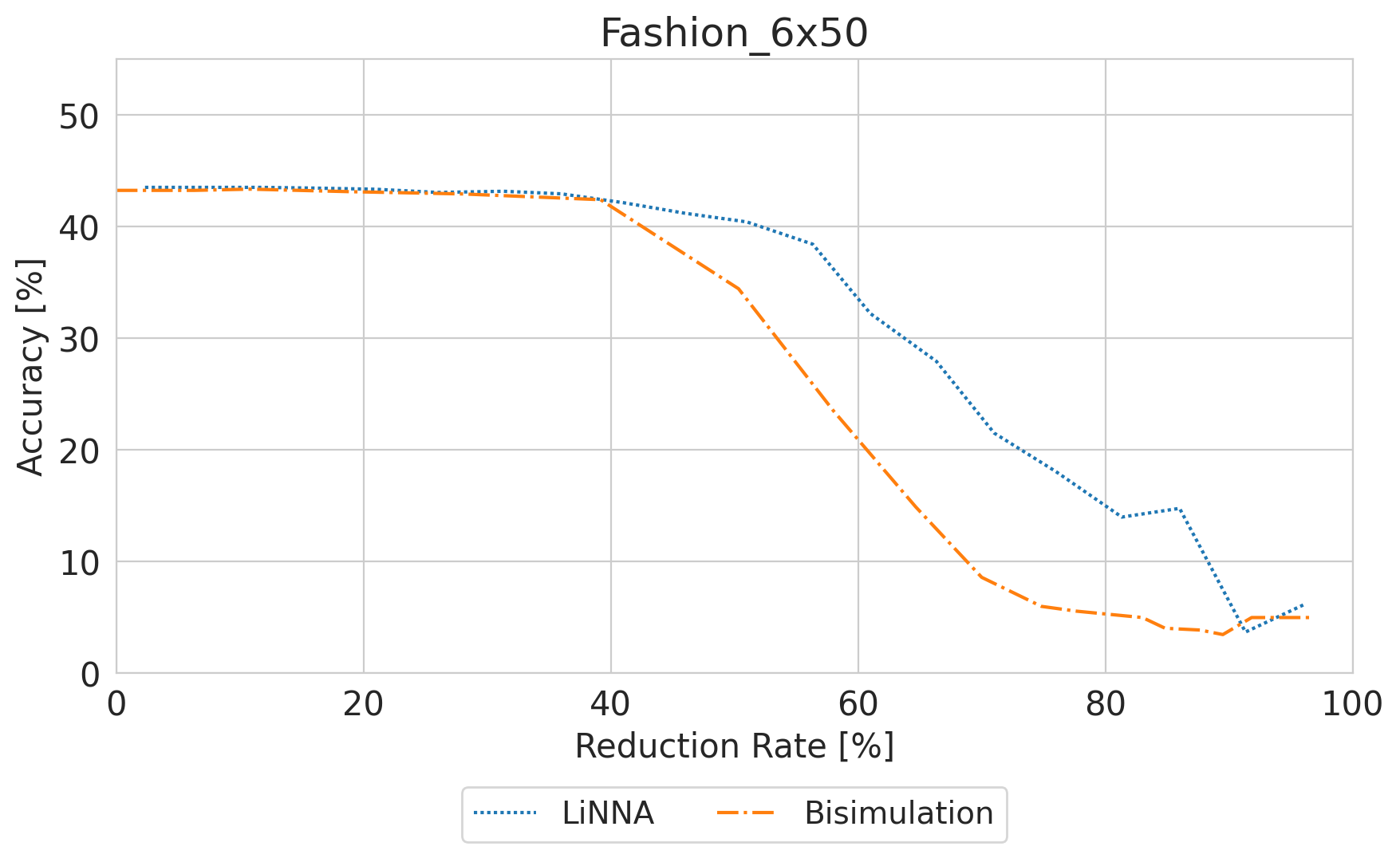}
	\end{subfigure}\hfill
	\begin{subfigure}[]{0.48\textwidth}
		\includegraphics[width=\textwidth]{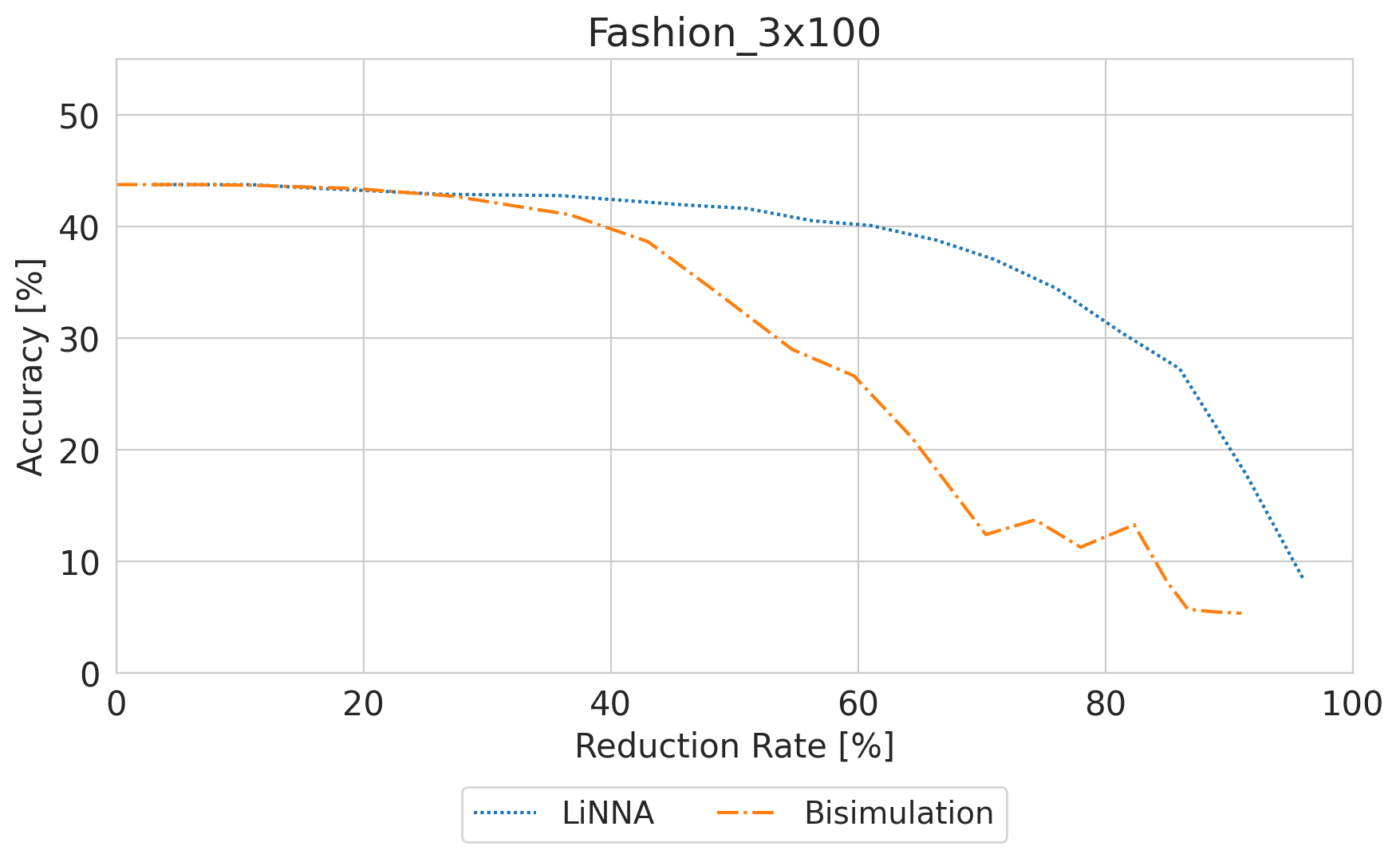}
	\end{subfigure}\hfill
	
	\begin{subfigure}[]{0.48\textwidth}
		\includegraphics[width=\textwidth]{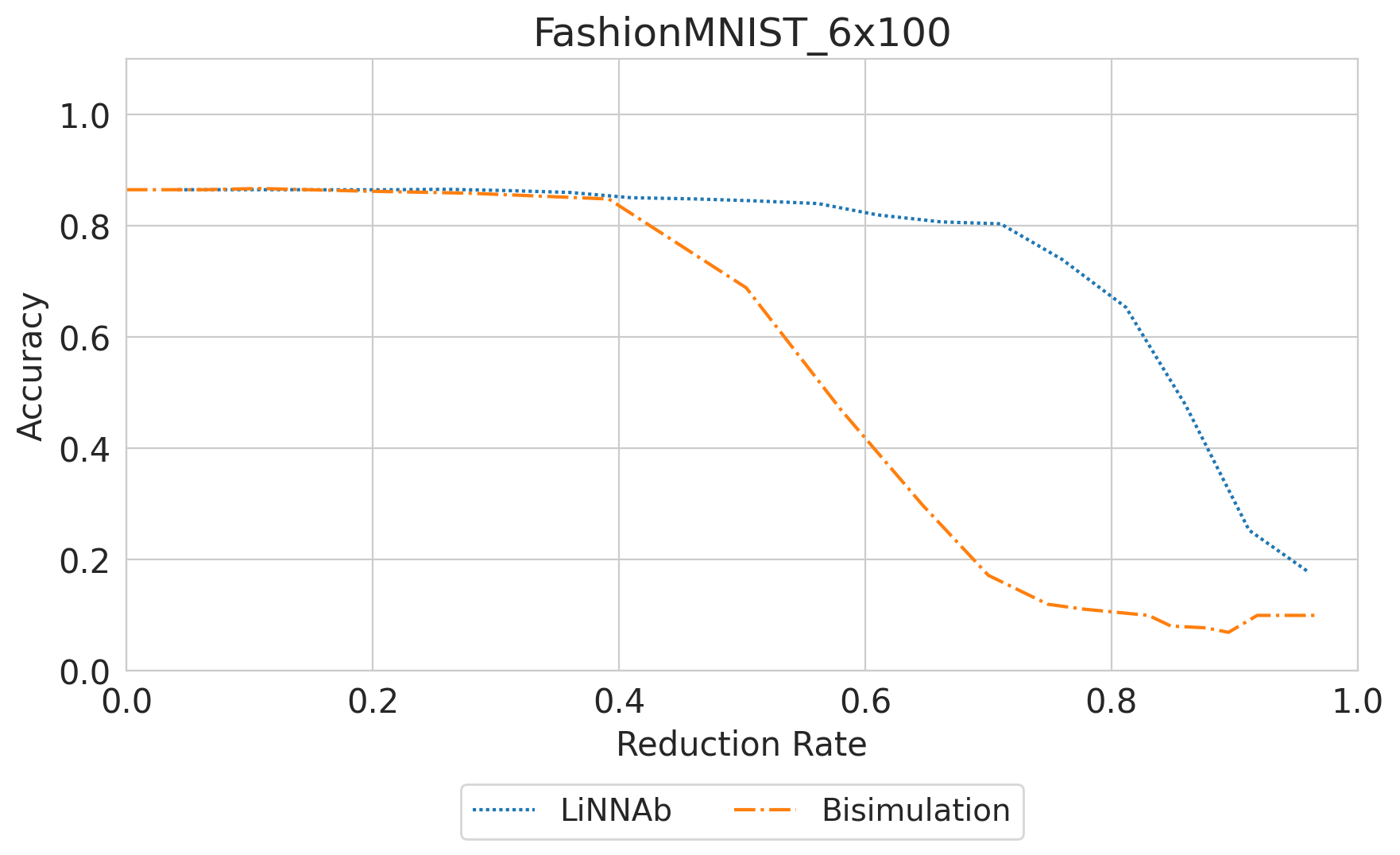}
	\end{subfigure}\hfill
	\begin{subfigure}[]{0.48\textwidth}
		\includegraphics[width=\textwidth]{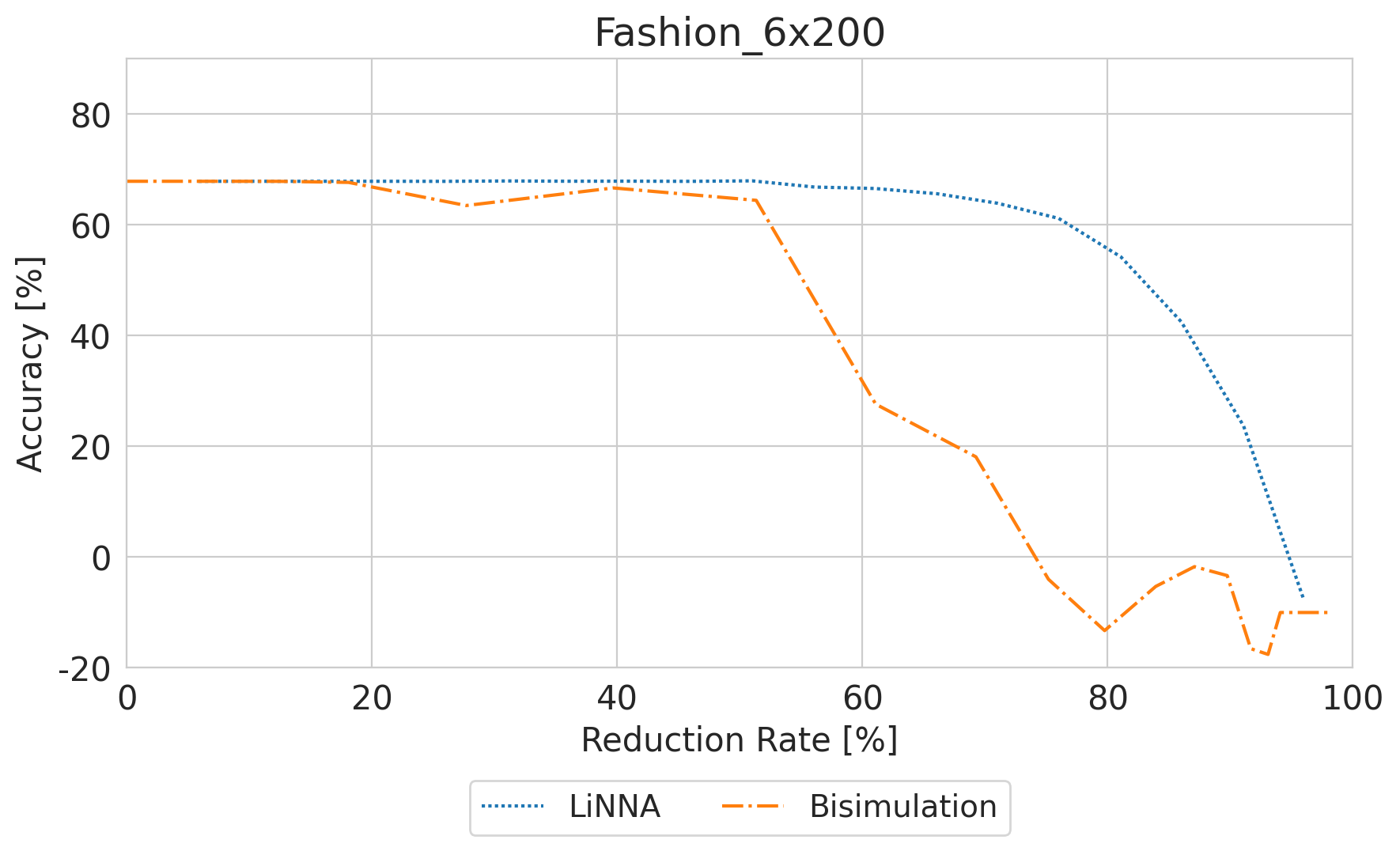}
	\end{subfigure}\hfill
	
	\begin{subfigure}[]{0.48\textwidth}
		\includegraphics[width=\textwidth]{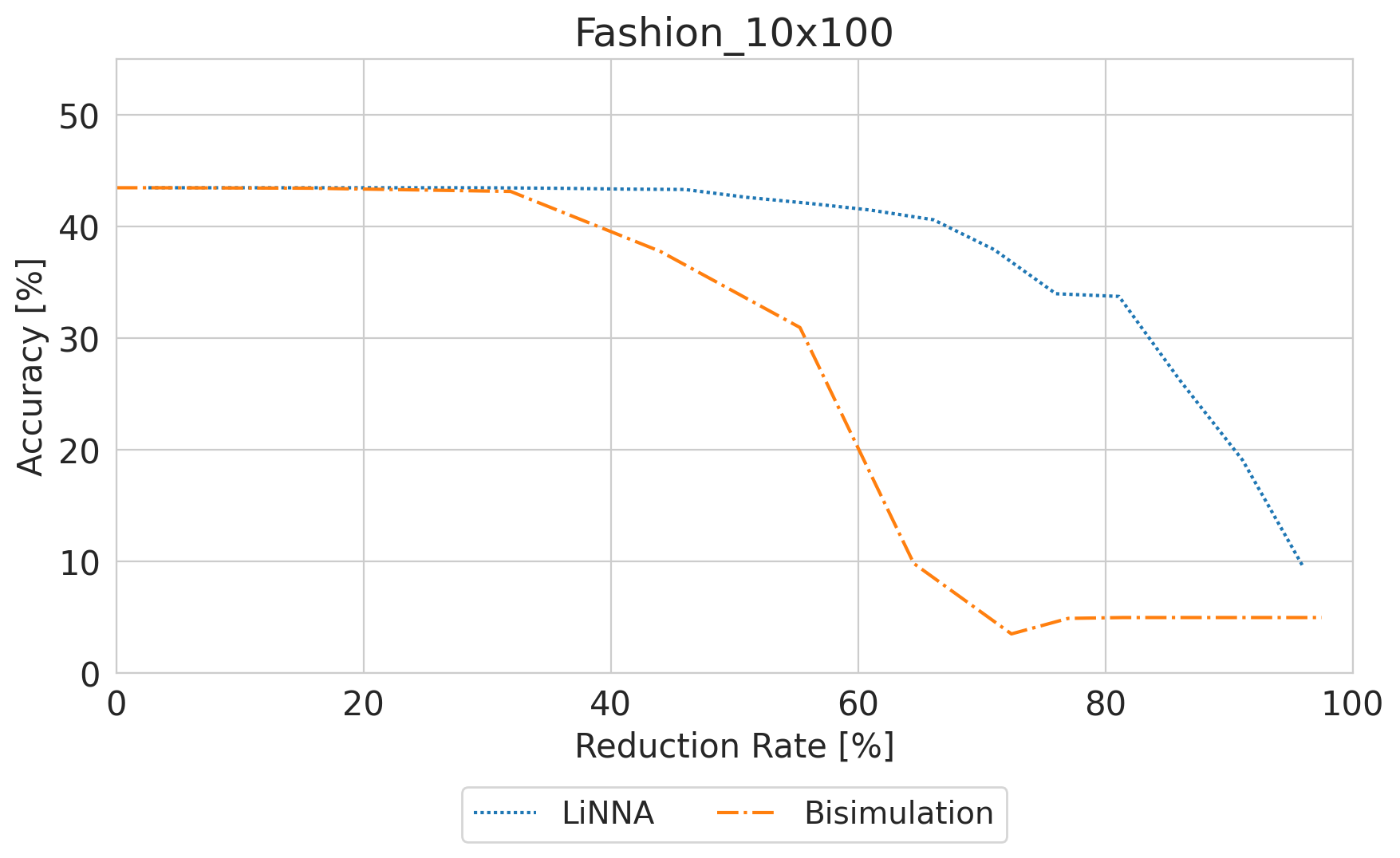}
	\end{subfigure}\hfill
	\caption{Comparison of bisimulation and heuristic LiNNA on FashionMNIST}
	\label{fig:fmnist-bisim-linna}
\end{figure}

In \cref{fig:fmnist-related-work}, we show the comparison of LiNNA, greedy and variance-based, in comparison with DeepAbstract and the bisimulation on a FashionMNIST 3x100 network in terms of accuracy. 
We can see that it looks very similar to the MNIST plots: LiNNA greedy outperforms all other approaches, closely followed by LiNNA with heuristic. DeepAbstract performs better than the bisimulation, which shows a rapid decrease in the accuracy already at 40\% reduction rate, whereas LiNNA can keep the accuracy stable up until 60\%.

Since the greedy approach of LiNNA and DeepAbstract can take quite long, we performed only a comparison of the bisimulation and the heuristic based LiNNA on more networks, see \cref{fig:fmnist-bisim-linna}. The plots differ slightly, but their overall message is the same: LiNNA performs better in terms of accuracy.
Additionally, we can see that it is easier to reduce networks that were bigger to begin with. Take, for example, the 6x50 and the 6x200 network in comparison. 
LiNNA can reduce up to 50\% on the first without a relevant decrease in the accuracy, but up to 70\% on the bigger network.
This can be explained by the fact that bigger networks can contain more redundant information that our approach can detect and remove.

\begin{figure}[!h]
	\centering
	\begin{subfigure}[]{0.48\textwidth}
		\includegraphics[width=\textwidth]{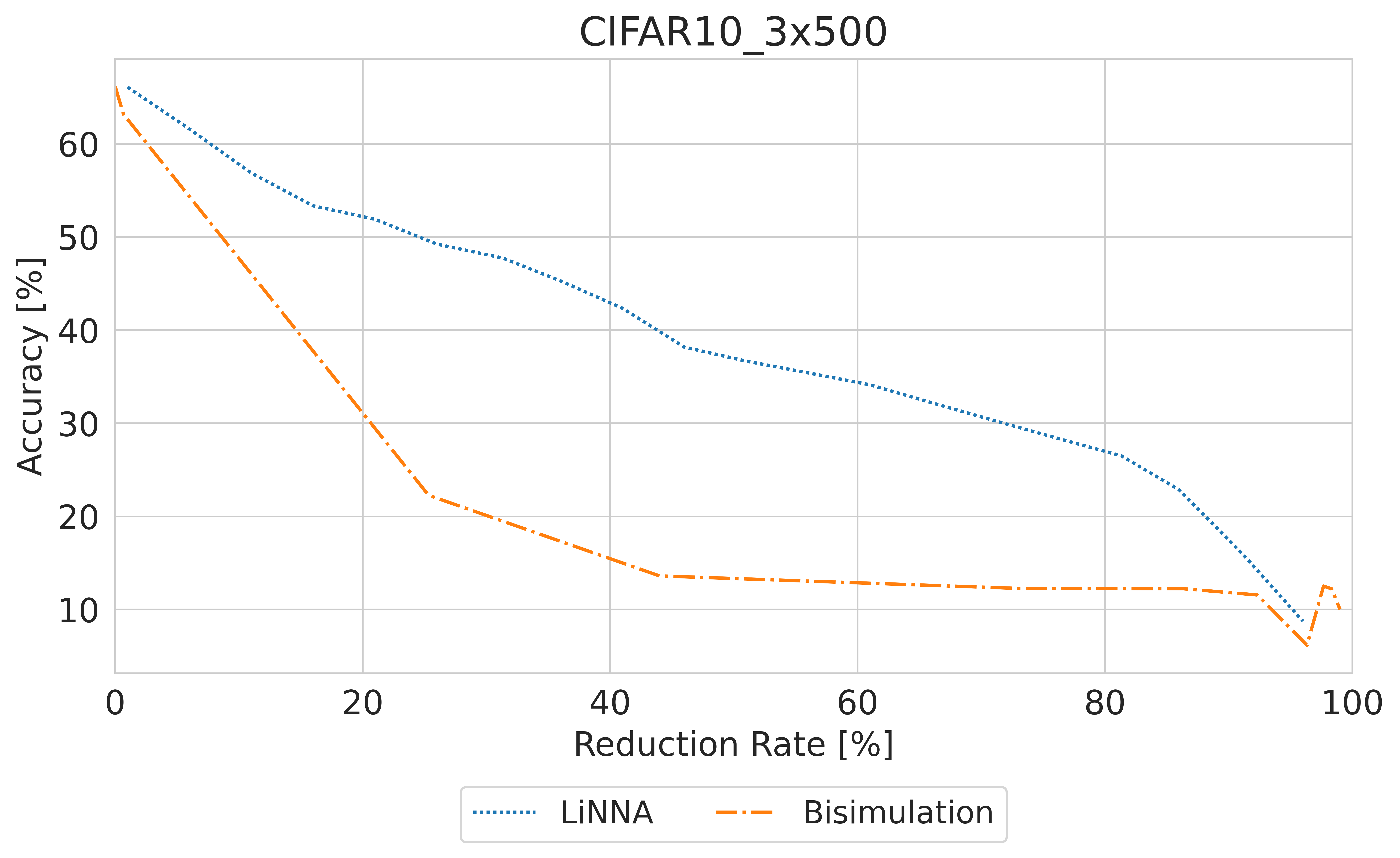}
	\end{subfigure}\hfill
	\begin{subfigure}[]{0.48\textwidth}
		\includegraphics[width=\textwidth]{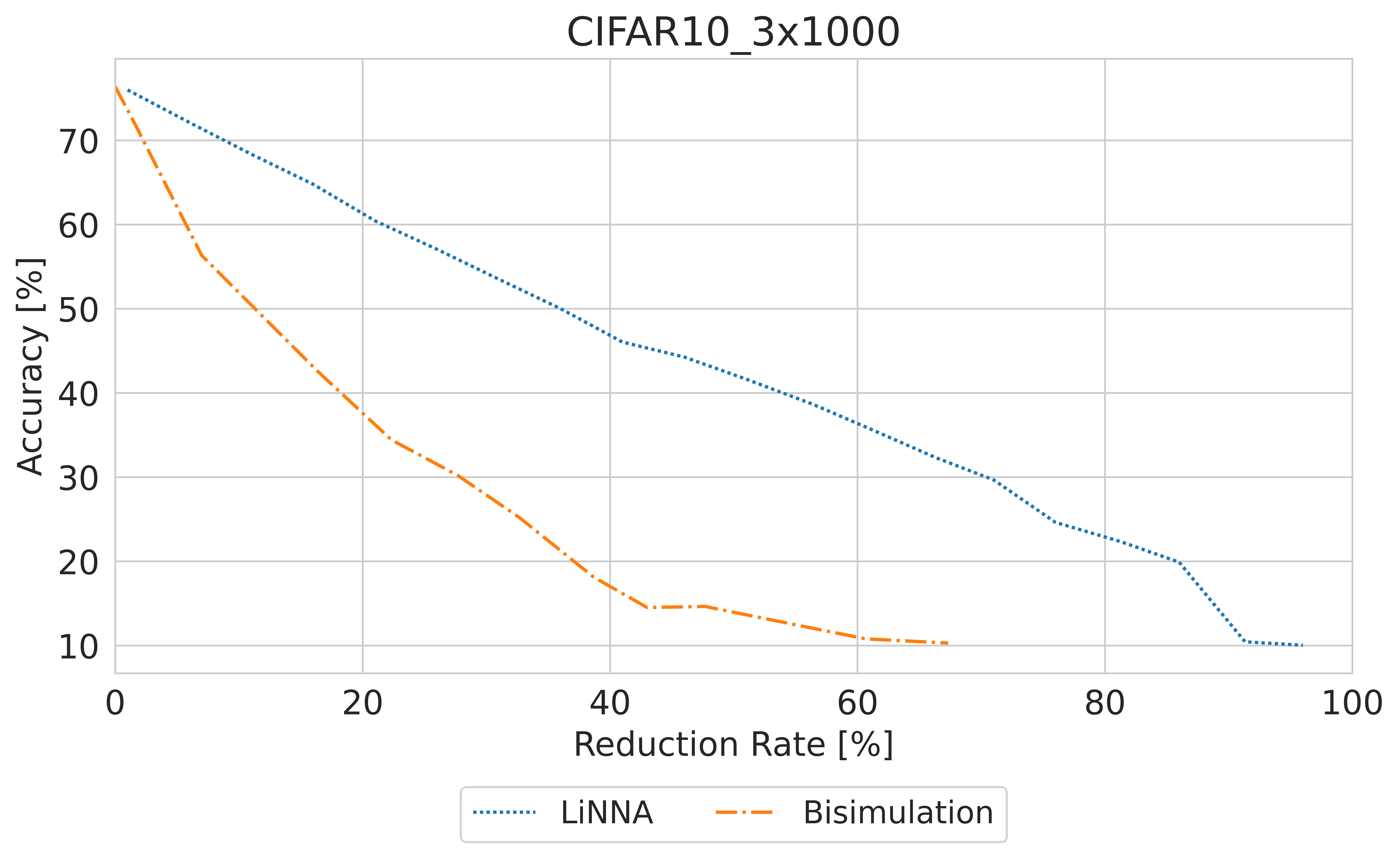}
	\end{subfigure}\hfill
	
	\begin{subfigure}[]{0.48\textwidth}
		\includegraphics[width=\textwidth]{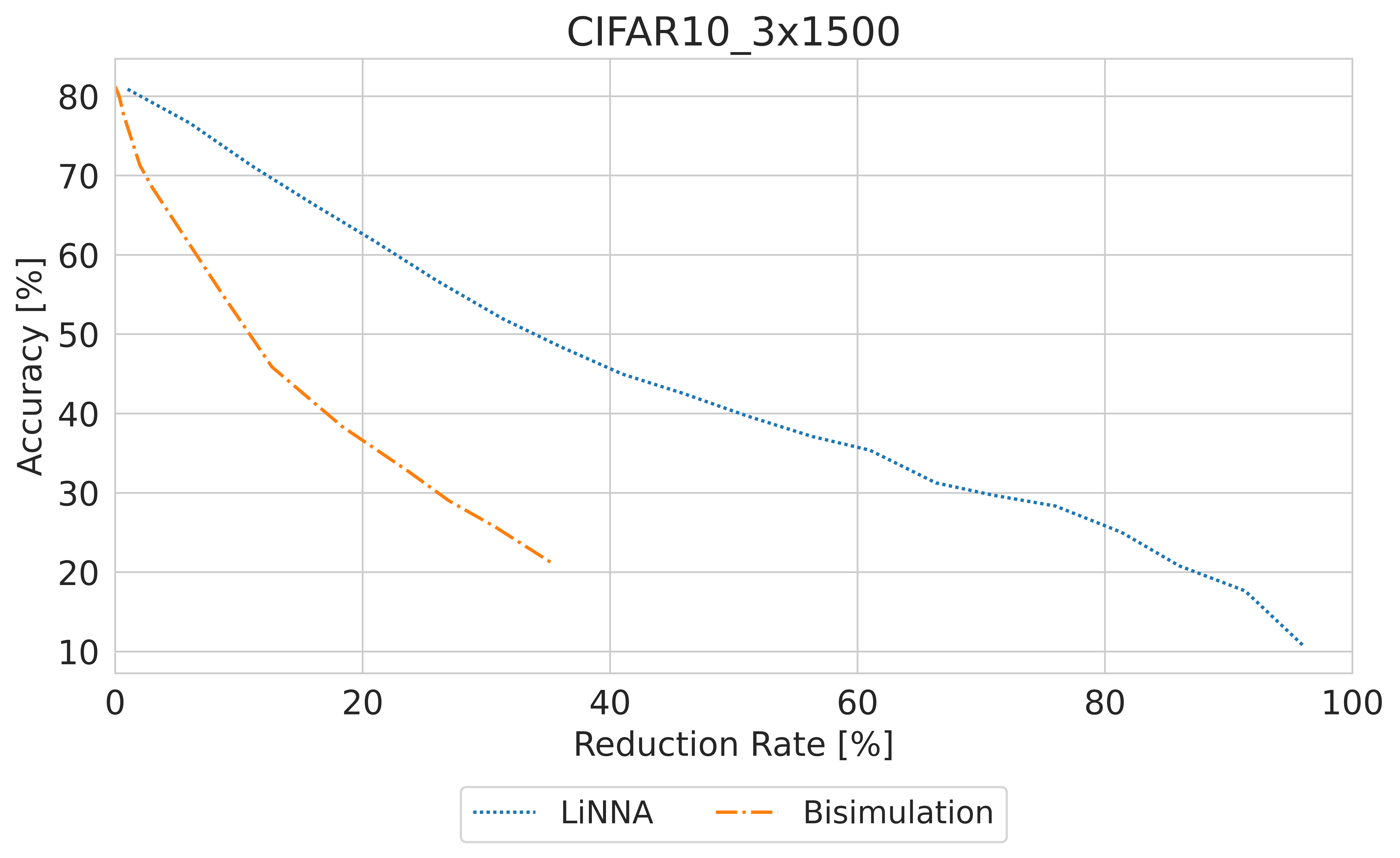}
	\end{subfigure}\hfill
	\begin{subfigure}[]{0.48\textwidth}
		\includegraphics[width=\textwidth]{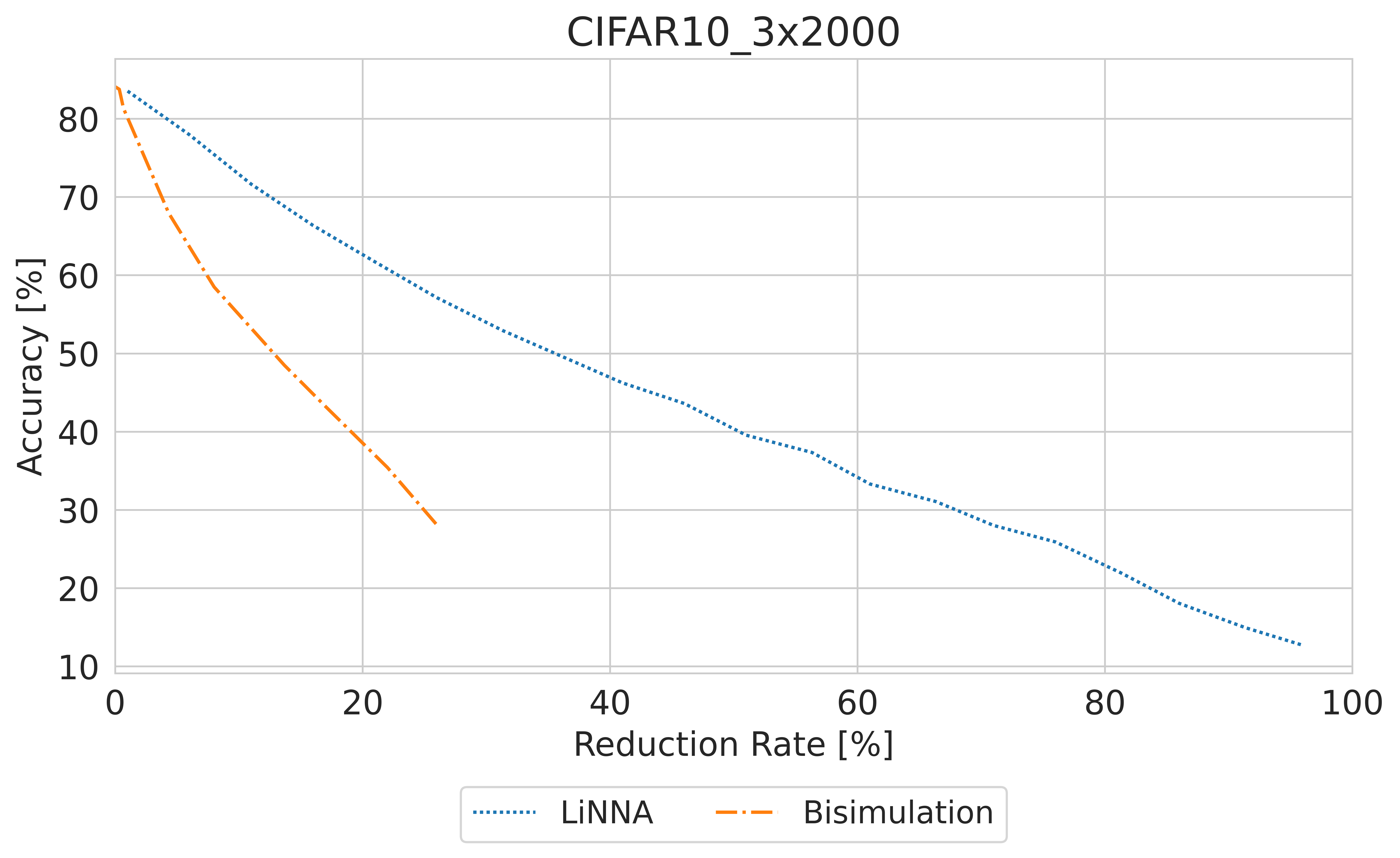}
	\end{subfigure}\hfill
	
	\begin{subfigure}[]{0.48\textwidth}
		\includegraphics[width=\textwidth]{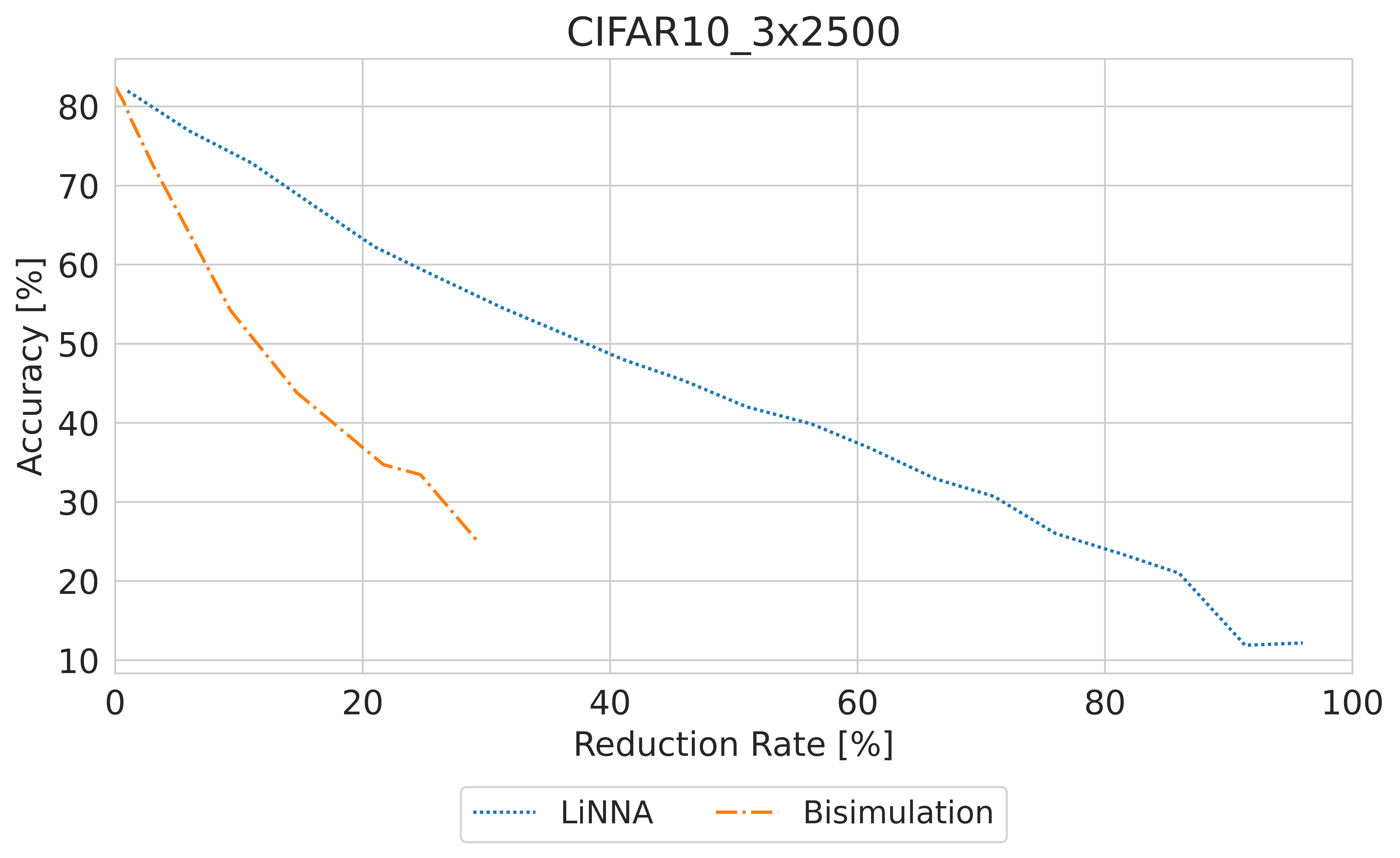}
	\end{subfigure}\hfill
	\begin{subfigure}[]{0.48\textwidth}
		\includegraphics[width=\textwidth]{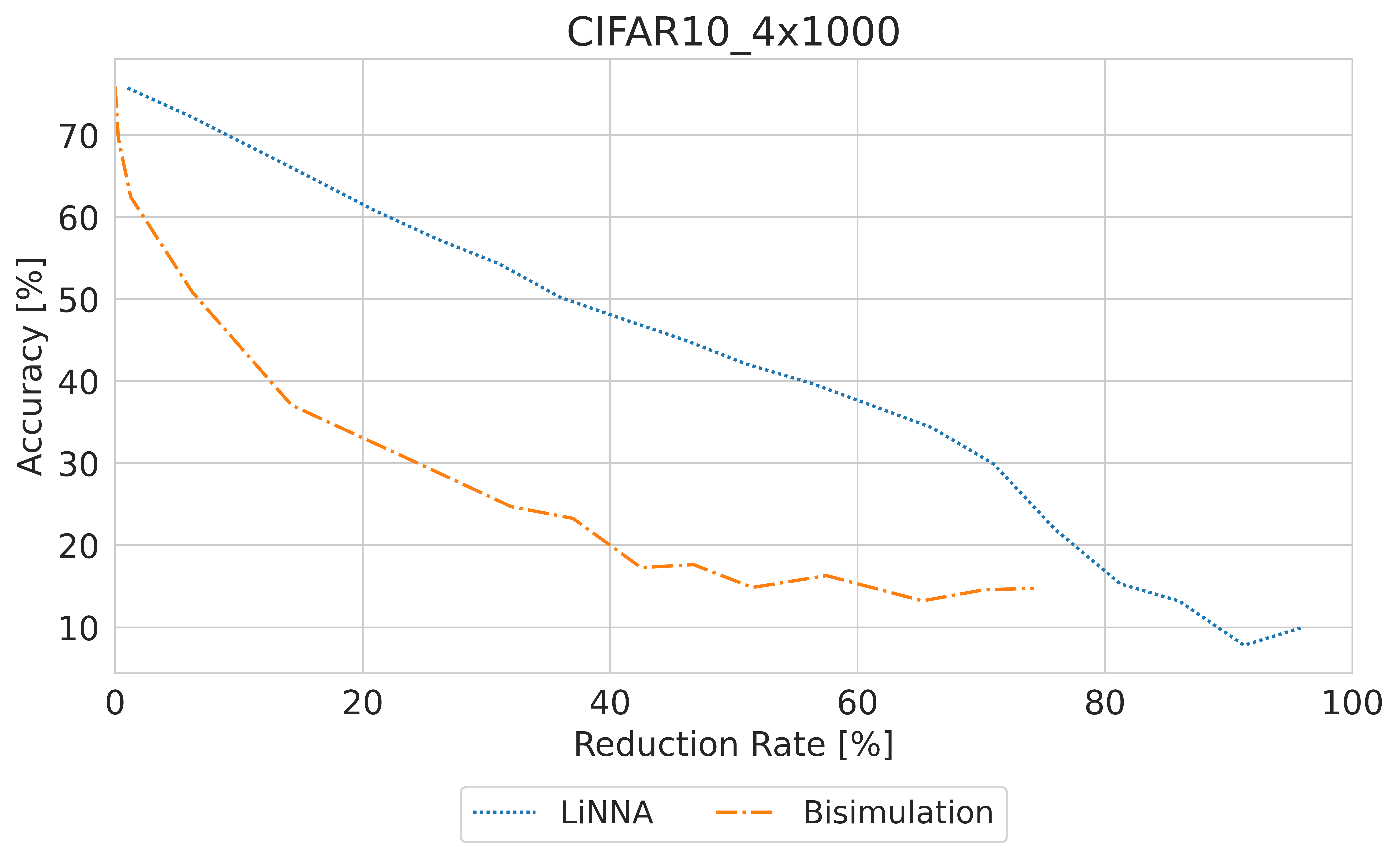}
	\end{subfigure}\hfill
	
	\begin{subfigure}[]{0.48\textwidth}
		\includegraphics[width=\textwidth]{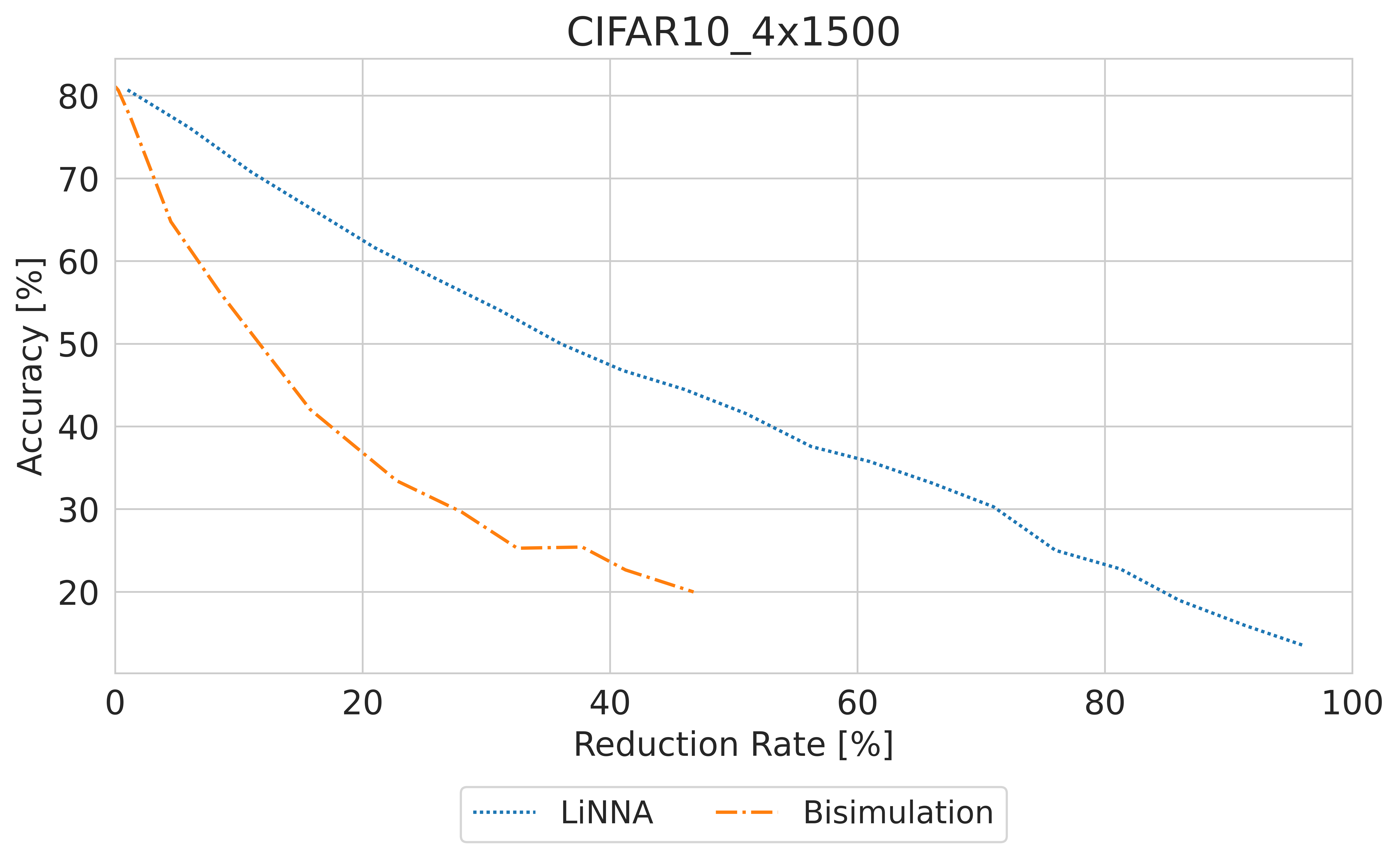}
	\end{subfigure}\hfill
	
	\caption{Comparison of bisimulation and heuristic LiNNA on CIFAR-10}
	\label{fig:cifar-bisim-linna}
\end{figure}

In \cref{fig:cifar-bisim-linna}, we provide a comparison of the bisimulation and LiNNA based on the variance heuristic on some more networks that were trained on CIFAR-10.
Some values for the bisimulation were not possible to provide, because it was difficult to find suitable $\delta$-values for the bisimulation.
Nevertheless, we can still see that LiNNA performs much better than the bisimulation. Its abstractions have always a higher accuracy as the ones resulting from the bisimulation.

\newpage

\section{Supplemental Experiments on Syntactic VS Semantic}\label{sec:synVSsem}
This section contains more plots on the comparison of the syntactic- and semantic-based abstraction, similar to \cref{fig:comp-syn-sem}.
\begin{figure}[!h]
	\centering
	\begin{minipage}[hbt]{0.48\textwidth}
		\includegraphics[width=\textwidth]{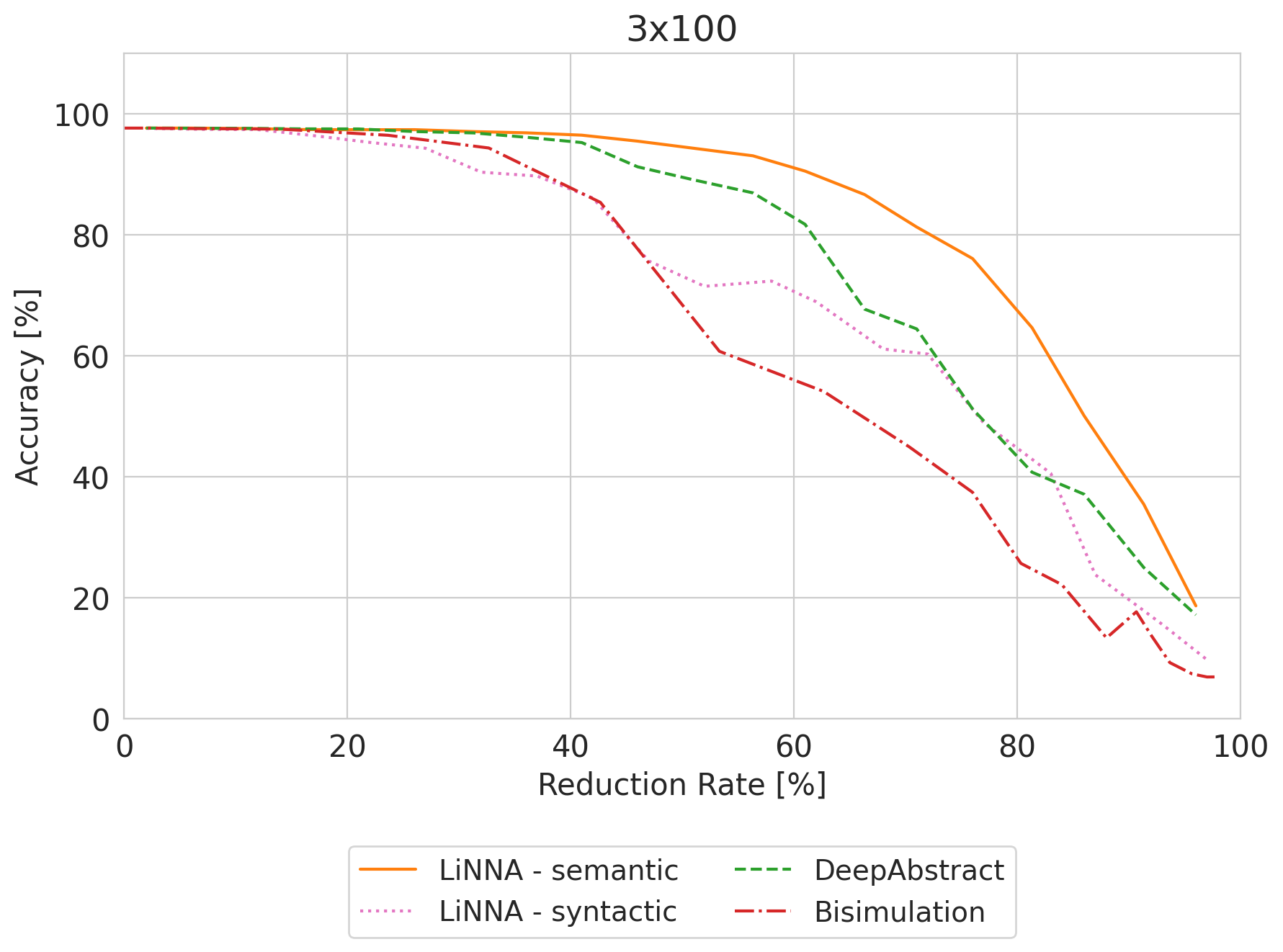}
	\end{minipage}\hfill
	\begin{minipage}[hbt]{0.48\textwidth}
		\includegraphics[width=\textwidth]{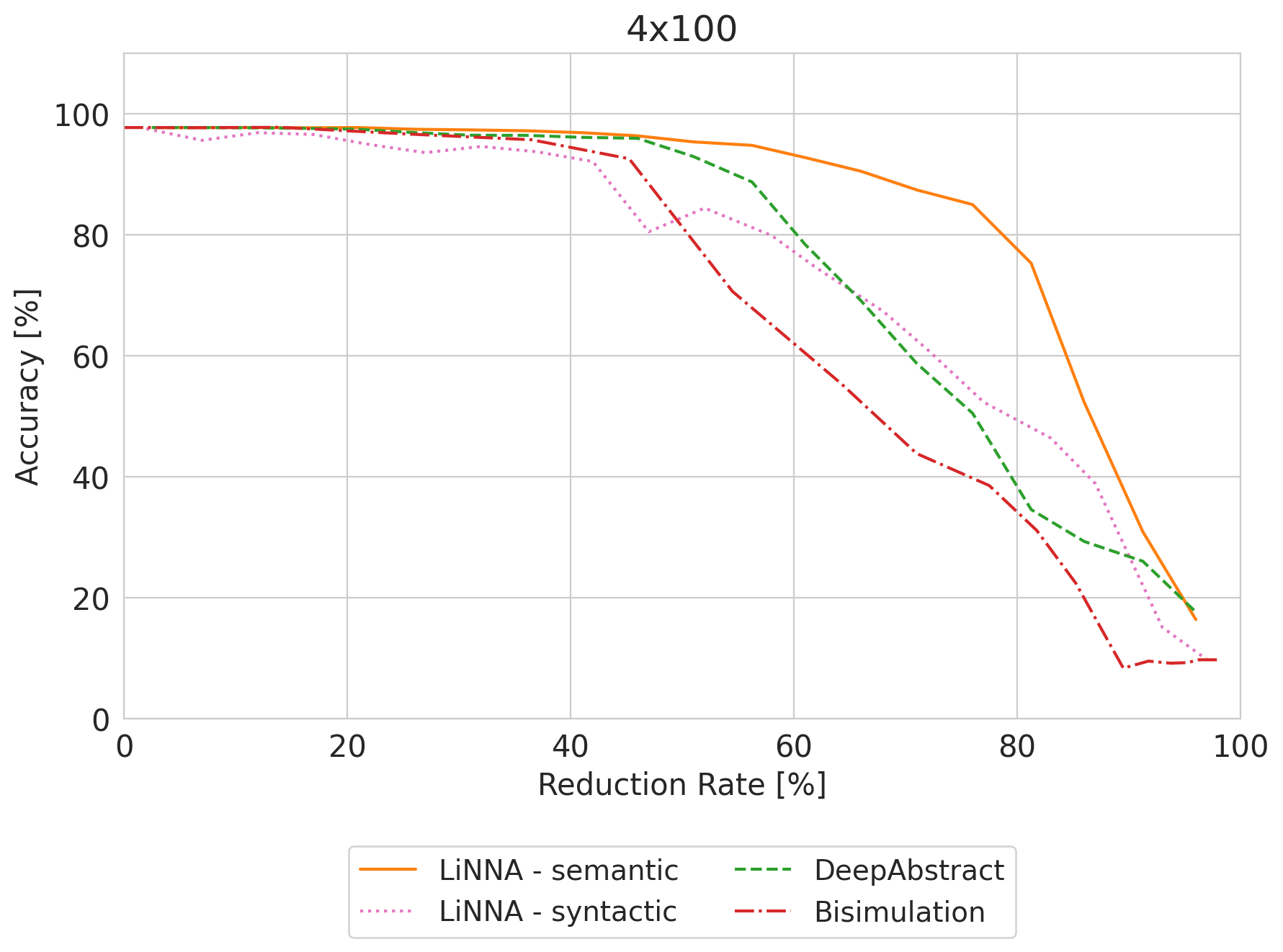}
	\end{minipage}
	
	\begin{minipage}[hbt]{0.48\textwidth}
		\includegraphics[width=\textwidth]{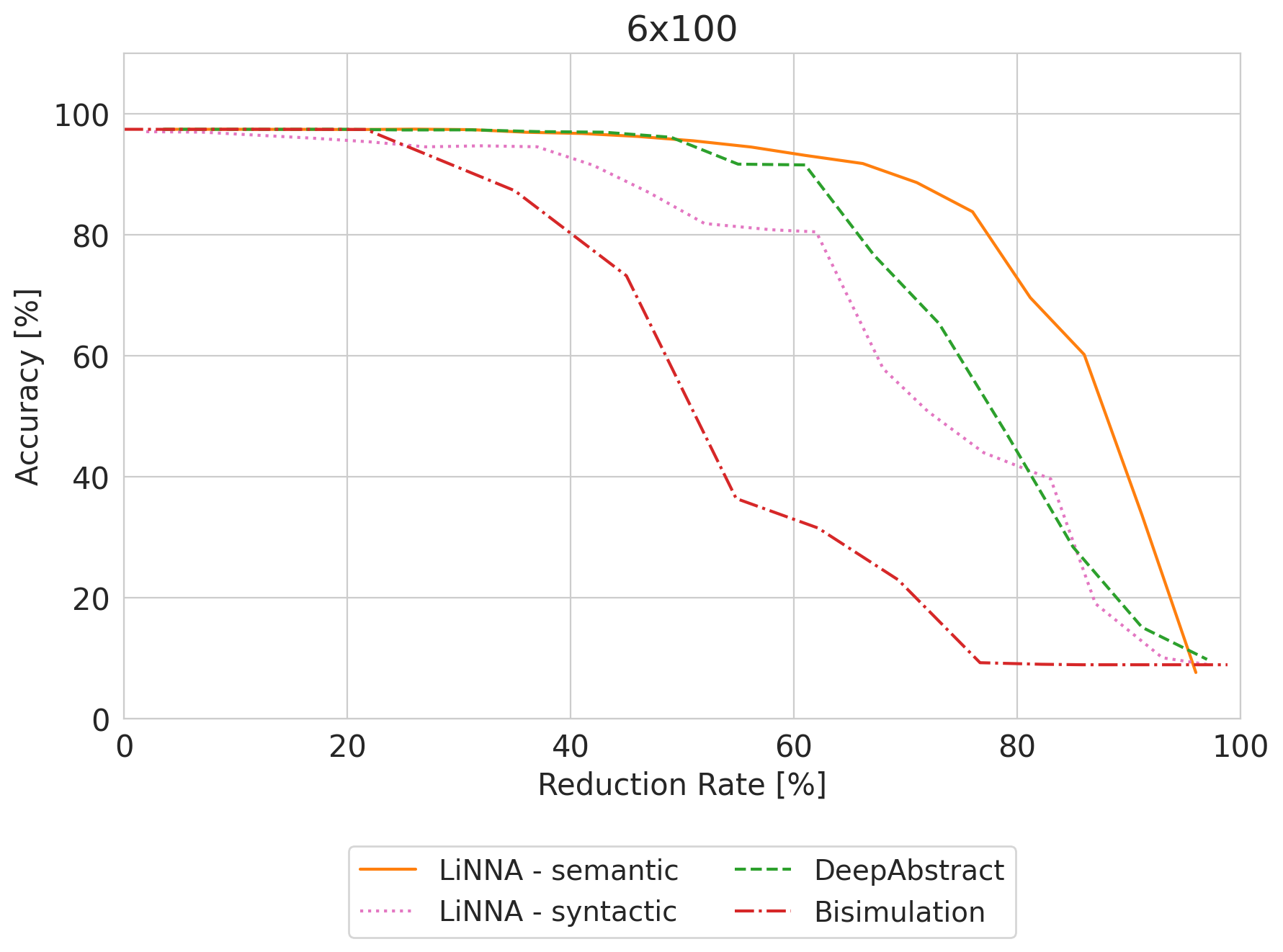}
	\end{minipage}\hfill
	\begin{minipage}[hbt]{0.48\textwidth}
		\includegraphics[width=\textwidth]{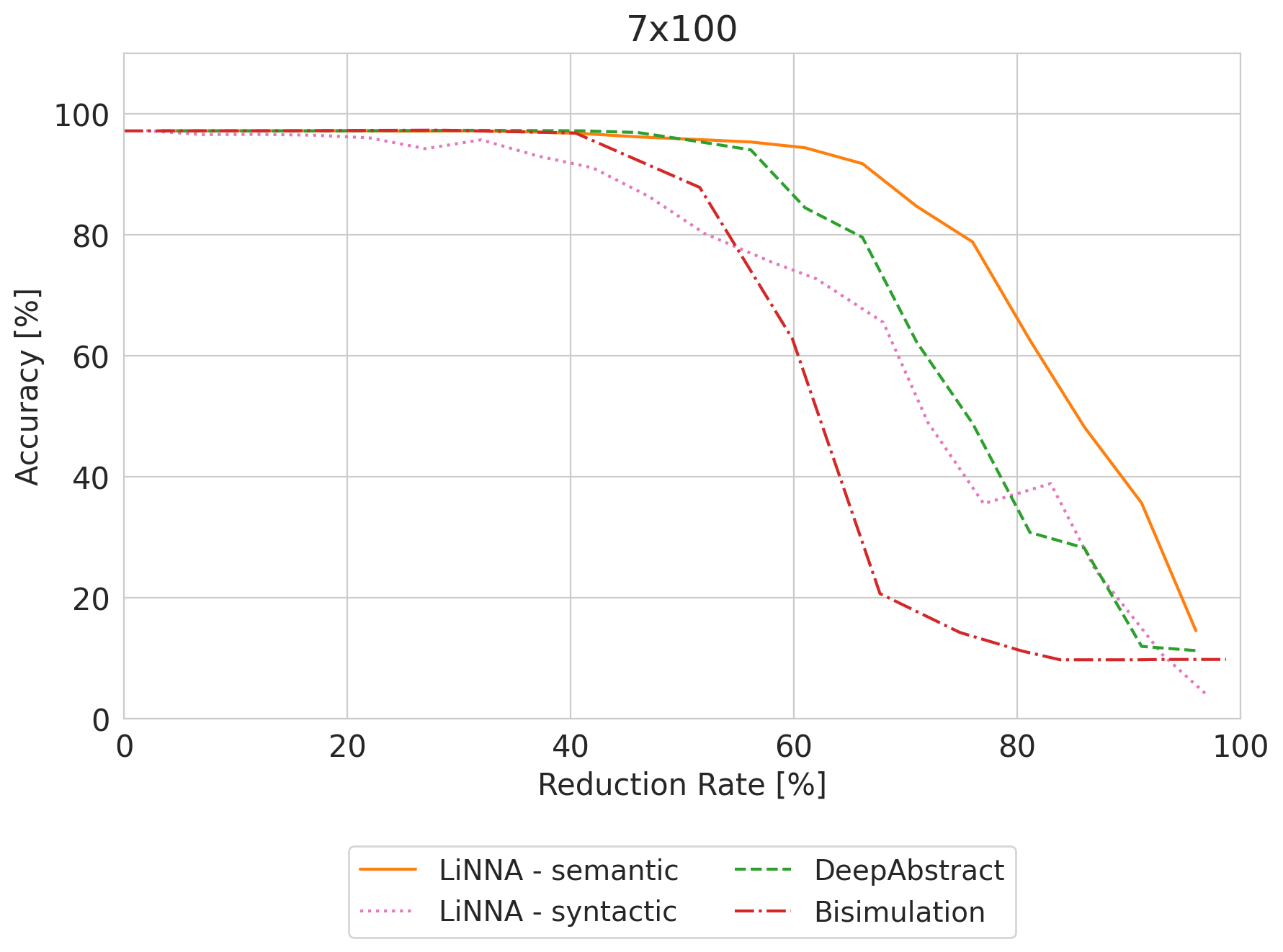}
	\end{minipage}
	
	\begin{minipage}[]{0.48\textwidth}
		\includegraphics[width=\textwidth]{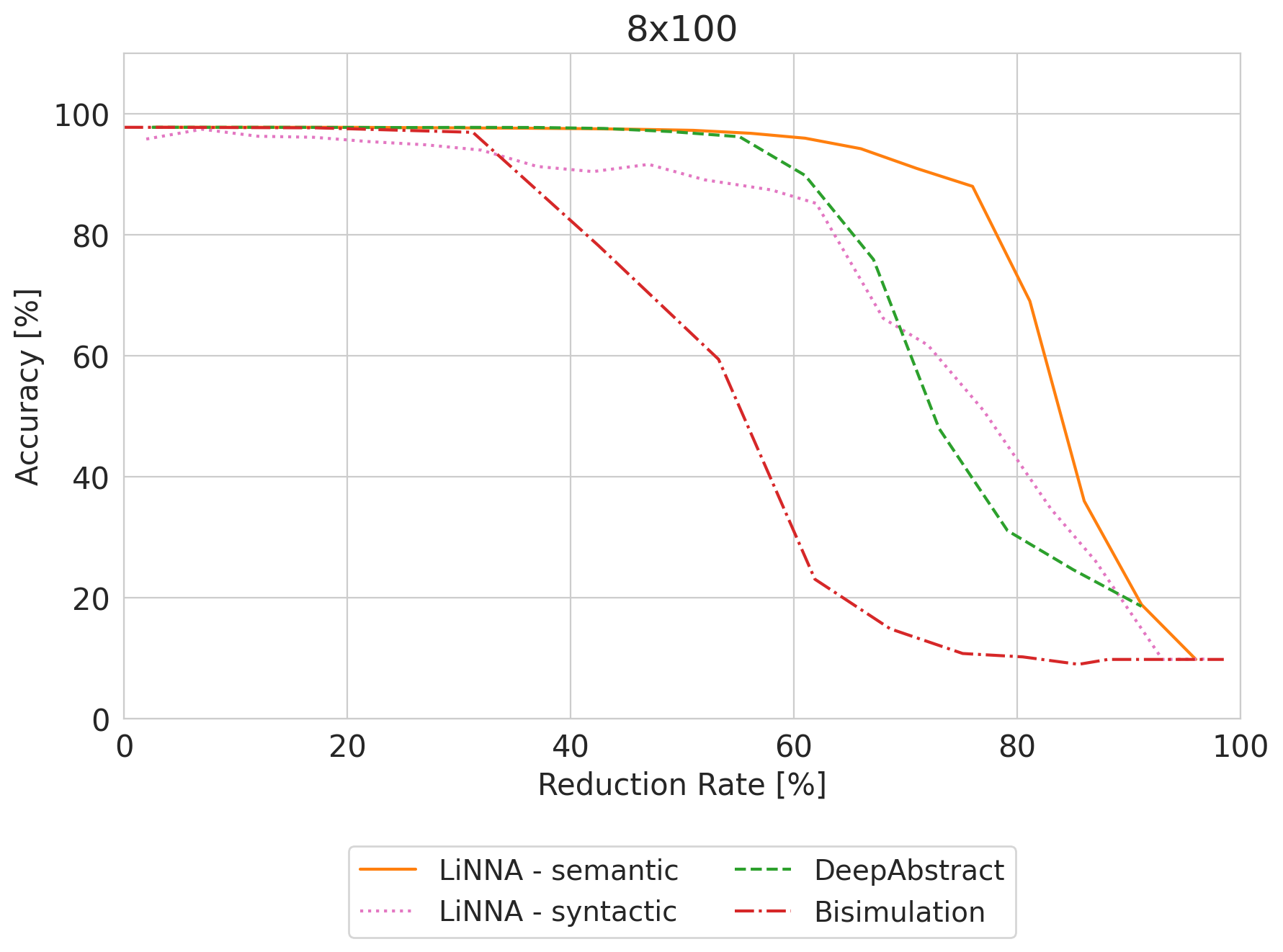}
	\end{minipage}	
	\caption{\emph{Syntactic VS. Semantic} - These plots show the difference of using semantic resp. syntactic information for the abstraction on different networks, trained on MNIST}
	\label{fig:comp-syn-sem2}
\end{figure}
\newpage
\section{Time Comparison of the Different Approaches}\label{sec:time-tables}
This section contains tables with the runtimes of the approaches. 
Each neural network was abstracted with increasing reduction rates.
In the tables, we report the mean, median, minimum and maximum runtimes for the different approaches over all reduction rates.
This should offer the possibility to have a more detailed insight into the computation times.
\begin{table}[!h]
	\centering
	\caption{MNIST3x100 - Comparison of computation times [s]}
	\begin{tabular}{lrrrr}
		\toprule
		type &  median &    mean &    min &      max \\
		\midrule
		LiNNA greedy (OP) &   55.70 &   47.03 &  12.87 &    64.47 \\
		LiNNA heuristic (OP) &    1.48 &    1.47 &   1.32 &    1.53 \\
		LiNNA greedy (LP) & 5807.21 & 5129.88 & 774.40 &  7420.29 \\
		LiNNA heuristic (LP) &   17.50 &   16.49 &   2.42 &    27.90 \\
		Bisimulation &    1.07 &    1.10 &   1.00 &     1.44 \\
		DeepAbstract &  187.03 &  183.30 & 147.54 &   208.97 \\
		
		\bottomrule
	\end{tabular}
	\label{tab:time-3x100}
\end{table}
\begin{table}[!h]
	\centering
	\caption{MNIST4x100 - Comparison of computation times [s]}
	\begin{tabular}{lrrrr}
		\toprule
		type &  median &    mean &    min &      max \\
		\midrule
		LiNNA greedy (OP) &   92.21 &   77.93 &  20.30 &   102.08 \\
		LiNNA heuristic (OP) &    1.81 &   1.84 &   1.70 &   2.05 \\
		Bisimulation &    1.25 &    1.27 &   1.12 &     1.43 \\
		DeepAbstract &  335.73 &  329.88 & 223.52 &   367.45 \\
		\bottomrule
	\end{tabular}
\end{table}
\begin{table}[!h]
	\centering
	\caption{MNIST5x100 - Comparison of computation times [s]}
	\begin{tabular}{lrrrr}
		\toprule
		type &  median &    mean &    min &      max \\
		\midrule
		LiNNA greedy (OP) &  120.41 &  106.81 &  25.76 &   140.67 \\
		LiNNA heuristic (OP) &    2.21 &   2.20 &  1.97 &   2.35 \\	
		Bisimulation &    1.38 &    1.42 &   1.23 &     1.73 \\
		DeepAbstract & 1142.42 & 1269.52 & 541.95 &  2205.43 \\
		\bottomrule
	\end{tabular}
\end{table}
\begin{table}[!th]
	\centering
	\caption{MNIST6x100 - Comparison of computation times [s]}
	\begin{tabular}{lrrrr}
		\toprule
		type &  median &   mean &   min &    max \\
		\midrule
		LiNNA greedy (OP) &  166.15 & 146.38 & 35.42 & 201.71 \\
		LiNNA heuristic (OP) &    2.55 &   2.51 &  2.24 &   2.66 \\
		Bisimulation &    1.61 &   1.66 &  1.41 &   2.18 \\
		DeepAbstract &  756.35 & 695.91 & 64.04 & 832.05 \\
		\bottomrule
	\end{tabular}
\end{table}
\begin{table}[!th]
	\centering
	\caption{MNIST7x100 - Comparison of computation times [s]}
	\begin{tabular}{lrrrr}
		\toprule
		type &  median &    mean &    min &     max \\
		\midrule
		LiNNA greedy (OP) &  229.85 &  199.01 &  48.14 &  257.41 \\
		LiNNA heuristic (OP) &    3.01 &    3.00 &   2.60 &    3.54 \\
		Bisimulation &    1.84 &    1.86 &   1.60 &    2.17 \\
		DeepAbstract & 2418.94 & 2181.16 & 151.61 & 2626.65 \\
		\bottomrule
	\end{tabular}
\end{table}
\vfill
\begin{table}[!th]
	\centering
	\caption{MNIST4x50 - Comparison of computation times [s]}
	\begin{tabular}{lrrrr}
		\toprule
		type &  median &  mean &  min &  max \\
		\midrule
		Bisimulation &    0.86 &  0.87 & 0.82 & 0.93 \\
		LiNNA heuristic (OP) &    1.44 &  1.47 & 1.33 & 2.00 \\
		\bottomrule
	\end{tabular}
\end{table}

\begin{table}[!th]
	\centering
	\caption{MNIST4x150 - Comparison of computation times [s]}
	\begin{tabular}{lrrrr}
		\toprule
		type &  median &  mean &  min &  max \\
		\midrule
		Bisimulation &    2.02 &  2.04 & 1.74 & 2.37 \\
		LiNNA heuristic (OP) &    1.74 &  1.77 & 1.52 & 2.31 \\
		\bottomrule
	\end{tabular}
\end{table}

\begin{table}[!th]
	\centering
	\caption{MNIST4x200 - Comparison of computation times [s]}
	\begin{tabular}{lrrrr}
		\toprule
		type &  median &  mean &  min &  max \\
		\midrule
		Bisimulation &    2.53 &  2.48 & 2.10 & 2.65 \\
		LiNNA heuristic (OP) &    1.95 &  2.05 & 1.79 & 2.61 \\
		\bottomrule
	\end{tabular}
\end{table}

\begin{table}[!th]
	\centering
	\caption{MNIST4x250 - Comparison of computation times [s]}
	\begin{tabular}{lrrrr}
		\toprule
		type &  median &  mean &  min &  max \\
		\midrule
		Bisimulation &    3.12 &  3.04 & 2.56 & 3.28 \\
		LiNNA heuristic (OP) &    2.18 &  2.09 & 1.77 & 2.52 \\
		\bottomrule
	\end{tabular}
\end{table}

\begin{table}[!th]
	\centering
	\caption{MNIST4x300 - Comparison of computation times [s]}
	\begin{tabular}{lrrrr}
		\toprule
		type &  median &  mean &  min &  max \\
		\midrule
		Bisimulation &    4.08 &  4.05 & 3.48 & 4.72 \\
		LiNNA heuristic (OP) &    2.07 &  2.09 & 1.70 & 2.38 \\
		\bottomrule
	\end{tabular}
\end{table}

\begin{table}[!th]
	\centering
	\caption{MNIST4x350 - Comparison of computation times [s]}
	\begin{tabular}{lrrrr}
		\toprule
		type &  median &  mean &  min &  max \\
		\midrule
		Bisimulation &    5.26 &  5.21 & 4.52 & 5.62 \\
		LiNNA heuristic (OP) &    2.56 &  2.59 & 1.90 & 3.22 \\
		\bottomrule
	\end{tabular}
\end{table}

\begin{table}[!th]
	\centering
	\caption{MNIST4x400 - Comparison of computation times [s]}
	\begin{tabular}{lrrrr}
		\toprule
		type &  median &  mean &  min &  max \\
		\midrule
		Bisimulation &    5.84 &  5.91 & 5.31 & 6.54 \\
		LiNNA heuristic (OP) &    2.55 &  2.46 & 1.62 & 2.89 \\
		\bottomrule
	\end{tabular}
\end{table}

\begin{table}[!th]
	\centering
	\caption{MNIST4x450 - Comparison of computation times [s]}
	\begin{tabular}{lrrrr}
		\toprule
		type &  median &  mean &  min &  max \\
		\midrule
		Bisimulation &    7.38 &  7.30 & 6.51 & 8.11 \\
		LiNNA heuristic (OP) &    2.79 &  2.66 & 1.72 & 2.90 \\
		\bottomrule
	\end{tabular}
\end{table}

\begin{table}[!th]
	\centering
	\caption{MNIST4x500 - Comparison of computation times [s]}
	\begin{tabular}{lrrrr}
		\toprule
		type &  median &  mean &  min &  max \\
		\midrule
		Bisimulation &    8.29 &  8.52 & 7.39 & 9.39 \\
		LiNNA heuristic (OP) &    3.10 &  2.95 & 1.81 & 3.28 \\
		\bottomrule
	\end{tabular}
\end{table}

\section{Details on Refinement}\label{sec:appendix-refinement}
This section contains some more supplementary material for the refinement.
We have a table to give a more detailed insight into the computation time. 
Note, however, that there appeared to be a malfunction in the difference-based approach on the 3x100 network.
The high maximum number most likely occurred due to some issues with the machine it was run on.
\begin{table}[!th]
	\centering
	\caption{Refinement - Comparison of computation times [s]}
	\label{tab:refinement-time-comp}
	\begin{tabular}{ll|rrrr}
		\toprule
		NN &      method &  mean &  median &  min &      max \\
		\midrule
		3x100 &  Difference &  0.77 &    0.02 & 0.02 &  7901.53 \\
		&    Gradient &  0.45 &    0.05 & 0.04 &    11.27 \\
		&   Lookahead &  5.19 &    1.11 & 0.16 &  8724.71 \\\midrule
		4x100 &  Difference &  0.20 &    0.02 & 0.02 &     8.13 \\
		&    Gradient &  0.56 &    0.07 & 0.02 &    13.45 \\
		&   Lookahead &  6.85 &    1.53 & 0.22 & 10719.30 \\\midrule
		5x100 &  Difference &  0.21 &    0.03 & 0.02 &     9.87 \\
		&    Gradient &  1.93 &    0.07 & 0.02 & 11499.10 \\
		&   Lookahead &  1.32 &    1.94 & 0.30 &     9.31 \\\midrule
		6x100 &  Difference &  0.24 &    0.03 & 0.03 &     7.14 \\
		&    Gradient &  2.91 &    0.08 & 0.02 & 16172.10 \\
		&   Lookahead &  1.56 &    2.37 & 0.34 &     5.45 \\
		\bottomrule
	\end{tabular}
	
\end{table}

Additionally, we have the plot from \cref{fig:barplot-refinement} with two more networks to show more on the evolution of the refinement approaches.

\begin{figure}[h!]
	\includegraphics[width=\textwidth]{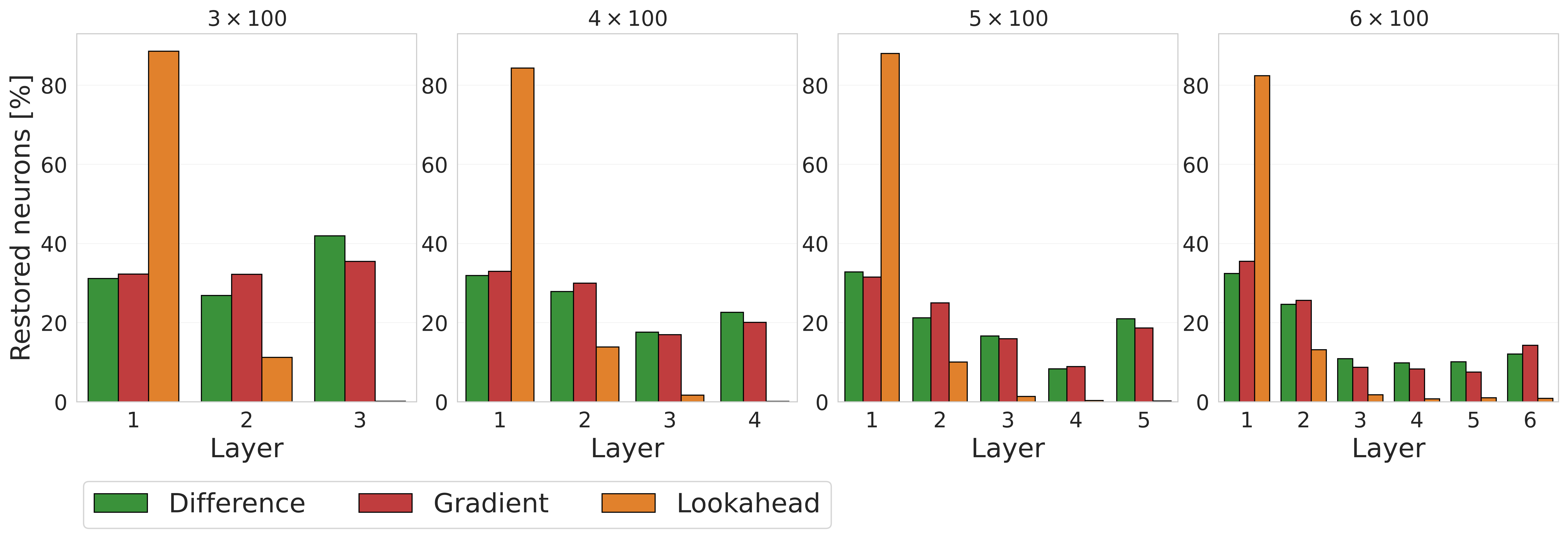}
	\caption{We considered abstractions that were obtained with a $50\%$ reduction rate and fixed $1000$ counterexamples. The plots depict the percentage of restored neurons in the layers of the different MNIST networks.}
	\label{fig:full-refinement-barplot}
\end{figure}

\section{Experiments on the Error}\label{sec:appendix-error}
\begin{figure}[t!]
	\centering
	\begin{minipage}[t]{0.3\textwidth}
		\centering
		\includegraphics[width=0.9\textwidth]{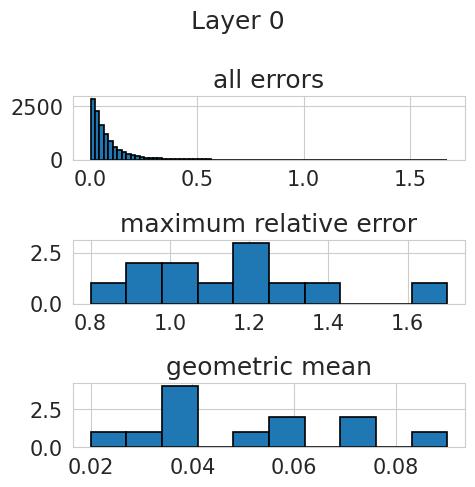}
	\end{minipage}
	\begin{minipage}[t]{0.3\textwidth}
		\centering
		\includegraphics[width=0.9\textwidth]{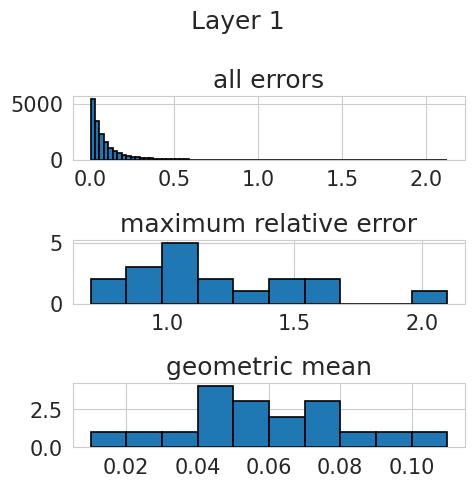}
	\end{minipage}
	\begin{minipage}[t]{0.3\textwidth}
		\centering
		\includegraphics[width=0.9\textwidth]{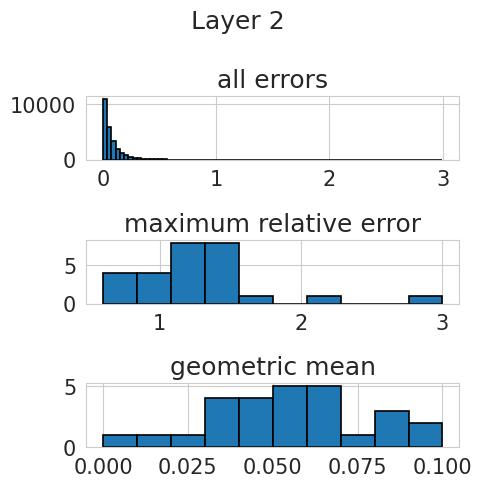}
	\end{minipage}
	
	\begin{minipage}[t]{0.3\textwidth}
		\centering
		\includegraphics[width=0.9\textwidth]{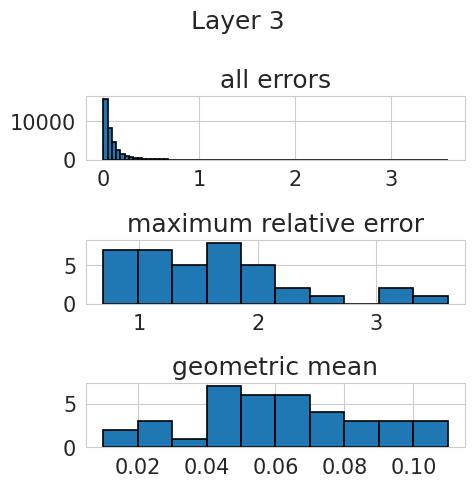}
	\end{minipage}
	\begin{minipage}[t]{0.3\textwidth}
		\centering
		\includegraphics[width=0.9\textwidth]{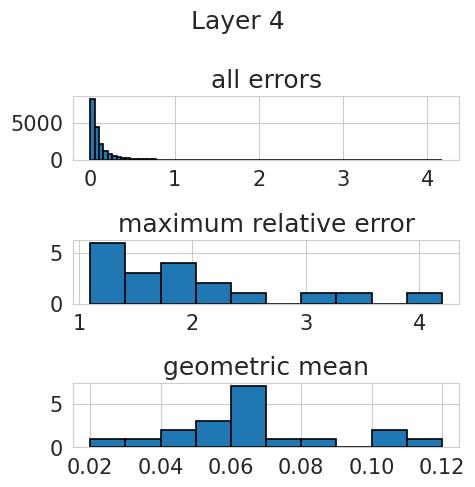}
	\end{minipage}
	\caption{Histograms of the relative error of a MNIST5x100 network that was reduced by 30\%. The first row depicts a histogram over all relative errors for all replaced neurons on 1000 images of the test set. 
		The second row shows the maximum relative error of each neuron that occurred for some input from the test set. The las row plots the geometric mean of the relative error of each neuron over 100 images of the test set.}
	\label{fig:error5}
\end{figure}

\begin{figure}[t!]
	\centering
	\begin{minipage}[t]{0.3\textwidth}
		\centering
		\includegraphics[width=0.9\textwidth]{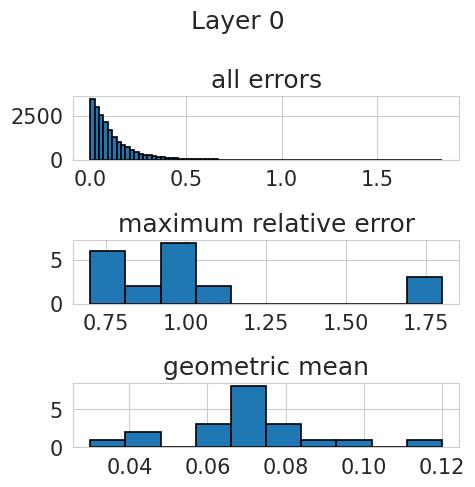}
	\end{minipage}
	\begin{minipage}[t]{0.3\textwidth}
		\centering
		\includegraphics[width=0.9\textwidth]{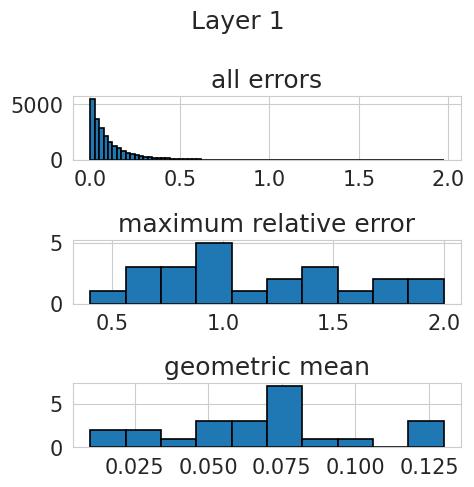}
	\end{minipage}
	\begin{minipage}[t]{0.3\textwidth}
		\centering
		\includegraphics[width=0.9\textwidth]{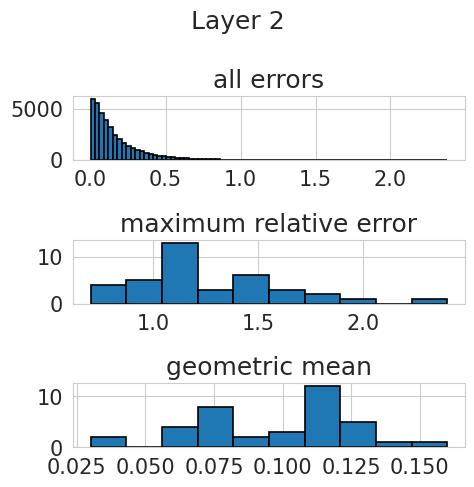}
	\end{minipage}
	\caption{Histograms of the relative error of a MNIST3x100 network that was reduced by 30\%. The first row depicts a histogram over all relative errors for all replaced neurons on 1000 images of the test set. 
		The second row shows the maximum relative error of each neuron that occurred for some input from the test set. The las row plots the geometric mean of the relative error of each neuron over 100 images of the test set.}
	\label{fig:error3}
\end{figure}
In addition to the plots that we have already seen in \cref{fig:error}, we have a plot the shows the histogram over all errors of all replaced neurons in a layer. 
The values are usually very close to 0.

\begin{figure}[t!]
	\centering
	\begin{subfigure}[t]{\textwidth}
		\centering
		\includegraphics[width=\textwidth]{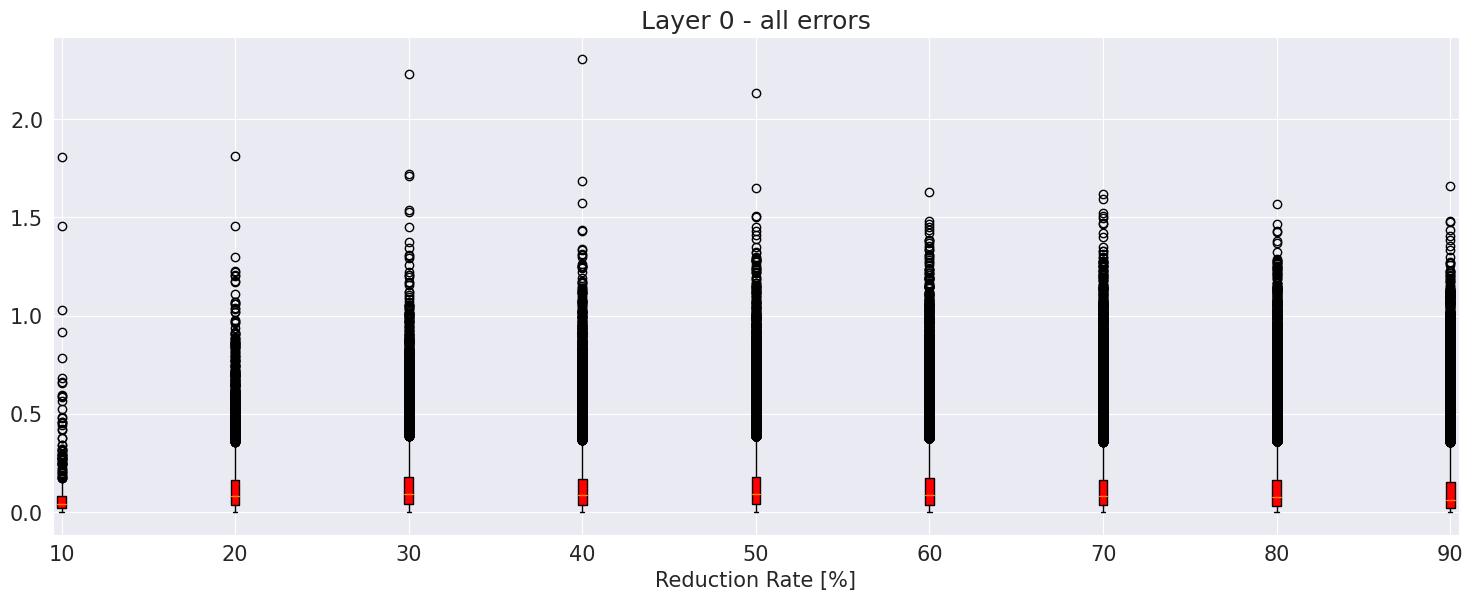}
		\caption{\color{white}.}
	\end{subfigure}\vspace{0.5cm}
	
	\begin{subfigure}[t]{\textwidth}
		\centering
		\includegraphics[width=\textwidth]{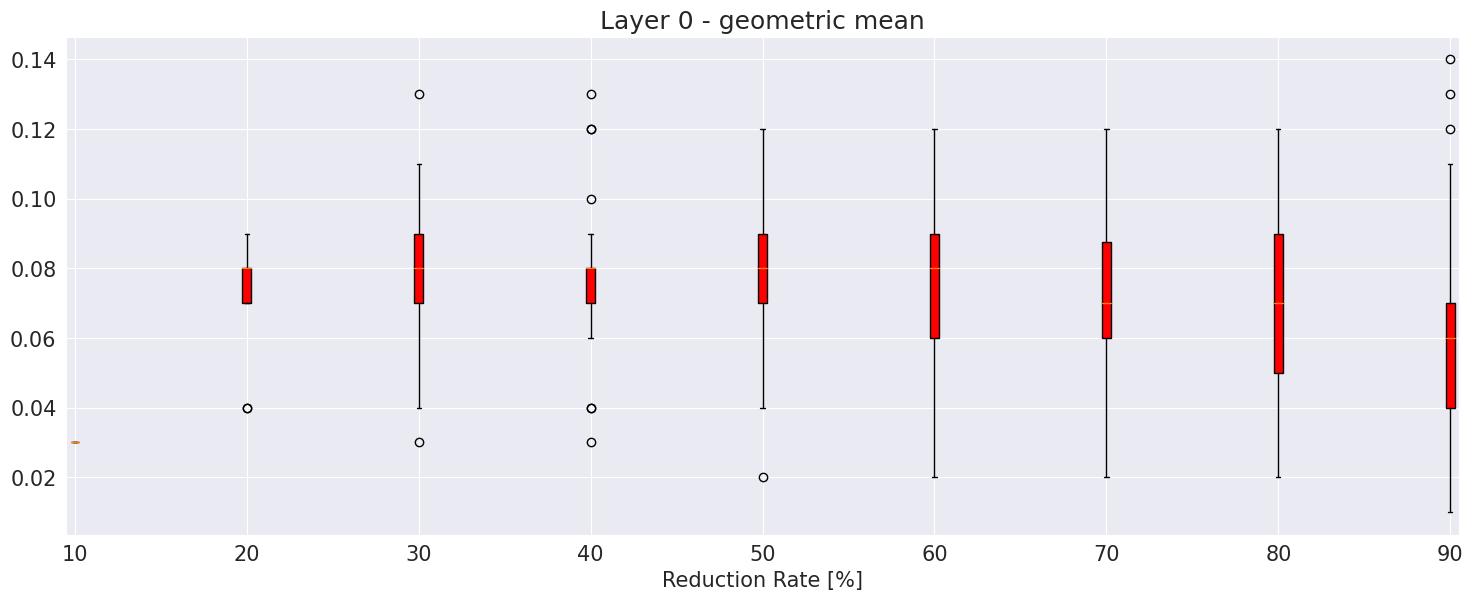}
		\caption{\color{white}.}
	\end{subfigure}\vspace{0.5cm}
	
	\begin{subfigure}[t]{\textwidth}
		\centering
		\includegraphics[width=\textwidth]{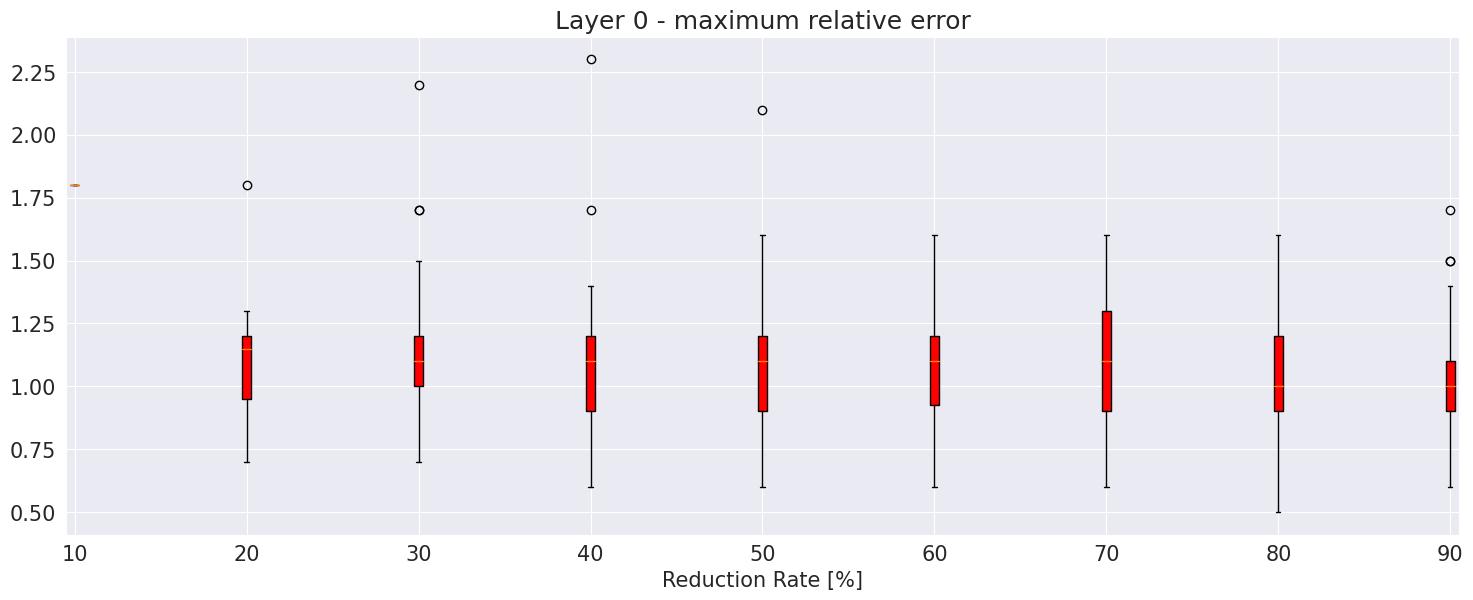}
		\caption{\color{white}.}
	\end{subfigure}
	\caption{Evolution of the relative error for different reduction rates. The network is MNIST3x100 and the zeroth layer. We see for each reduction rate in [10-90] a boxplot for a) all errors, b) the geometric mean, c) the maximum error.}
	\label{fig:error4-mn3100-l0}
\end{figure}

Additionally, we show how the error evolves when reducing the network more. 
To this end, we have in \cref{fig:error4-mn3100-l0}, boxplots for a) all appearing relative errors, b) the maximum relative error, c) the geometric mean of the relative error on a MNIST network with 3x100 neurons for different reduction rates.
We can see that the error looks most stable in the first layer and least stable in the last layer.
However, even for a reduction of 90\%, the geometric mean of the relative error is still below 0.3 for all cases but one. This could indicate that the number of cases where the abstraction fails increase only slightly.
The maximum relative error seems to increase steadily, which could mean that whenever the abstraction fails, it fails even more. 

\begin{figure}[t!]
	\centering
	\begin{subfigure}[t]{\textwidth}
		\centering
		\includegraphics[width=\textwidth]{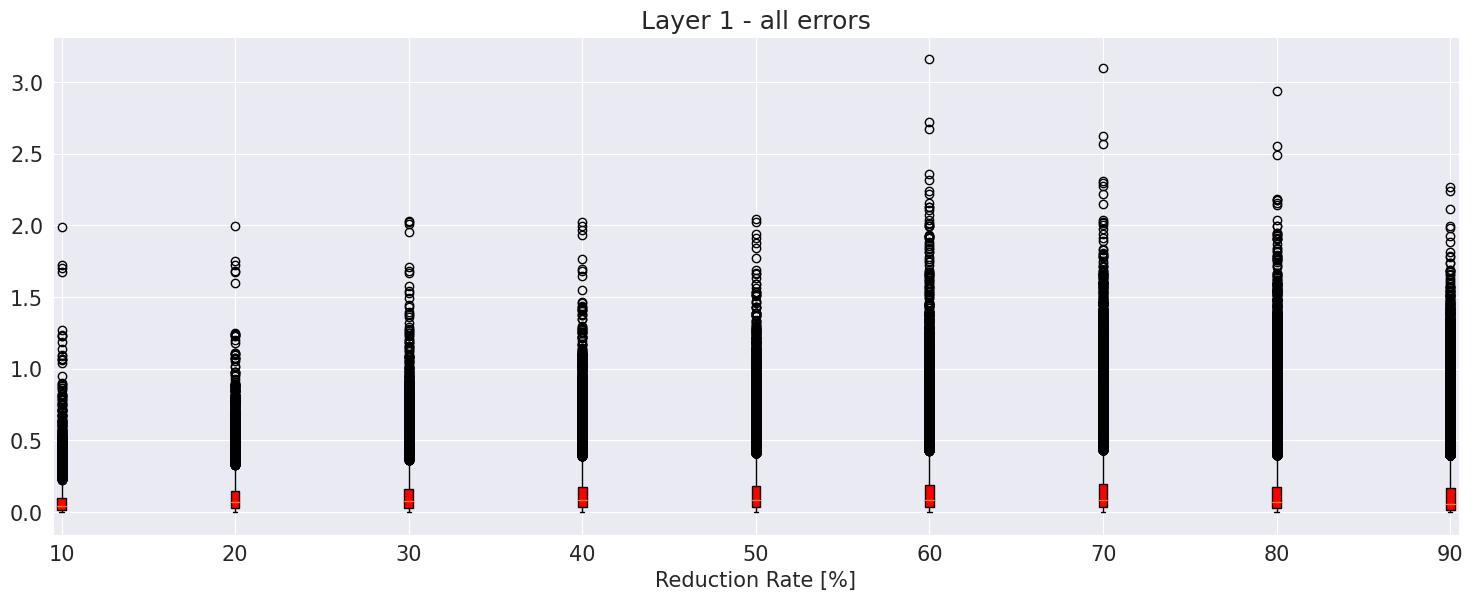}
		\caption{\color{white}.}
	\end{subfigure}\vspace{0.5cm}
	
	\begin{subfigure}[t]{\textwidth}
		\centering
		\includegraphics[width=\textwidth]{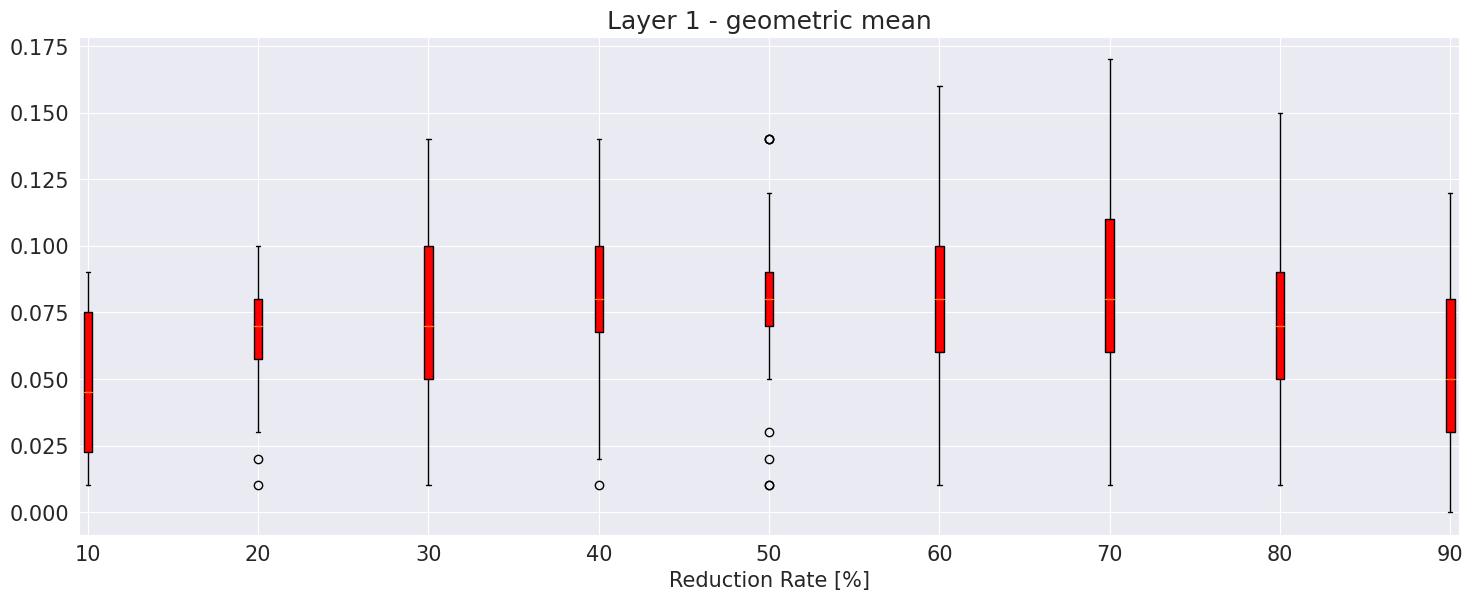}
		\caption{\color{white}.}
	\end{subfigure}\vspace{0.5cm}
	
	\begin{subfigure}[t]{\textwidth}
		\centering
		\includegraphics[width=\textwidth]{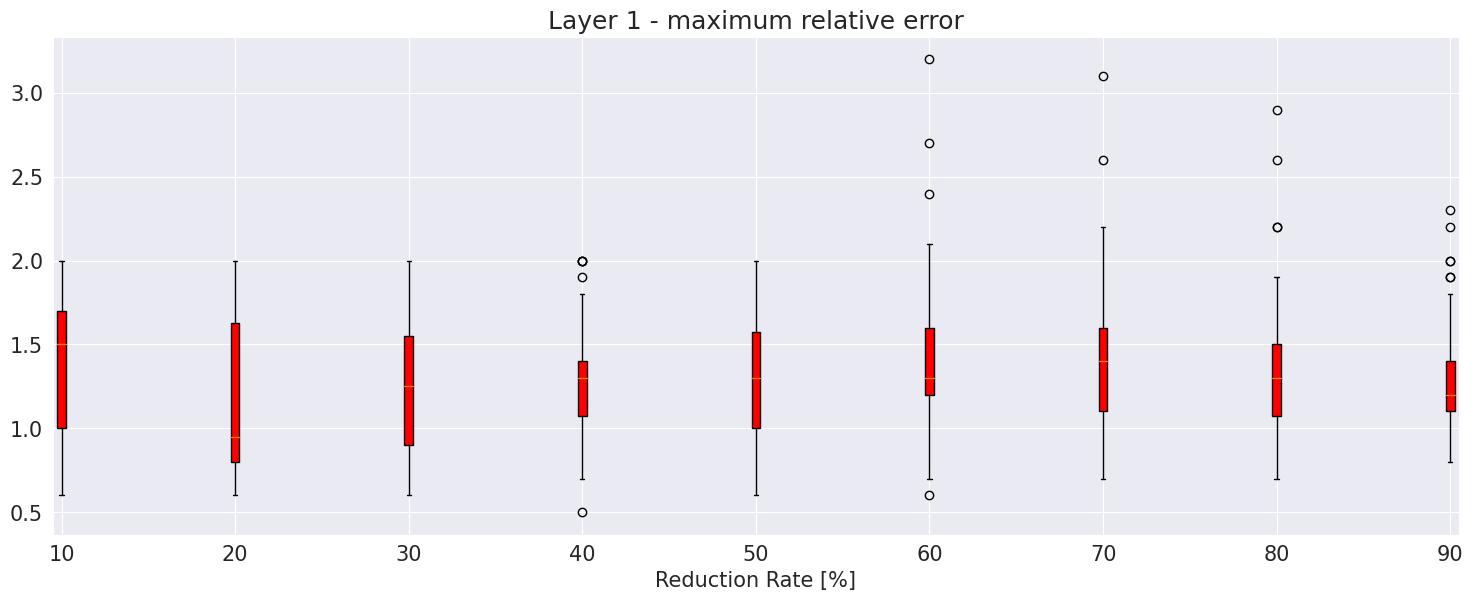}
		\caption{\color{white}.}
	\end{subfigure}
	\caption{Evolution of the relative error for different reduction rates. The network is MNIST3x100 and the first layer. We see for each reduction rate in [10-90] a boxplot for a) all errors, b) the geometric mean, c) the maximum error.}
	\label{fig:error4-mn3100-l1}
\end{figure}

\begin{figure}[t!]
	\centering
	\begin{subfigure}[t]{\textwidth}
		\centering
		\includegraphics[width=\textwidth]{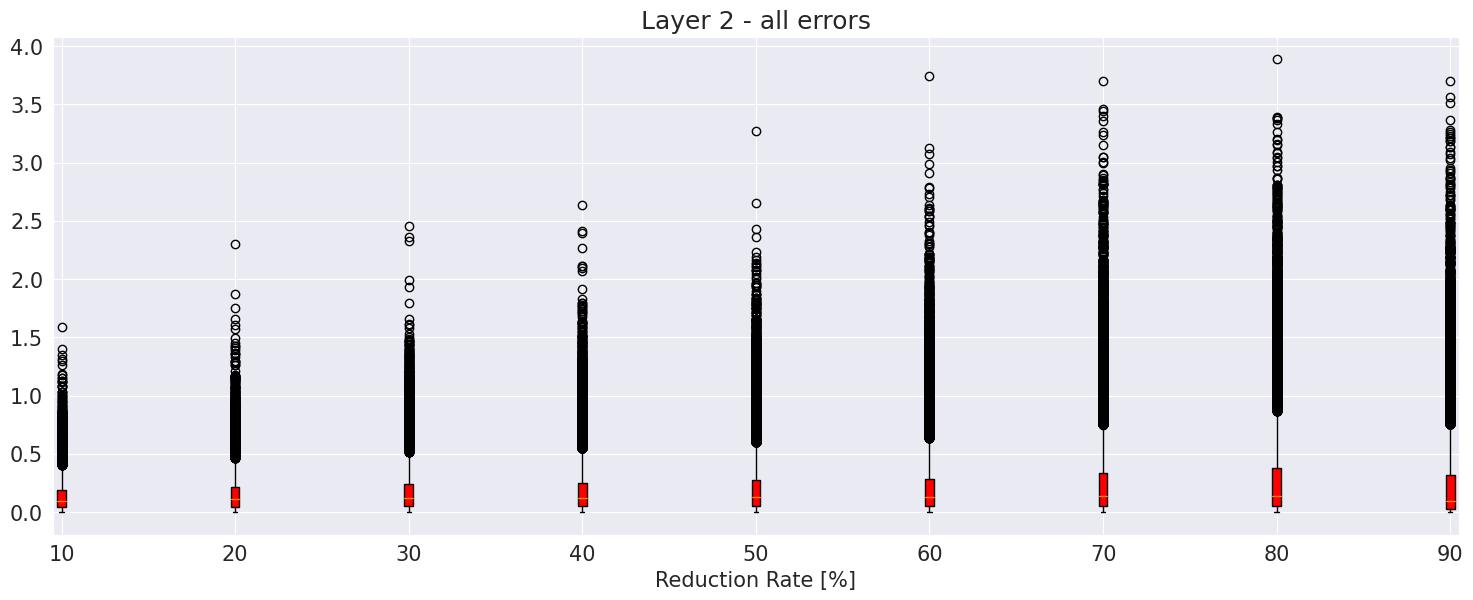}
		\caption{\color{white}.}
	\end{subfigure}\vspace{0.5cm}
	
	\begin{subfigure}[t]{\textwidth}
		\centering
		\includegraphics[width=\textwidth]{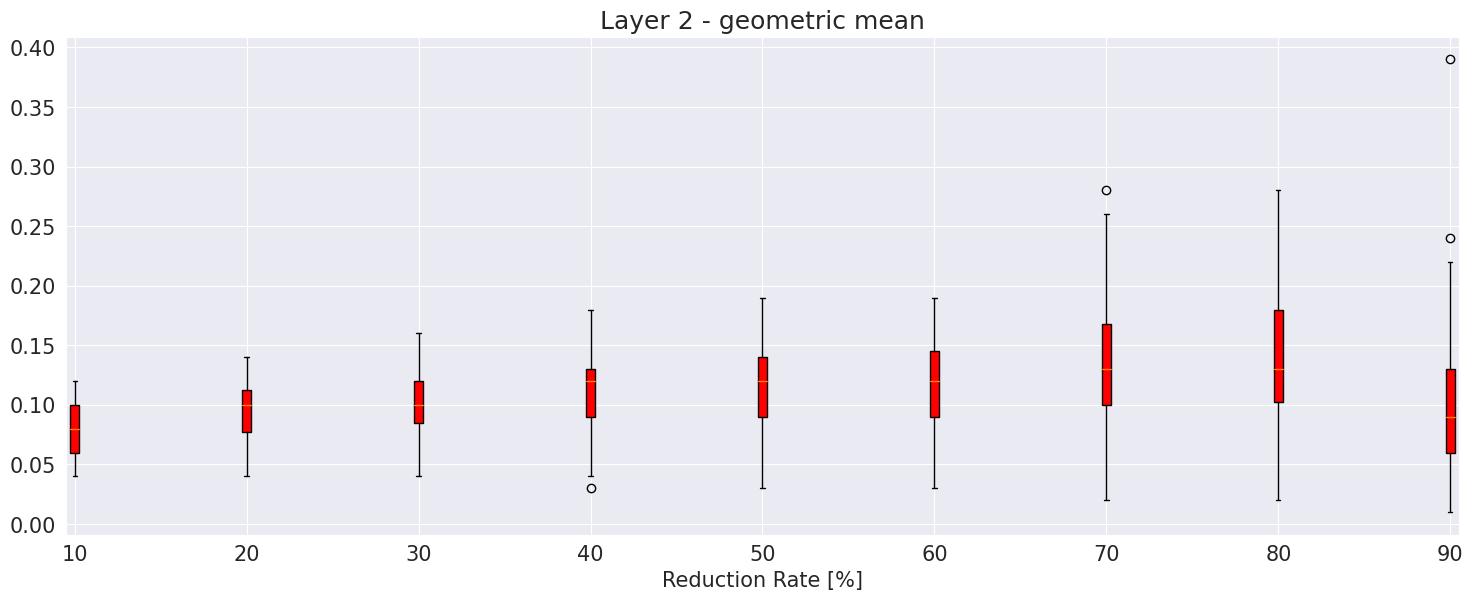}
		\caption{\color{white}.}
	\end{subfigure}\vspace{0.5cm}
	
	\begin{subfigure}[t]{\textwidth}
		\centering
		\includegraphics[width=\textwidth]{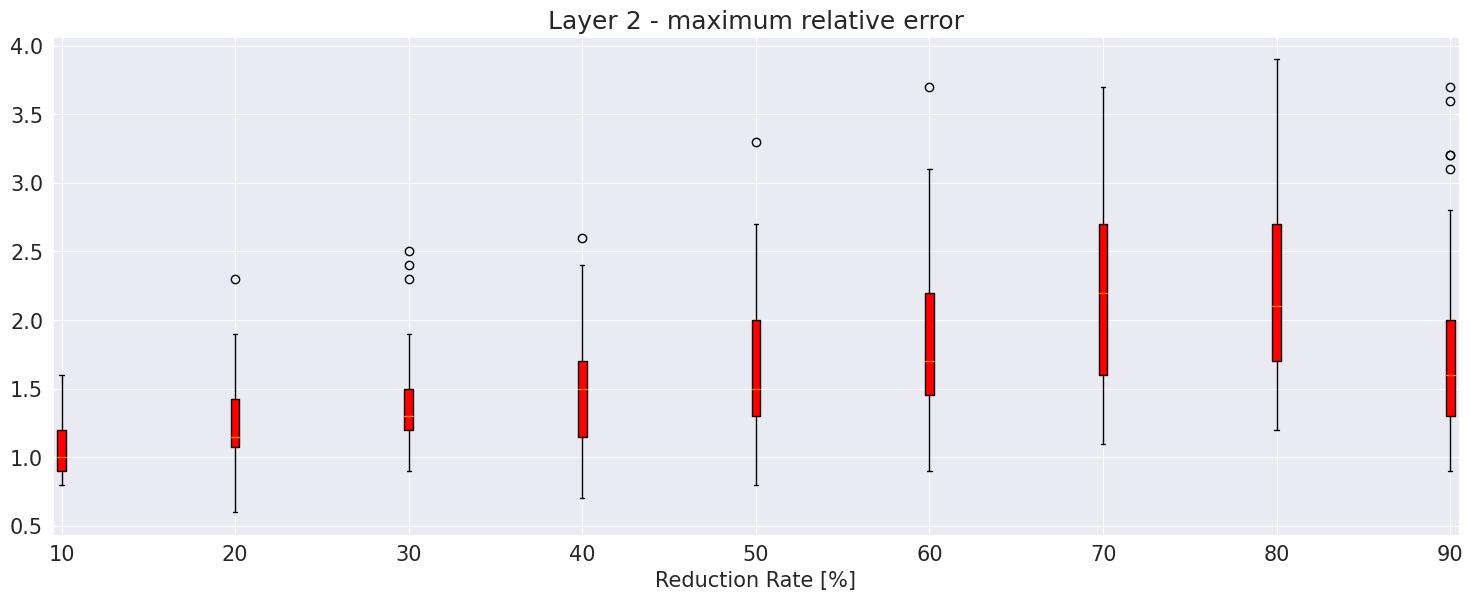}
		\caption{\color{white}.}
	\end{subfigure}
	\caption{Evolution of the relative error for different reduction rates. The network is MNIST3x100 and the second layer. We see for each reduction rate in [10-90] a boxplot for a) all errors, b) the geometric mean, c) the maximum error.}
	\label{fig:error4-mn3100-l2}
\end{figure}

\end{document}